\RequirePackage{lineno}
\documentclass[twocolumn,showpacs,preprintnumbers,amsmath,amssymb,superscriptaddress]{revtex4}

\usepackage{graphicx}  
\usepackage{dcolumn}   
\usepackage{bm}        
\usepackage{axodraw}
\usepackage{setspace}
\usepackage{comma}

\newcommand{\hqetlbsemic}{\ensuremath{7.6}}
\newcommand{\hqetlbhadc}{\ensuremath{0.54}}
\newcommand{\hqetlbhade}{\ensuremath{0.18}}
\newcommand{\chenglbhadc}{\ensuremath{0.50}}
\newcommand{\chenglbhade}{\ensuremath{0.17}}

\newcommand{\brlbsemic}{\ensuremath{7.3}}
\newcommand{\brlbsemie}{\ensuremath{1.4}}
\newcommand{\brlbsemistate}{\ensuremath{0.8}}
\newcommand{\brlbsemisyste}{\ensuremath{1.1}}

\newcommand{\bmesondecaywidthc}{\ensuremath{7.33}}
\newcommand{\bmesondecaywidthe}{\ensuremath{0.16}}
\newcommand{\lifetimeRc}{\ensuremath{0.99}}
\newcommand{\lifetimeRe}{\ensuremath{0.10}}

\newcommand{\rwendyRc}{\ensuremath{1.09}}
\newcommand{\rwendyRe}{\ensuremath{0.21}}

\newcommand{\rxsecc}{\ensuremath{0.61}}
\newcommand{\rxsececdf}{\ensuremath{0.22}}
\newcommand{\rxsecedelphi}{\ensuremath{0.12}}
\newcommand{\rxsecept}{\ensuremath{{+0.21 \atop -0.12}}}

\newcommand{\brdhadc}{\ensuremath{0.268}}
\newcommand{\brdhade}{\ensuremath{0.013}}

\newcommand{\brlbhadc}{\ensuremath{0.36}}
\newcommand{\brlbhade}{\ensuremath{{+0.23 \atop -0.15}}}
\newcommand{\brlbhadhqee}{\ensuremath{0.07}}
\newcommand{\brlbhadpte}{\ensuremath{{+0.05 \atop -0.07}}}
\newcommand{\brlbhadecombine}{\ensuremath{{+(0.24) \atop -(0.18)}}}

\newcommand{\dmitrione}{\ensuremath{0.126}}
\newcommand{\dmitrionestat}{\ensuremath{0.033}}
\newcommand{\dmitrionesyst}{\ensuremath{{+0.047 \atop -0.038}}}

\newcommand{\dmitritwo}{\ensuremath{0.210}}
\newcommand{\dmitritwostat}{\ensuremath{0.042}}
\newcommand{\dmitritwosyst}{\ensuremath{{+0.071 \atop -0.050}}}

\newcommand{\dmitrithree}{\ensuremath{0.054}}
\newcommand{\dmitrithreestat}{\ensuremath{0.022}}
\newcommand{\dmitrithreesyst}{\ensuremath{{+0.021 \atop -0.018}}}

\newcommand{\brlcstarc}{\ensuremath{0.9}}
\newcommand{\brlcstare}{\ensuremath{0.4}}
\newcommand{\brlcsstarc}{\ensuremath{1.5}}
\newcommand{\brlcsstare}{\ensuremath{0.6}}
\newcommand{\brsigcc}{\ensuremath{0.39}}
\newcommand{\brsigce}{\ensuremath{0.23}}
\newcommand{\brlcsemitau}{\ensuremath{2.0}}
\newcommand{\brfzero}{\ensuremath{0.32}}
\newcommand{\brlcpipi}{\ensuremath{0.64}}

\newcommand{\brtautomuc}{\ensuremath{17.36}}
\newcommand{\brtautomue}{\ensuremath{0.05}}

\newcommand{\brinclbsemic}{\ensuremath{9.9}}
\newcommand{\brinclbsemie}{\ensuremath{2.6}}
\newcommand{\diffbr}{\ensuremath{1.3}}

\newcommand{\fracxib}{\ensuremath{0.2}}
\newcommand{\fracomegab}{\ensuremath{0.03}}

\newcommand{\bpxsecc}{\ensuremath{2.78}}
\newcommand{\bpxsece}{\ensuremath{0.24}}

\newcommand{\lbhadbbccc}{\ensuremath{1.3}}
\newcommand{\lbhadbbcce}{\ensuremath{0.4}}
\newcommand{\bbrc}{\ensuremath{0.016}}
\newcommand{\bbre}{\ensuremath{0.008}}
\newcommand{\ccrc}{\ensuremath{0.0018}}
\newcommand{\ccre}{\ensuremath{0.0009}}

\newcommand{\fracsignal}{\ensuremath{72.5}}
\newcommand{\fracbbcc}{\ensuremath{0.3}}
\newcommand{\fracfakemu}{\ensuremath{3.2}}
\newcommand{\fracphysb}{\ensuremath{24.0}}
\newcommand{\lbbbccinputc}{\ensuremath{0.017}}
\newcommand{\lbbbccinpute}{\ensuremath{0.009}}
\newcommand{\lbphysinputc}{\ensuremath{1.660}}
\newcommand{\lbphysinpute}{\ensuremath{0.018}}

\newcommand{\lbeffratioc}{\ensuremath{0.303}}
\newcommand{\lbeffratioe}{\ensuremath{0.004}}

\newcommand{\rlbc}{\ensuremath{16.6}}
\newcommand{\rlbe}{\ensuremath{3.0}}
\newcommand{\rlbsyste}{\ensuremath{1.0}}
\newcommand{\rlbbre}{\ensuremath{{+2.6 \atop -3.4}}}
\newcommand{\rlbubre}{\ensuremath{0.3}}
\newcommand{\erlbrlcbr}{\ensuremath{{+1.2 \atop -2.0}}}
\newcommand{\erlbrrxsec}{\ensuremath{\pm 1.9}}

\newcommand{\bdbbrc}{\ensuremath{0.08}}
\newcommand{\bdbbre}{\ensuremath{0.01}}
\newcommand{\bdccrc}{\ensuremath{0.05}}
\newcommand{\bdccre}{\ensuremath{0.01}}
\newcommand{\bdphysinputc}{\ensuremath{3.10}}
\newcommand{\bdphysinpute}{\ensuremath{0.03}}
\newcommand{\bdfracsignal}{\ensuremath{55.5}}
\newcommand{\bdfracbbcc}{\ensuremath{1.6}}
\newcommand{\bdfracfakemu}{\ensuremath{4.9}}
\newcommand{\bdfracphysb}{\ensuremath{38.0}}

\newcommand{\rdc}{\ensuremath{9.9}}
\newcommand{\rde}{\ensuremath{1.0}}
\newcommand{\rdsyste}{\ensuremath{0.6}}
\newcommand{\rdbre}{\ensuremath{0.4}}
\newcommand{\rdubre}{\ensuremath{0.5}}
\newcommand{\rdcombinee}{\ensuremath{1.3}}
\newcommand{\rdpdgc}{\ensuremath{8.1}}
\newcommand{\rdpdge}{\ensuremath{0.6}}
\newcommand{\agreebd}{\ensuremath{1.3}}

\newcommand{\bdstarbbrc}{\ensuremath{0.08}}
\newcommand{\bdstarbbre}{\ensuremath{0.01}}
\newcommand{\bdstarccrc}{\ensuremath{0.05}}
\newcommand{\bdstarccre}{\ensuremath{0.01}}
\newcommand{\bdstarphysinputc}{\ensuremath{2.08}}
\newcommand{\bdstarphysinpute}{\ensuremath{0.02}}
\newcommand{\bdstarfracsignal}{\ensuremath{73.7}}
\newcommand{\bdstarfracbbcc}{\ensuremath{1.3}}
\newcommand{\bdstarfracfakemu}{\ensuremath{4.2}}
\newcommand{\bdstarfracphysb}{\ensuremath{20.8}}

\newcommand{\rdstarc}{\ensuremath{16.5}}
\newcommand{\rdstare}{\ensuremath{2.3}}
\newcommand{\rdstarsyste}{\ensuremath{0.6}}
\newcommand{\rdstarbre}{\ensuremath{0.5}}
\newcommand{\rdstarubre}{\ensuremath{0.8}}
\newcommand{\rdstarcombinee}{\ensuremath{2.6}}
\newcommand{\rdstarpdgc}{\ensuremath{18.7}}
\newcommand{\rdstarpdge}{\ensuremath{1.0}}
\newcommand{\agreebdstar}{\ensuremath{0.8}}

\newcommand{\ndstarhadc}{\ensuremath{106}}
\newcommand{\ndstarhade}{\ensuremath{11}}

\newcommand{\ndhadc}{\ensuremath{579}}
\newcommand{\ndhade}{\ensuremath{30}}

\newcommand{\nlbhadc}{\ensuremath{179}}
\newcommand{\nlbhade}{\ensuremath{19}}

\newcommand{\nlbhad}{\ensuremath{\nlbhadc\ \pm \nlbhade}}

\newcommand{\nlbsemic}{\ensuremath{1237}}
\newcommand{\nlbsemie}{\ensuremath{97}}

\newcommand{\ndsemic}{\ensuremath{4720}}
\newcommand{\ndsemie}{\ensuremath{100}}

\newcommand{\ndstarsemic}{\ensuremath{1059}}
\newcommand{\ndstarsemie}{\ensuremath{33}}

\newcommand{\nlbsemi}{\ensuremath{\nlbsemic\ \pm \nlbsemie}}

\newcommand{\eff}{\ensuremath{\epsilon}}
\newcommand{\bsemi}{\ensuremath{{\cal B}_\mathrm{exclsemi}}}
\newcommand{\bhad}{\ensuremath{{\cal B}_\mathrm{had}}}
\newcommand{\effsemi}{\ensuremath{\eff_\mathrm{exclsemi}}}
\newcommand{\effhad}{\ensuremath{\eff_\mathrm{had}}}

\newcommand{\nhad}{\ensuremath{N_\mathrm{had}}}
\newcommand{\nincsemi}{\ensuremath{N_\mathrm{inclsemi}}}
\newcommand{\nsemi}{\ensuremath{N_\mathrm{exclsemi}}}
\newcommand{\nbg}{\ensuremath{N_\mathrm{semibg}}}
\newcommand{\nphys}{\ensuremath{N_\mathrm{feed}}}
\newcommand{\nfake}{\ensuremath{N_\mathrm{false\mu}}}
\newcommand{\nbbcc}{\ensuremath{N_{b\bar{b},c\bar{c}}}}
\newcommand{\fig}{Fig.}
\newcommand{\figc}{Figure}
\newcommand{\eq}{Eq.}
\newcommand{\eqc}{Equation}
\newcommand{\bgen}{{\sc bgenerator}}
\newcommand{\pythia}{{\sc pythia}}
\newcommand{\evtgen}{{\sc evtgen}}
\newcommand{\geant}{{\sc geant}}
\newcommand{\etal}{{\it et~al.}}
\newcommand{\eg}{{\it e.g.}}

\newcommand{\rphi}{\ensuremath{r}-\ensuremath{\phi}}
\newcommand{\chixy}{\ensuremath{\chi^2_{r\phi}}}
\newcommand{\lxy}{\ensuremath{L_{r\phi}}}
\newcommand{\pt}{\ensuremath{p_T}}
\newcommand{\ctau}{\ensuremath{ct}}

\newcommand{\gevc}{\ensuremath{{\rm GeV}/c}}
\newcommand{\gevcsq}{\ensuremath{{\rm GeV}/c^{2}}}
\newcommand{\mev}{\ensuremath{\rm MeV}}
\newcommand{\mevcsq}{\ensuremath{{\rm MeV}/c^{2}}}

\newcommand{\Lb}{\ensuremath{\Lambda_b^0}}
\newcommand{\Lc}{\ensuremath{\Lambda_c^+}}
\newcommand{\Lcstar}{\ensuremath{\Lambda_c\left(2595\right)^+}}
\newcommand{\Lcsstar}{\ensuremath{\Lambda_c\left(2625\right)^+}}
\newcommand{\Sigczero}{\ensuremath{\Sigma_c(2455)^0}}
\newcommand{\Sigcpp}{\ensuremath{\Sigma_c(2455)^{++}}}
\newcommand{\Sigcp}{\ensuremath{\Sigma_c(2455)^+}}
\newcommand{\Sigmac}{\ensuremath{\Sigma_c(2455)}}
\newcommand{\Lamc}{\ensuremath{\Lambda_c}}
\newcommand{\Bd}{\ensuremath{\bar{B}^0}}
\newcommand{\Bs}{\ensuremath{\bar{B}_s^0}}
\newcommand{\Bu}{\ensuremath{{B}^-}}
\newcommand{\phip}{\ensuremath{\phi(1020)}}

\newcommand{\dstar}{\ensuremath{D^*(2010)^+}}
\newcommand{\dplus}{\ensuremath{D^+}}
\newcommand{\dzero}{\ensuremath{D^0}}

\newcommand{\incdstarsemi}{\ensuremath{\dstar\mu^- X}}
\newcommand{\dstarsemi}{\ensuremath{\bar{B}^0 \rightarrow \dstar\mu^-\bar{\nu}_{\mu}}}
\newcommand{\dstarhad}{\ensuremath{\bar{B}^0 \rightarrow \dstar\pi^-}}
\newcommand{\incdsemi}{\ensuremath{D^{+}\mu^- X}}
\newcommand{\dsemi}{\ensuremath{\bar{B}^0 \rightarrow D^{+}\mu^-\bar{\nu}_{\mu}}}
\newcommand{\dhad}{\ensuremath{\bar{B}^0 \rightarrow D^{+}\pi^-}}
\newcommand{\inclbsemi}{\ensuremath{\Lc\mu^- X}}
\newcommand{\lbsemi}{\ensuremath{\Lb \rightarrow \Lc\mu^-\bar{\nu}_{\mu}}}
\newcommand{\lbsemil}{\ensuremath{\Lb \rightarrow \Lc\ell^-\bar{\nu}_{\ell}}}
\newcommand{\lbhad}{\ensuremath{\Lb \rightarrow \Lc\pi^-}}

\newcommand{\yile}{\ensuremath{\frac{\sigma_{\Lb}(\pt>6.0){\cal B}(\lbhad)}{\sigma_{\Bd}(\pt>6.0){\cal B}(\dhad)}}}
\newcommand{\rxsec}{\ensuremath{\sigma_{\Lb}/\sigma_{\Bd}}}

\newcommand{\dmitridecayone}{\ensuremath{\Lb \rightarrow \Lcstar \mu^- \bar{\nu}_{\mu}}}
\newcommand{\dmitridecaytwo}{\ensuremath{\Lb \rightarrow \Lcsstar \mu^- \bar{\nu}_{\mu}}} 
\newcommand{\dmitridecaythree}{\ensuremath{\Lb \rightarrow \Sigczero \pi^+ \mu^- \bar{\nu}_{\mu}}}
\newcommand{\dmitridecayfour}{\ensuremath{\Lb \rightarrow \Sigcpp \pi^- \mu^- \bar{\nu}_{\mu}}}
\newcommand{\dmitridecaythreefour}{\ensuremath{\Lb \rightarrow \Sigmac \pi \mu^- \bar{\nu}_{\mu}}}
\newcommand{\lblcfzero}{\ensuremath{\Lb \rightarrow \Lc f_0(980)\mu^- \bar{\nu}_{\mu}}}
\newcommand{\lblcpizero}{\ensuremath{\Lb \rightarrow \Lc \pi^0\pi^0\mu^- \bar{\nu}_{\mu}}}
\newcommand{\lblcpim}{\ensuremath{\Lb \rightarrow \Lc \pi^+\pi^-\mu^- \bar{\nu}_{\mu}}}
\newcommand{\lblctau}{\ensuremath{\Lb \rightarrow \Lc \tau^- \bar{\nu}_{\tau}}}
\newcommand{\lbsigmacp}{\ensuremath{\Lb \rightarrow \Sigcp \pi^0 \mu^- \bar{\nu}_{\mu}}}

\newcommand{\bddpizeromunu}{\ensuremath{\bar{B}^0\rightarrow D^{+}\pi^0\mu^-\bar{\nu}_{\mu}}}
\newcommand{\bpdpimunu}{\ensuremath{B^-\rightarrow D^{+}\pi^-\mu^-\bar{\nu}_{\mu}}}
\newcommand{\bddstarpizeromunu}{\ensuremath{\bar{B}^0\rightarrow \dstar\pi^0\mu^-\bar{\nu}_{\mu}}}
\newcommand{\bpdstarpimunu}{\ensuremath{B^-\rightarrow \dstar\pi^-\mu^-\bar{\nu}_{\mu}}}
\newcommand{\bpdonezeromunu}{\ensuremath{B^-\rightarrow D_1(2420)^{0}\mu^-\bar{\nu}_{\mu}}}
\newcommand{\bpdpronezeromunu}{\ensuremath{B^-\rightarrow D_1^{\prime}(2430)^0\mu^-\bar{\nu}_{\mu}}}
\newcommand{\bddonemunu}{\ensuremath{\bar{B}^0\rightarrow D_1(2420)^{+}\mu^-\bar{\nu}_{\mu}}}
\newcommand{\bddpronemunu}{\ensuremath{\bar{B}^0\rightarrow D_1^{\prime}(2430)^+\mu^-\bar{\nu}_{\mu}}}
\newcommand{\bddtau}{\ensuremath{\bar{B}^0\rightarrow D^{+}\tau^-\bar{\nu}_{\tau}}}
\newcommand{\bddstartau}{\ensuremath{\bar{B}^0\rightarrow \dstar\tau^-\bar{\nu}_{\tau}}}
\newcommand{\bsdkzero}{\ensuremath{\bar{B}_s^0\rightarrow D^{+}K^0\mu^-\bar{\nu}_{\mu}}}

\begin{document}
\title{First Measurement of the Ratio of Branching Fractions
   ${\cal B}\left(\Lb\rightarrow\Lc\mu^-\bar{\nu}_{\mu}\right)/
         {\cal B}\left(\Lb\rightarrow\Lc\pi^-\right)$ 
   }

\affiliation{Institute of Physics, Academia Sinica, Taipei, Taiwan 11529, Republic of China} 
\affiliation{Argonne National Laboratory, Argonne, Illinois 60439} 
\affiliation{University of Athens, 157 71 Athens, Greece} 
\affiliation{Institut de Fisica d'Altes Energies, Universitat Autonoma de Barcelona, E-08193, Bellaterra (Barcelona), Spain} 
\affiliation{Baylor University, Waco, Texas  76798} 
\affiliation{Istituto Nazionale di Fisica Nucleare Bologna, $^v$University of Bologna, I-40127 Bologna, Italy} 
\affiliation{Brandeis University, Waltham, Massachusetts 02254} 
\affiliation{University of California, Davis, Davis, California  95616} 
\affiliation{University of California, Los Angeles, Los Angeles, California  90024} 
\affiliation{University of California, San Diego, La Jolla, California  92093} 
\affiliation{University of California, Santa Barbara, Santa Barbara, California 93106} 
\affiliation{Instituto de Fisica de Cantabria, CSIC-University of Cantabria, 39005 Santander, Spain} 
\affiliation{Carnegie Mellon University, Pittsburgh, PA  15213} 
\affiliation{Enrico Fermi Institute, University of Chicago, Chicago, Illinois 60637} 
\affiliation{Comenius University, 842 48 Bratislava, Slovakia; Institute of Experimental Physics, 040 01 Kosice, Slovakia} 
\affiliation{Joint Institute for Nuclear Research, RU-141980 Dubna, Russia} 
\affiliation{Duke University, Durham, North Carolina  27708} 
\affiliation{Fermi National Accelerator Laboratory, Batavia, Illinois 60510} 
\affiliation{University of Florida, Gainesville, Florida  32611} 
\affiliation{Laboratori Nazionali di Frascati, Istituto Nazionale di Fisica Nucleare, I-00044 Frascati, Italy} 
\affiliation{University of Geneva, CH-1211 Geneva 4, Switzerland} 
\affiliation{Glasgow University, Glasgow G12 8QQ, United Kingdom} 
\affiliation{Harvard University, Cambridge, Massachusetts 02138} 
\affiliation{Division of High Energy Physics, Department of Physics, University of Helsinki and Helsinki Institute of Physics, FIN-00014, Helsinki, Finland} 
\affiliation{University of Illinois, Urbana, Illinois 61801} 
\affiliation{The Johns Hopkins University, Baltimore, Maryland 21218} 
\affiliation{Institut f\"{u}r Experimentelle Kernphysik, Universit\"{a}t Karlsruhe, 76128 Karlsruhe, Germany} 
\affiliation{Center for High Energy Physics: Kyungpook National University, Daegu 702-701, Korea; Seoul National University, Seoul 151-742, Korea; Sungkyunkwan University, Suwon 440-746, Korea; Korea Institute of Science and Technology Information, Daejeon, 305-806, Korea; Chonnam National University, Gwangju, 500-757, Korea} 
\affiliation{Ernest Orlando Lawrence Berkeley National Laboratory, Berkeley, California 94720} 
\affiliation{University of Liverpool, Liverpool L69 7ZE, United Kingdom} 
\affiliation{University College London, London WC1E 6BT, United Kingdom} 
\affiliation{Centro de Investigaciones Energeticas Medioambientales y Tecnologicas, E-28040 Madrid, Spain} 
\affiliation{Massachusetts Institute of Technology, Cambridge, Massachusetts  02139} 
\affiliation{Institute of Particle Physics: McGill University, Montr\'{e}al, Qu\'{e}bec, Canada H3A~2T8; Simon Fraser University, Burnaby, British Columbia, Canada V5A~1S6; University of Toronto, Toronto, Ontario, Canada M5S~1A7; and TRIUMF, Vancouver, British Columbia, Canada V6T~2A3} 
\affiliation{University of Michigan, Ann Arbor, Michigan 48109} 
\affiliation{Michigan State University, East Lansing, Michigan  48824}
\affiliation{Institution for Theoretical and Experimental Physics, ITEP, Moscow 117259, Russia} 
\affiliation{University of New Mexico, Albuquerque, New Mexico 87131} 
\affiliation{Northwestern University, Evanston, Illinois  60208} 
\affiliation{The Ohio State University, Columbus, Ohio  43210} 
\affiliation{Okayama University, Okayama 700-8530, Japan} 
\affiliation{Osaka City University, Osaka 588, Japan} 
\affiliation{University of Oxford, Oxford OX1 3RH, United Kingdom} 
\affiliation{Istituto Nazionale di Fisica Nucleare, Sezione di Padova-Trento, $^w$University of Padova, I-35131 Padova, Italy} 
\affiliation{LPNHE, Universite Pierre et Marie Curie/IN2P3-CNRS, UMR7585, Paris, F-75252 France} 
\affiliation{University of Pennsylvania, Philadelphia, Pennsylvania 19104}
\affiliation{Istituto Nazionale di Fisica Nucleare Pisa, $^x$University of Pisa, $^y$University of Siena and $^z$Scuola Normale Superiore, I-56127 Pisa, Italy} 
\affiliation{University of Pittsburgh, Pittsburgh, Pennsylvania 15260} 
\affiliation{Purdue University, West Lafayette, Indiana 47907} 
\affiliation{University of Rochester, Rochester, New York 14627} 
\affiliation{The Rockefeller University, New York, New York 10021} 
\affiliation{Istituto Nazionale di Fisica Nucleare, Sezione di Roma 1, $^{aa}$Sapienza Universit\`{a} di Roma, I-00185 Roma, Italy} 

\affiliation{Rutgers University, Piscataway, New Jersey 08855} 
\affiliation{Texas A\&M University, College Station, Texas 77843} 
\affiliation{Istituto Nazionale di Fisica Nucleare Trieste/Udine, $^{bb}$University of Trieste/Udine, Italy} 
\affiliation{University of Tsukuba, Tsukuba, Ibaraki 305, Japan} 
\affiliation{Tufts University, Medford, Massachusetts 02155} 
\affiliation{Waseda University, Tokyo 169, Japan} 
\affiliation{Wayne State University, Detroit, Michigan  48201} 
\affiliation{University of Wisconsin, Madison, Wisconsin 53706} 
\affiliation{Yale University, New Haven, Connecticut 06520} 
\author{T.~Aaltonen}
\affiliation{Division of High Energy Physics, Department of Physics, University of Helsinki and Helsinki Institute of Physics, FIN-00014, Helsinki, Finland}
\author{J.~Adelman}
\affiliation{Enrico Fermi Institute, University of Chicago, Chicago, Illinois 60637}
\author{T.~Akimoto}
\affiliation{University of Tsukuba, Tsukuba, Ibaraki 305, Japan}
\author{M.G.~Albrow}
\affiliation{Fermi National Accelerator Laboratory, Batavia, Illinois 60510}
\author{B.~\'{A}lvarez~Gonz\'{a}lez}
\affiliation{Instituto de Fisica de Cantabria, CSIC-University of Cantabria, 39005 Santander, Spain}
\author{S.~Amerio$^w$}
\affiliation{Istituto Nazionale di Fisica Nucleare, Sezione di Padova-Trento, $^w$University of Padova, I-35131 Padova, Italy} 

\author{D.~Amidei}
\affiliation{University of Michigan, Ann Arbor, Michigan 48109}
\author{A.~Anastassov}
\affiliation{Northwestern University, Evanston, Illinois  60208}
\author{A.~Annovi}
\affiliation{Laboratori Nazionali di Frascati, Istituto Nazionale di Fisica Nucleare, I-00044 Frascati, Italy}
\author{J.~Antos}
\affiliation{Comenius University, 842 48 Bratislava, Slovakia; Institute of Experimental Physics, 040 01 Kosice, Slovakia}
% ...

\collaboration{CDF Collaboration\footnote{With visitors from $^a$University of Massachusetts Amherst, Amherst, Massachusetts 01003,
$^b$Universiteit Antwerpen, B-2610 Antwerp, Belgium, 
$^c$University of Bristol, Bristol BS8 1TL, United Kingdom,
$^d$Chinese Academy of Sciences, Beijing 100864, China, 
$^e$Istituto Nazionale di Fisica Nucleare, Sezione di Cagliari, 09042 Monserrato (Cagliari), Italy,
$^f$University of California Irvine, Irvine, CA  92697, 
$^g$University of California Santa Cruz, Santa Cruz, CA  95064, 
$^h$Cornell University, Ithaca, NY  14853, 
$^i$University of Cyprus, Nicosia CY-1678, Cyprus, 
$^j$University College Dublin, Dublin 4, Ireland,
$^k$Royal Society of Edinburgh/Scottish Executive Support Research Fellow,
$^l$University of Edinburgh, Edinburgh EH9 3JZ, United Kingdom, 
$^m$Universidad Iberoamericana, Mexico D.F., Mexico,
$^n$Queen Mary, University of London, London, E1 4NS, England,
$^o$University of Manchester, Manchester M13 9PL, England, 
$^p$Nagasaki Institute of Applied Science, Nagasaki, Japan, 
$^q$University of Notre Dame, Notre Dame, IN 46556,
$^r$University de Oviedo, E-33007 Oviedo, Spain, 
$^s$Texas Tech University, Lubbock, TX  79409, 
$^t$IFIC(CSIC-Universitat de Valencia), 46071 Valencia, Spain,
$^u$University of Virginia, Charlottesville, VA  22904,
$^{cc}$On leave from J.~Stefan Institute, Ljubljana, Slovenia, 
}}
\noaffiliation


\begin{abstract}
This article presents the first measurement of the ratio of branching 
fractions ${\cal B}\left(\Lb \rightarrow \Lc\mu^-\bar{\nu}_{\mu}\right)/{\cal B}\left(\Lb \rightarrow \Lc\pi^-\right)$. 
Measurements in two control samples using the same technique, 
${\cal B}\left(\bar{B}^0 \rightarrow D^{+}\mu^-\bar{\nu}_{\mu}\right)/{\cal B}\left(\bar{B}^0 \rightarrow D^{+}\pi^-\right)$ 
and ${\cal B}\left(\bar{B}^0 \rightarrow \dstar\mu^-\bar{\nu}_{\mu}\right)/{\cal B}\left(\bar{B}^0 \rightarrow \dstar\pi^-\right)$, are also 
reported. 
The analysis uses data from an integrated luminosity of approximately 
172~$\rm pb^{-1}$ of $p\bar{p}$ collisions at $\sqrt{s}=1.96$ TeV, collected 
with the CDF II detector at the Fermilab Tevatron. 
The relative branching fractions are measured to be: 
\begin{eqnarray*}
\frac{{\cal B}\left(\Lb \rightarrow \Lc\mu^-\bar{\nu}_{\mu}\right)}{{\cal B}\left(\Lb \rightarrow \Lc\pi^-\right)} &= &
	16.6 \pm 3.0 \left(\rm{stat}\right) \pm 1.0 \left(\rm{syst}\right)
	         {+2.6 \atop -3.4} \left(\rm{PDG}\right) \pm 0.3 \left(\rm{EBR}\right), \\
\frac{{\cal B}\left(\bar{B}^0 \rightarrow D^{+}\mu^-\bar{\nu}_{\mu}\right)}{{\cal B}\left(\bar{B}^0 \rightarrow D^{+}\pi^-\right)} & = &
	9.9 \pm 1.0 \left(\rm{stat}\right) \pm 0.6 \left(\rm{syst}\right)
	         \pm 0.4 \left(\rm{PDG}\right) \pm 0.5 \left(\rm{EBR}\right),\\
\frac{{\cal B}\left(\bar{B}^0 \rightarrow \dstar\mu^-\bar{\nu}_{\mu}\right)}{{\cal B}\left(\bar{B}^0 \rightarrow \dstar\pi^-\right)} & = &
	16.5 \pm 2.3 \left(\rm{stat}\right) \pm 0.6 \left(\rm{syst}\right)
	         \pm 0.5 \left(\rm{PDG}\right) \pm 0.8 \left(\rm{EBR}\right). \\
\end{eqnarray*}
The uncertainties are from statistics $\left(\rm{stat}\right)$, internal 
systematics $\left(\rm{syst}\right)$, world averages of measurements published 
by the Particle Data Group or subsidiary measurements in this analysis
$\left(\rm{PDG}\right)$, and unmeasured branching 
fractions estimated from theory $\left(\rm{EBR}\right)$, respectively. 
This article also presents measurements of the branching fractions of 
four new $\Lb$ semileptonic decays: 
$\Lb \rightarrow \Lcstar \mu^- \bar{\nu}_{\mu}$, 
$\Lb \rightarrow \Lcsstar \mu^- \bar{\nu}_{\mu}$, 
$\Lb \rightarrow \Sigczero \pi^+ \mu^- \bar{\nu}_{\mu}$, and 
$\Lb \rightarrow \Sigcpp \pi^- \mu^- \bar{\nu}_{\mu}$, 
 relative to the branching fraction of the $\Lb \rightarrow \Lc\mu^-\bar{\nu}_{\mu}$ decay. Finally, the transverse-momentum distribution of $\Lb$ baryons produced in 
$p\bar{p}$ collisions is measured and found to be significantly different 
from that of $\bar{B}^0$ mesons,  
which results in a modification in the production cross-section ratio 
$\sigma_{\Lb}/\sigma_{\Bd}$ with respect to the CDF I measurement. 

\end{abstract}


\maketitle

\tableofcontents

\section{Introduction \label{sec:intro}}
Amplitudes for the weak decays of $b$ hadrons are described by the 
product of Cabibbo-Kobayashi-Maskawa (CKM) matrix 
elements~\cite{Cabibbo:1963yz,Kobayashi:1973fv} and dynamical factors. The 
CKM matrix elements represent 
the coupling strength of the weak decays and are fundamental 
parameters of the standard model of particle physics. In order to extract 
values of the CKM elements, knowledge of the dynamical factors is
needed either from experiment or theory. Calculation of the 
dynamical factors, in the case of $b$-hadron decays, is aided by
heavy quark effective theory (HQET)~\cite{Manohar:2000dt,Godfrey:1985xj,Isgur:1988gb}. 
HQET is an approximation relying on the large mass of the $b$ quark 
($m_b \approx 4000~\mevcsq$), as compared with the Quantum 
Chromodynamics (QCD) energy scale $\left(\Lambda_\mathrm{QCD}\approx 200~\mev\right)$, 
to imply a spin-independent interaction between the $b$ quark and the 
light degrees of freedom. 
In baryon spectroscopy, the light degrees of freedom 
are in a relative spin-0 state for all $\Lambda$-type baryons;  
there is no spin-related 
interaction between the $b$ quark and the light degrees of freedom. 
Therefore, the subleading order corrections to the heavy 
quark limit are simpler than those mesons which contain a $b$ quark ($b$ mesons)~\cite{Georgi:1990ei}. 
Measurements of \Lb-baryon branching fractions may be compared with predictions
 by HQET and test the calculation of dynamical factors to subleading order.
 However, in contrast to the $b$ mesons, little is known about the \Lb\ baryon.
 At the time of writing this article, only five decay modes of the \Lb\ have 
been observed, with large uncertainties on their branching fraction 
measurements~\cite{pdg}. On the theoretical side, 
combining measurements of the CKM matrix element, 
$\left|V_{cb}\right|$, and the world average of the \Lb\ lifetime~\cite{theory,pdg}, 
the branching fraction predicted by HQET 
for \lbsemi\ is \hqetlbsemic\%\  by Huang~\etal~\cite{Huang:2005me}, and 
that for \lbhad\ is \hqetlbhadc\%\ by Leibovich~\etal~\cite{Leibovich:2003tw}.
An independent prediction of ${\cal B}(\lbhad)$ by Cheng using the 
non-relativistic quark model yields \chenglbhadc\%~\cite{Cheng:1996cs}.

Presented here is the first measurement of the ratio of the \Lb\ branching 
fractions, ${\cal B}\left(\lbsemi\right)/{\cal B}\left(\lbhad\right)$. This 
measurement is based on data from an integrated luminosity of approximately 
172~$\rm pb^{-1}$ of $p\bar{p}$ collisions at 
$\sqrt{s}=1.96$ TeV, collected with the CDF~II detector at Fermilab. 
Taking advantage of the relatively long lifetime of $b$ hadrons 
$\left(c\tau\approx 400~\mu\rm{m}\right)$, all $b$-hadron decays described in 
this article are reconstructed from data satisfying an online event selection 
(trigger) which requires two charged tracks forming a vertex displaced from 
the location of $p\bar{p}$ collisions (two-track trigger). The \Lc\ is 
reconstructed using the three-body decay $\Lc \rightarrow pK^-\pi^+$, 
therefore both the \lbhad\ and \lbsemi\ decays result in four charged particles 
which are observable in the detector and have a similar topology (\fig~\ref{fig:bigd0}).
 Since both decays have a similar topology and satisfy the same trigger, 
most systematic uncertainties from the detector, trigger, and reconstruction 
efficiencies cancel in the measurement of the ratio of branching fractions. 
Throughout this article, the inclusion of charge conjugate decays is 
implied. 

The ratio of the branching fraction of the \Lb\ exclusive semileptonic decay 
relative to that of the \Lb\ hadronic decay, $\bsemi/\bhad$, is extracted from 
the ratio of signal yields $\left(\nsemi/\nhad\right)$ divided by the ratio of 
acceptance times efficiency $\left(\effsemi/\effhad\right)$:
\begin{eqnarray}
\label{eq:bigpicture}
 \lefteqn{\frac{{\cal B}\left(\lbsemi\right)}{{\cal B}\left(\lbhad\right)}
  \equiv  \frac{\bsemi}{\bhad}} \nonumber \\ 
 & = &  \left(\frac{\nsemi}{\nhad}\right)\left/
 \left(\frac{\effsemi}{\effhad}\right)\right. \nonumber \\
 &=&
\left(\frac{\nincsemi-\nbg}{\nhad}\right)
 \frac{\effhad}{\effsemi}. 
\end{eqnarray}
The analysis procedure can be summarized in four steps. 
First, the hadronic ($\Lc\pi^-$) and inclusive semileptonic ($\Lc\mu^-X$) 
candidates are reconstructed. Second, the yields, \nhad\ and \nincsemi, are 
determined by fitting the mass distributions. 
Third, the contribution of backgrounds which produce a $\Lc\mu^-$ in the 
final state is either measured or estimated and combined into \nbg. 
The estimate of \nbg\ requires a modification of the 
production cross-section ratio, \rxsec, with respect to the CDF I 
measurement~\cite{Affolder:1999iq}. The dominant backgrounds which contribute 
to \nbg, \dmitridecayone, \dmitridecaytwo, \dmitridecaythree, and 
\dmitridecayfour, have also been reconstructed in the data for the first 
time. Measurements of their branching fractions relative to the branching 
fraction of the \lbsemi\ decay will be used in the estimate of \nbg. 
Fourth, the ratio of the products of detector acceptance and reconstruction 
efficiency, $\effhad/\effsemi$, is estimated from simulation. 

The analysis method described above is tested by performing the 
same measurements in \Bd\ decays which have a similar event topology. 
Specifically, the following ratios of branching fractions are measured:
\(
{\cal B}\left(\dsemi\right)/{\cal B}\left(\dhad\right)\) 
 where $\dplus \rightarrow K^-\pi^+\pi^+$, and 
\({\cal B}\left(\dstarsemi\right)/{\cal B}\left(\dstarhad\right)\)
 where $\dstar \rightarrow \dzero\pi^+, \dzero \rightarrow K^-\pi^+$. 
The results of the \Bd\ measurements are compared with previous results 
from the $B$ factories~\cite{pdg} to check the techniques used in this 
analysis.

This article is structured as follows. 
Section~\ref{sec:detector} describes the relevant parts of 
the CDF II detector and trigger. Section~\ref{sec:reconstruction} details the 
event selections for the $\Lc\pi^-$ and $\Lc\mu^-X$ samples. 
Section~\ref{sec:simulation} describes the simulations used in this analysis. 
Section~\ref{sec:fitting} gives an account of the determination of the yields 
\nhad\ and \nincsemi. In Section~\ref{sec:slbackgrounds}, $\nbg$ is estimated. 
Section~\ref{sec:lbbr} includes measurements and estimates of the 
branching fractions of other \Lb\ semileptonic decays which may contribute 
to \nbg, and an estimate of ${\cal B}(\lbhad)$ 
derived from a modification of the CDF I measurement of \rxsec. 
Section~\ref{sec:systematics} summarizes the systematic uncertainties. 
Section~\ref{sec:bmeson} shows the measurements with the \Bd\ control samples 
using the same analysis technique. 
Section~\ref{sec:results} compares the results of the \Lb\ and \Bd\ 
relative branching fractions with the predictions from HQET and the world 
averages, respectively. 
Finally, Section~\ref{sec:conclusion} gives the conclusion. 
Unless stated otherwise, branching fractions, fragmentation fractions, and 
lifetimes are obtained from the Particle Data Group world averages~\cite{pdg}.
 The symbols ``$H_c$'' and ``$H_b$'' are used to generically denote 
hadrons containing charm and bottom quarks, ``$c$ hadrons''
and ``$b$ hadrons'', respectively. The symbol ``MC'', which stands for 
``Monte Carlo'', is used to generically denote simulation.

\section{\label{sec:detector}The CDF II Detector and Trigger}
The CDF II detector is a cylindrically symmetric apparatus described 
in detail elsewhere~\cite{CDF}. Only the parts of the detector relevant for 
this analysis are summarized here. The crucial features of the detector for 
this measurement are the tracking and muon systems. The tracking system, which 
enables reconstruction of the trajectories of charged particles, is contained 
in a superconducting solenoid which generates a 1.4~Tesla magnetic field in 
the $-z$ direction~\cite{coor}.  
The 96-cm long silicon vertex detector (SVX~II)~\cite{SVXII} 
consists of six equal subsections in $z$ and five concentric layers of 
double-sided silicon sensors from 
$r=2.45$~cm to $r=10.60$~cm. 
The 310-cm long central outer tracker (COT)~\cite{COT}, an open-cell wire 
drift chamber, consists of 96 sense wire layers from $r=40$~cm 
to $r=137$~cm which are grouped into alternating axial and $\pm 2^\circ$ 
stereo superlayers. The SVX~II and COT provide both \rphi\ and $z$ 
position measurements in the pseudorapidity region of $\left|\eta\right| <2$ 
and $\left|\eta\right| < 1$~\cite{eta}, respectively. The 452-cm long central 
muon detector 
(CMU)~\cite{CMU}, a set of drift chambers mounted outside of the central hadron 
calorimeter at $r=347$~cm, contains 4 sense wire layers which allow the 
formation of short track segments (stubs) and identify the muon candidates in 
the region of $\left|\eta\right|<0.6$. 

\begin{figure}[tbp]
     \begin{center}
        \includegraphics[width=250pt, angle=0]
	{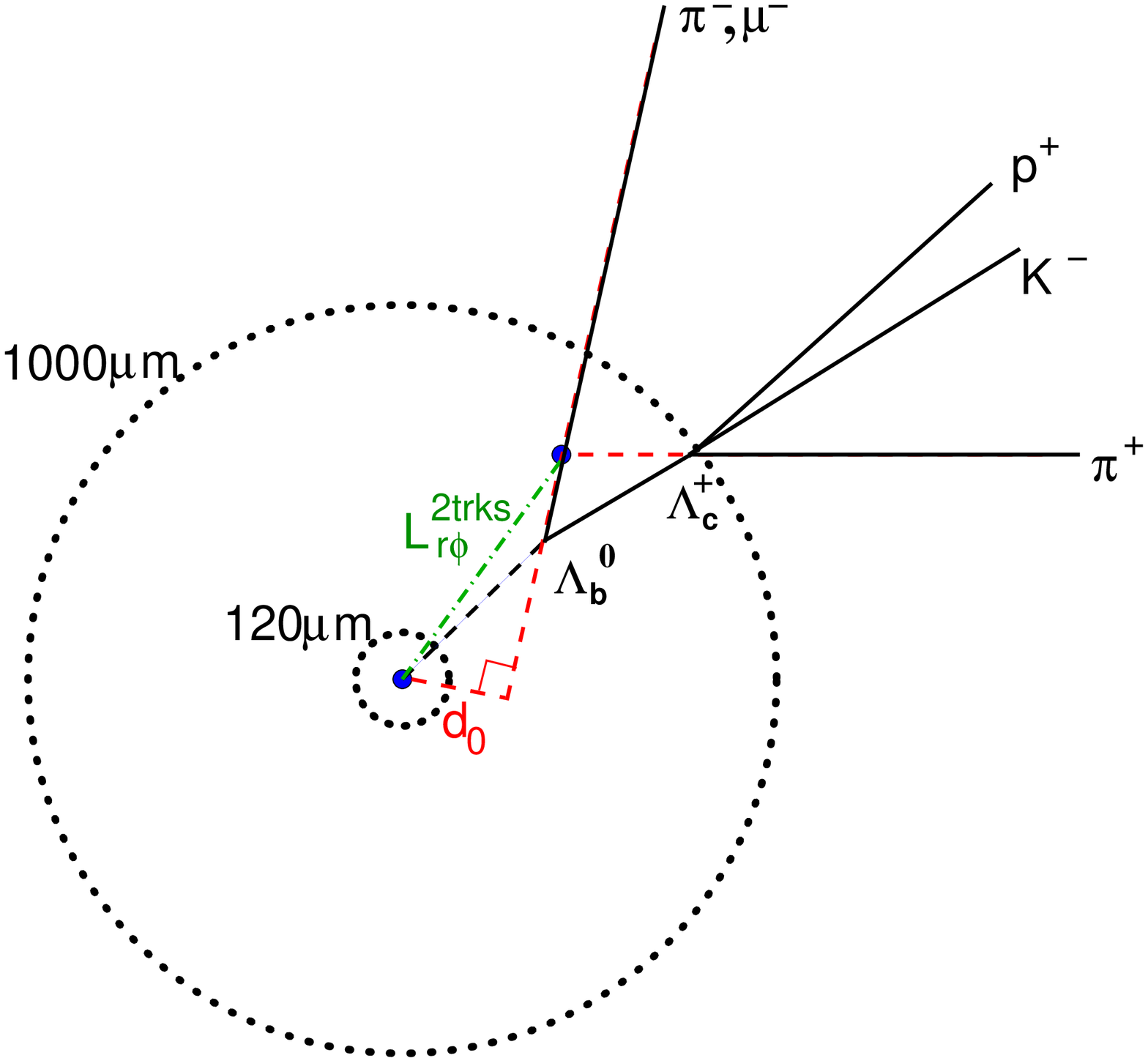}
     \end{center}
        \caption  		 
	{\label{fig:bigd0} 
	An \rphi\ view of a \lbhad\ $\left(\lbsemi\right)$ decay with a 
	two-prong \Lb\ decay vertex 
	and a three-prong \Lc\ decay vertex. In this case, the $d_0$ of each pion 
	(the pion and the muon) and the $\lxy^\mathrm{2trks}$ of the 
	two-pion vertex (pion-muon vertex) satisfy the trigger requirements.
	} 	
  \end{figure}

The data for this analysis are collected with a three-level, two-track trigger 
which selects events with a displaced vertex. Consequently, data 
satisfying this trigger are rich in heavy flavor with a low background from 
the combination of random tracks (combinatorial~background). 
 A schematic diagram of the event topology and trigger requirements is shown 
in \fig~\ref{fig:bigd0}. 
The strategy of the two-track trigger is as follows: at the first trigger 
level, the 
extremely fast tracker (XFT)~\cite{Thomson:2002xp} finds two oppositely-charged 
tracks reconstructed in the COT, with a minimum transverse momentum (\pt)
 of 2.04~\gevc\ for each track. The scalar sum of the \pt\ 
from the two tracks is required to exceed 5.5~\gevc\ and the azimuthal angle 
between the two tracks $\left(\Delta\phi\right)$ to be less than $135^{\circ}$.
  At the second trigger level, the silicon vertex trigger (SVT)~\cite{Ashmanskas:2003gf} attaches hits measured with SVX~II to the tracks found by XFT. 
The SVT reapplies the \pt\ requirements made at level 1 and further requires 
that each track has a transverse impact parameter $\left(d_0\right)$, 
measured at the point of closest approach (POCA) with respect to the beam 
line~\cite{svtbeamline}, in the range 120~$\mu$m--1000~$\mu$m. 
In addition, $\Delta\phi$ between the two trigger tracks is required to be in 
the range 2$^{\circ}$~--~90$^{\circ}$. The intersection of 
the two tracks forms a displaced vertex. Finally, the quantity, 
$\lxy^\mathrm{2trks}$, defined 
as the projection of the vector from the primary vertex to the displaced vertex
 onto the vector of the total momentum of the two tracks
 in the \rphi\ plane, must be larger than 200~$\mu$m. The level 1 and 2
 triggers are implemented in hardware, while at the third level, a cluster of 
computers uses all detector information to perform a full reconstruction of 
the event~\cite{L3}. In addition to reinforcing the same requirements as 
applied at level 2, level 3 requires the difference in $z$ between the two 
tracks at the POCA to be less than 5 cm. 
The measurements presented in this article are based on an integrated 
luminosity of $\approx 172$~pb$^{-1}$ collected between February 2002 and 
September 2003, comprising $\approx 152$ million two-track trigger events.

%
\section{\label{sec:reconstruction}Event Reconstruction}
The final states $\Lc\pi^-$ and \inclbsemi, where $\Lc \rightarrow pK^-\pi^+$, are reconstructed in the data collected with the two-track trigger. 
The selection criteria for the hadronic and the inclusive semileptonic decay 
modes are kept as similar as possible, which reduces systematic uncertainties 
on the relative branching fractions. 

Both signal decays have a four-track topology. Therefore, events are 
selected which contain a minimum of four tracks, each with a minimum \pt\ of
 0.5~\gevc, $d_0$ less than 5000~$\mu$m (measured with the SVX~II), a minimum 
of 20 hits each in the COT axial and stereo layers~\cite{axialstereo}, and a 
minimum of three axial hits in the SVX~II. Each track is also required to be in
 the fiducial region of the COT and to traverse all 96 wire layers. Making 
these requirements on each track ensures good quality of the track 
reconstruction and good momentum resolution. In addition, the maximum 
requirement on $d_0$ suppresses background from daughters of $K_S^0$ and 
$\Lambda^0$, and from particles produced by inelastic collisions of beam 
products with the detector material.

The reconstruction begins by identifying the \Lc\ candidate. Only combinations
 of three tracks which satisfy the requirements described above are 
considered. Every combination must have two positively charged tracks and one 
negatively charged track. At least one of the three tracks must match a 
displaced track found by the SVT (SVT track~\cite{SVTtrk}). 
The proton mass is assigned to the positively charged track of higher \pt, 
the pion mass to the track of lower \pt, and the kaon mass to 
the negatively charged track. Assuming the proton track to be the higher \pt\ 
track reduces the combinatorial background by $\approx50\%$ while keeping 
$\approx90\%$ of the $\Lc$ signal. 
A three-track kinematic fit determines the \Lc\ decay vertex by varying the
 track parameters of the daughter particles simultaneously, within their 
uncertainties, so that the $\chi^2$ between the adjusted and the original track
 parameters is minimized. Only three-track candidates for which the fit 
converges and the invariant mass ($M_{pK\pi}$) is in the range 
2.18~--~2.38~\gevcsq\ are considered further.

Next, the selected \Lc\ candidate is combined with an additional 
negatively charged track to form a \Lb\ candidate. This fourth track must be 
matched to an SVT track. The combination is considered a \Lb\ semileptonic 
candidate and a muon mass is assumed for this track if the following two 
requirements are satisfied. First, a CMU muon stub must be present within 30~cm
 of the extrapolated track at the CMU radius ($r=347$~cm) in the \rphi\ view. 
Second, the matching $\chi^2$ between the track and the stub positions~\cite{cmuchi2} is less than 9. Otherwise, the combination is a \Lb\ hadronic candidate 
and a pion mass is assumed. Both the muon and the pion tracks from the \Lb\ 
decay must extrapolate to the fiducial region of the CMU. 
Making the same fiducial requirement for the hadronic and semileptonic modes 
ensures that the tracking efficiency of both modes cancel in the ratio. 

Once all four \Lb-candidate tracks are found, the two tracks which have been
 matched to SVT tracks (one track from the $\Lc$ candidate, the other is 
the fourth track) must pass the two-track trigger requirements as described in 
Section~\ref{sec:detector}. 
Then, a four-track kinematic fit is performed. This fit includes 
two constraints. First, the daughter tracks of the \Lc\ must originate from a 
common, tertiary vertex. Second, the trajectory of the \Lc\ candidate must 
intersect with that of the remaining \Lb-candidate track, in three dimensions;
 this intersection is the decay vertex of the \Lb\ candidate (defined as the 
secondary vertex). The secondary and tertiary vertices are determined 
in the four-track kinematic fit simultaneously. These constraints improve the 
precision of the \Lc\ decay vertex determination and the invariant mass of the 
\Lc\ candidate is recalculated. After the kinematic fit, the values of 
$M_{pK\pi}$ must be in the range: 2.269~--~2.302~\gevcsq\ (2$\sigma$ around 
the mean) for the hadronic candidates and 2.18~--~2.38~\gevcsq\ for the 
inclusive semileptonic candidates (see \fig~\ref{fig:allfit}). The wider \Lc\ 
mass window for the semileptonic candidates allows for the $M_{pK\pi}$ spectrum 
to be fit to extract the yield \nincsemi. 
Also for the semileptonic decays, the four-track invariant mass $M_{\Lamc\mu}$ 
must be in the range of 3.7~--~5.64~\gevcsq, where the minimum 
requirement on $M_{\Lamc\mu}$ reduces the background from other $c$-hadron and 
$b$-hadron decays. See Section~\ref{sec:slbackgrounds} for more details.

In order to reduce the combinatorial backgrounds further, the selection 
criteria on the following variables are optimized: \pt\ of the proton track, 
\pt\ of the fourth \Lb-candidate track [$\pt\left(\pi^-, \mu^-\right)$], 
\pt\ of \Lc, \pt\ of the four-track system, 
\chixy\ of the \Lc\ and the four-track kinematic fits, 
proper decay length $\left(\ctau\right)$ of the \Lc\ 
candidate, and (pseudo) proper decay length $\left(\ctau^*\right)$ of the \Lb\ 
candidate. The \chixy\ is the \rphi\ plane 
contribution to the $\chi^2$ returned by the kinematic fit. The 
\ctau\ is defined as:
\begin{eqnarray}
\ctau \equiv \lxy^c \frac{M_{\Lamc}}{\pt\left(\Lc\right)},
\end{eqnarray}
where 
$\lxy^c$ is the projection of the vector from the secondary to the tertiary 
vertex onto the momentum vector of \Lc\ in the \rphi\ plane, and $M_{\Lamc}$ is
 the world average of the \Lc\ mass~\cite{pdg}. The $\ctau^*$ has a similar 
definition: 
\begin{eqnarray}
\ctau^* \equiv \lxy^b \frac{M_{\Lb}}{\pt\left(4_\mathrm{trks}\right)},
\end{eqnarray}
where $\lxy^b$ is the projection of the vector from the primary to the 
secondary vertex onto the total momentum vector of the four tracks in the 
\rphi\ plane, $\pt\left(4_\mathrm{trks}\right)$ is the transverse component 
of the total momentum of the four tracks, and $M_{\Lb}$ is the world average of
 the \Lb\ mass~\cite{pdg}. Here, the primary vertex is estimated from the 
intersection of the beam line and the trajectory of the \Lb\ candidate. 

The optimization procedure maximizes the signal significance of the hadronic 
decays, $S/\sqrt{S+B}$, where $S$ is the number of \lbhad\ events in 
simulation multiplied by a data-to-MC scaling factor and $B$ is the number of 
background events estimated from the $\Lc\pi^-$ candidates in the data 
sideband. The data-to-MC scaling factor for $S$ is obtained by comparing 
the number of signal events in data and simulation with relaxed 
requirements. 
 The background $B$ is estimated by fitting the mass sideband region above the 
\Lb\ signal peak with an exponential function and then extrapolating from the 
sideband region to the 3$\sigma$ signal region around the peak. The optimized 
selection criteria are listed in Table~\ref{t:fanacut}. 
\figc~\ref{fig:allfit}(a) shows the reconstructed $M_{\Lamc\pi}$ 
spectrum from the hadronic data and \fig~\ref{fig:allfit}(b) shows the 
reconstructed $M_{pK\pi}$ spectrum from the inclusive semileptonic data, both 
after applying the optimized selections. The most significant peaks in 
\fig~\ref{fig:allfit} represent the signals for each decay mode. 
In order to obtain the correct signal yields, a good modeling of the mass 
spectra, which includes a description of signal and background, is needed. 
The mass spectrum shapes of backgrounds from partially-reconstructed or 
misidentified $b$-hadron decays are determined by fitting the mass 
distributions from simulation. The next section describes details 
of the simulations used in this analysis.

%

 \begin{table}[tbp]
   \caption{Optimized requirements for reconstructing the 
	\lbhad\ and \inclbsemi\ decays. \label{t:fanacut}}
   \begin{center}
   \begin{normalsize}
  \renewcommand{\arraystretch}{1.5}
  \renewcommand{\tabcolsep}{0.15in}
  \begin{tabular}{ll} 
   \hline  
   \hline  
    \multicolumn{2}{c}{\lbhad} \\
    \multicolumn{2}{c}{\inclbsemi} \\ 
   \hline  
   $\pt\left(p\right)$ & $>$ 2 \gevc \\
   $\pt\left(\pi^-, \mu^-\right)$ & $>$ 2 \gevc \\
   $\pt\left(\Lc\right)$ & $>$ 5 \gevc \\
   $\pt\left(4_\mathrm{trks}\right)$ & $>$ 6 \gevc \\   
   $\chixy\left(\Lc\right)$ & $<$ 14 \\ 
   $\chixy\left(4_\mathrm{trks}\right)$ & $<$ 15 \\  
   $\ctau\left(\Lc\right)$ & $>$ $-70$~$\mu$m \\ 
   $\ctau^*\left(\Lb\right)$ & $>$ 250~$\mu$m \\ 
   \hline  
  \hline
 \end{tabular}
 \end{normalsize}
 \end{center}
 \end{table}

\begin{figure*}[tbp]
 \begin{center}
  \renewcommand{\arraystretch}{10}
 \begin{tabular}{cc}
 \includegraphics[width=250pt, angle=0]
	{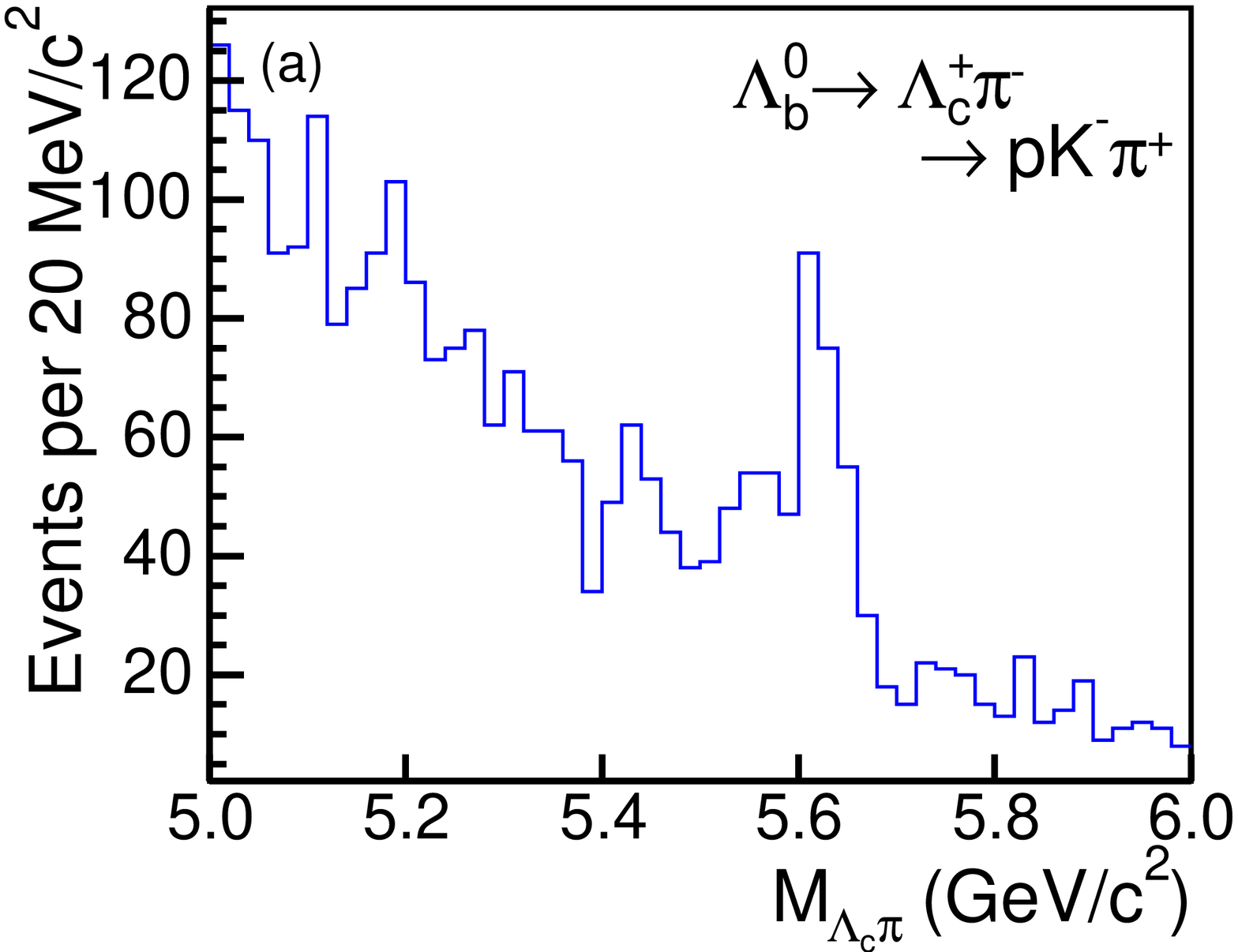} &
 \includegraphics[width=250pt, angle=0]
	{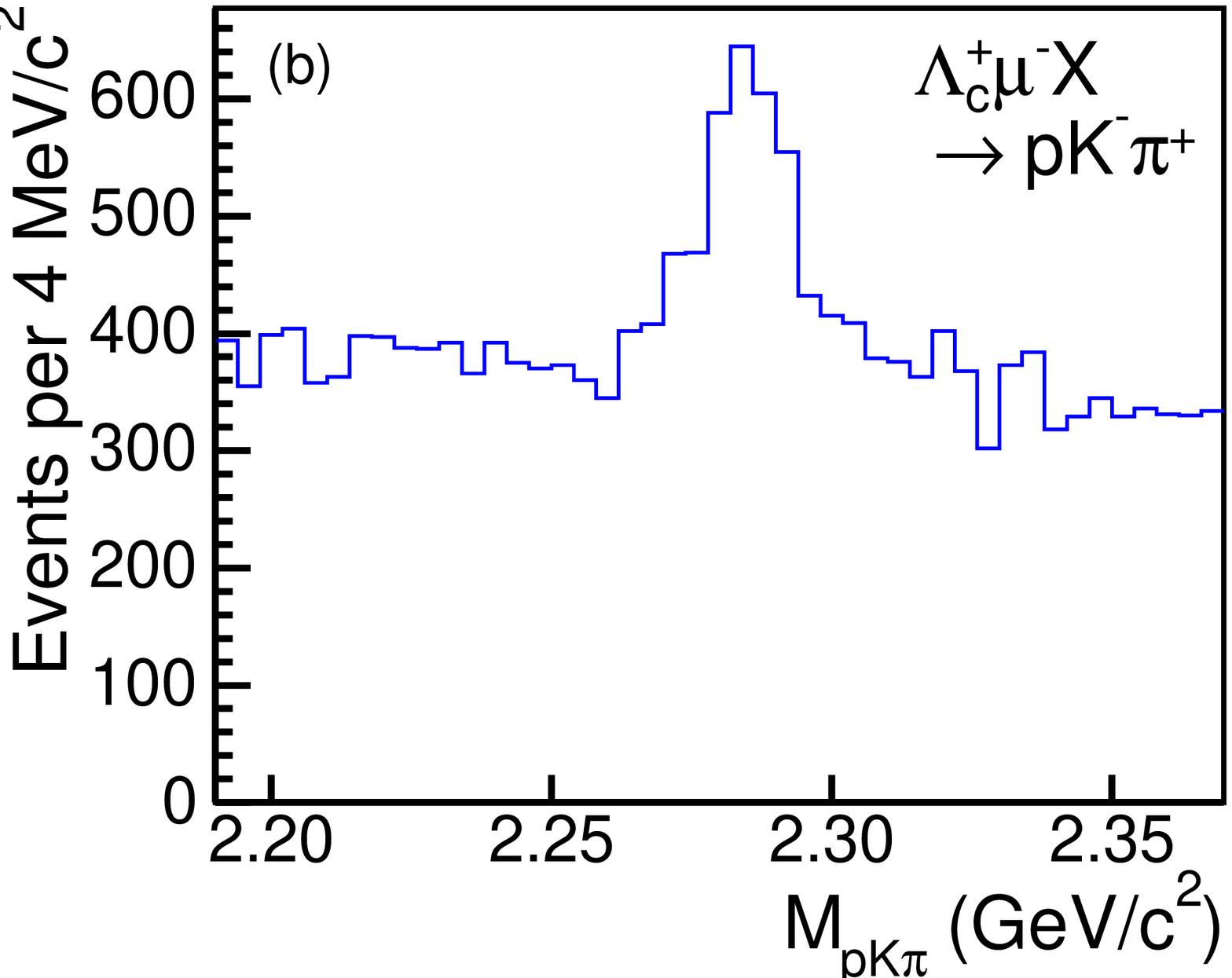} \\
 \end{tabular}
  \caption{The reconstructed invariant mass spectra in data after applying 
	all selection criteria: (a) the $M_{\Lamc\pi}$ spectrum of the \Lb\ 
	hadronic candidates; (b) the $M_{pK\pi}$ spectrum of the inclusive 
	semileptonic $\Lc\mu^-$ candidates. 
 \label{fig:allfit}}
\end{center}
 \end{figure*}

\section{\label{sec:simulation}Simulation}
In order to determine the mass spectrum shapes close to the signal peaks in 
\fig~\ref{fig:allfit} and to estimate the acceptance and efficiency of signal 
and background, both generator-level and full simulations are used. The 
generator-level simulation includes only the production and decay of $b$ 
hadrons, and the analysis requirements are applied to quantities immediately 
after generation. The full simulation includes simulation of the 
CDF~II detector and trigger, and track reconstruction. 
It was found that the efficiency ratios, $\effhad/\effsemi$, from a 
generator-level simulation and from a full simulation differ by only 
$\approx3\%$. The generator-level simulation is used to estimate the 
quantities which are found to be small or already have large uncertainties from
 other sources~\cite{whysmall}: the size of the background contribution where 
the \Lc\ and the $\mu^-$ originate from two different heavy-flavor hadrons 
produced by the fragmentation of $b\bar{b}$ or $c\bar{c}$ pairs 
(termed $b\bar{b}$/$c\bar{c}$ background), part of the \Lb\ 
systematic uncertainties (the semileptonic decay model and lifetime of \Lb, 
\Lb\ and \Lc\ polarizations, and \Lc\ Dalitz structure), and 
modification of the CDF I \rxsec\ result. Therefore, this 3$\%$
 difference has a negligible effect on the final measurement.  
 The following subsections describe the key components of the simulations used 
in this analysis. 

\subsection{Production and decay of $b$ hadrons\label{sec:production}}

Two different programs are used to simulate $b$-hadron production: \pythia\ 
version 6.2~\cite{pythia}, which simulates all of the strong interaction 
processes which are involved in $b$-hadron production, and \bgen~\cite{bgen}, which 
generates a single $b$ hadron in the event.
 Since \pythia\ simulates all of the products of the $p\bar{p}$ collision, it 
is computationally intensive to produce a given final state. Therefore, 
\pythia\ is used to estimate only the $b\bar{b}$/$c\bar{c}$ backgrounds in 
the inclusive semileptonic data (Appendix~\ref{sec:bbcc}). 
The \pythia\ generator simulates physics processes using leading-order matrix 
elements, supplemented by initial and final state radiation. The program also 
includes hadronization of the quarks and gluons in the final state and the 
beam remnants left when a parton undergoes high-momentum scattering. 
The \bgen\ program is very efficient at producing a large sample of a 
specific $b$-hadron under well-defined kinematic conditions.  
It is used to determine the acceptance and efficiency for signal and 
other backgrounds and to model the mass spectra. 
In the \bgen\ program, a single $b$ hadron is generated using the measured 
\pt\ spectra of $b$ hadrons as inputs. The \Lb\ and \Bd\ \pt\ spectra are 
derived from the fully-reconstructed \lbhad\ and \dhad\ 
decays in the two-track trigger data, after correcting for acceptance and 
efficiency.

After the event generation, the decays of the $b$ and $c$ hadrons and their 
daughters are simulated using the \evtgen\ package~\cite{evtgen}. For 
all other particles in the event, this is done by the \pythia\ program. 
The \evtgen\ program uses the dynamics from a full matrix-element calculation 
and is tuned to 
measurements, mainly results from experiments at the $\Upsilon\left(4S\right)$ 
resonance~\cite{babar:tdr,belle:tdr,Andrews:1982dp,Albrecht:1988vy} where the 
decay models for the $\Bd$ and the $B^-$ have been demonstrated to match data. 
As a full theoretical model for \Lb\ semileptonic decays is not yet implemented
 in \evtgen, a flat phase space (termed PHSP) simulation is used for \Lb\ 
decays. A correction is applied after generation to account for the proper \Lb\
 semileptonic decay dynamics. Details of this correction are given in 
Section~\ref{sec:scale}.

\subsection{\label{sec:cdfsim}Detector simulation and comparison of 
	kinematic distributions}
After an event has been simulated at the generator level, it is processed 
with a full simulation of the CDF II detector and trigger. The geometry and 
response of the active and passive detector components are simulated using the 
\geant\ software package~\cite{geant}. The events are then processed 
with a two-track trigger decision program and reconstructed using the same 
executable as that used to reconstruct the data. 
The resulting events have the same structure and format as the data and are 
analyzed in the framework described in 
Section~\ref{sec:reconstruction}.

\begin{figure*}[tbp]
     \begin{center}
        \includegraphics[width=250pt, angle=0]
	{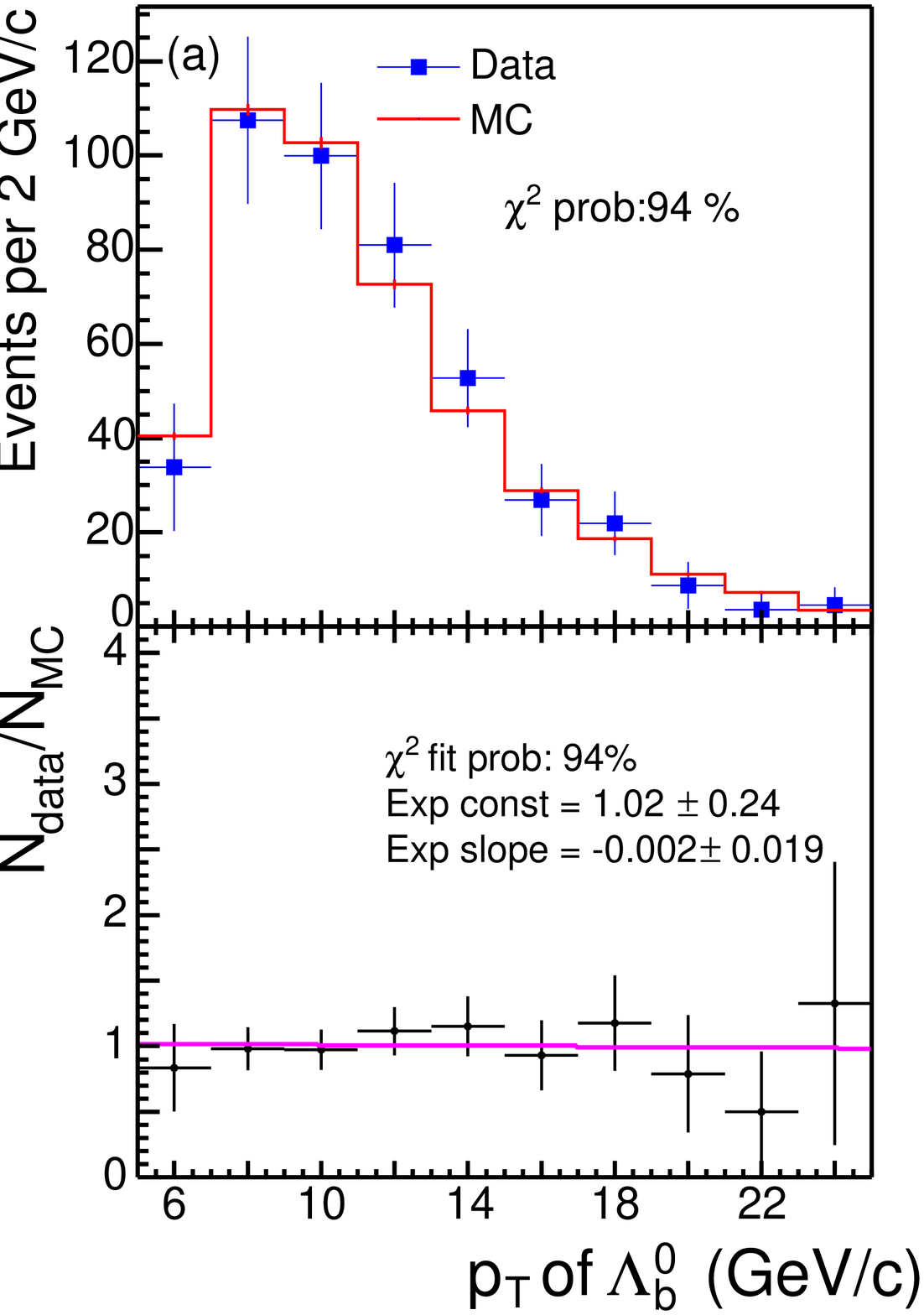}
        \includegraphics[width=250pt, angle=0]
	{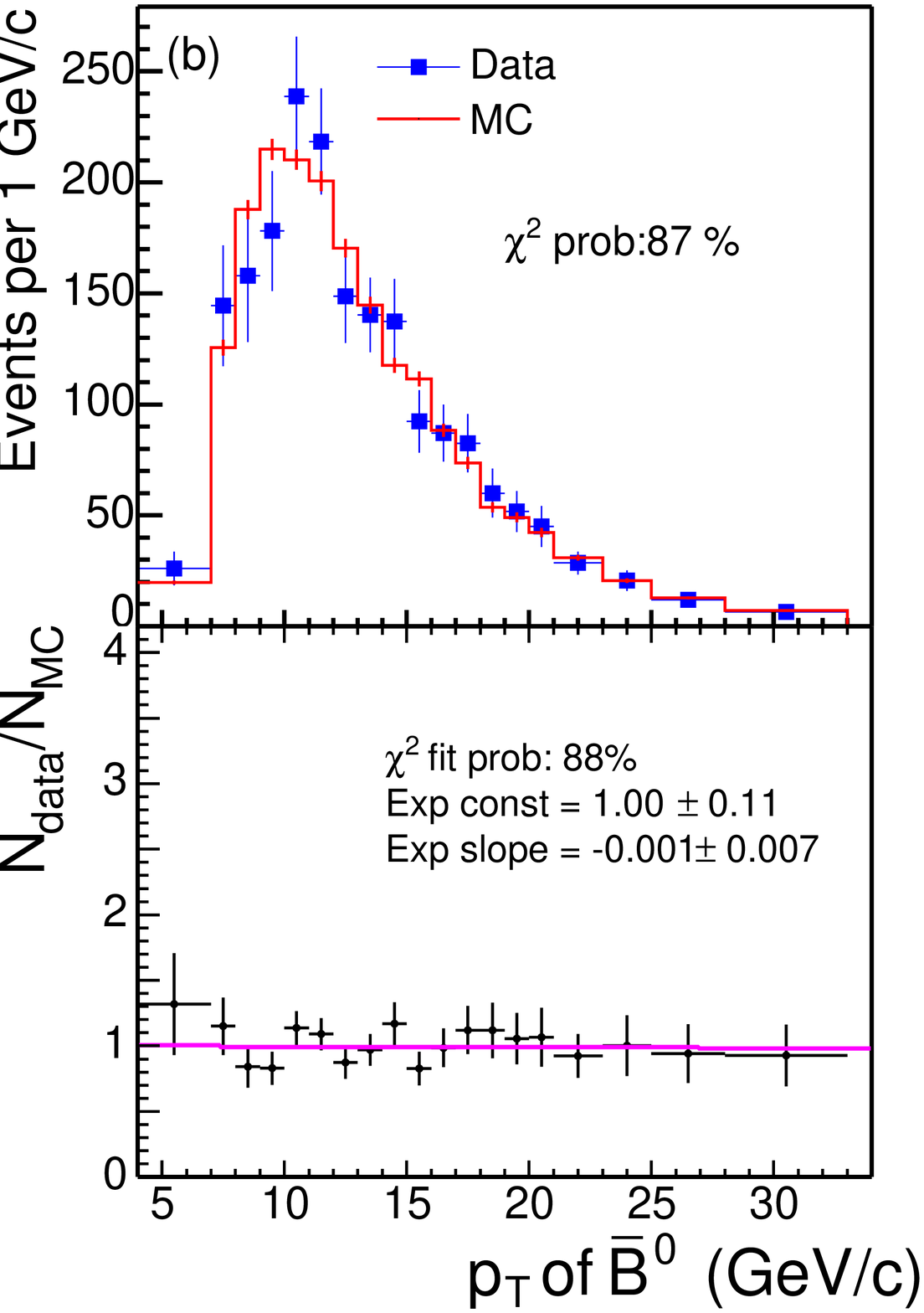}
        \caption
	{Comparison of reconstructed \pt\ spectra of $b$-hadrons 
	between the data and the full simulation: (a) $\Lb$ and (b) \Bd. 
	The top figures show the \pt\ distributions while the bottom 
         figures show the ratio of data to MC\@. The curves 
	in the bottom figures
         are the result of an exponential fit of the ratio.
     \label{fig:bptafter}}
     \end{center}
 \end{figure*}

Distributions of kinematic variables from the full simulation with 
\bgen\ input are compared with the same distributions from data. In order to 
compare the data and the simulation, the data distributions are 
background-subtracted. The agreement between the data and the simulation is 
quantified by a $\chi^2$ comparison probability and the ratio of 
spectra produced from the data and the simulation. 
All relevant distributions agree satisfactorily. 
\figc~\ref{fig:bptafter} shows good agreement between the data and the 
simulation in the $\pt\left(\Lb\right)$ and $\pt\left(\Bd\right)$ spectra. 
The \pt\ of the $b$ hadron is the most important kinematic variable in 
this analysis because the $b$-hadron momentum is distributed among three 
particles in the final state for the exclusive semileptonic decay and between 
two particles for the hadronic decay.

\subsection{Acceptance and efficiency scale factors \label{sec:scale}}
\begin{figure*}[tbp]
     \begin{center}
      \includegraphics[width=250pt, angle=0]
	{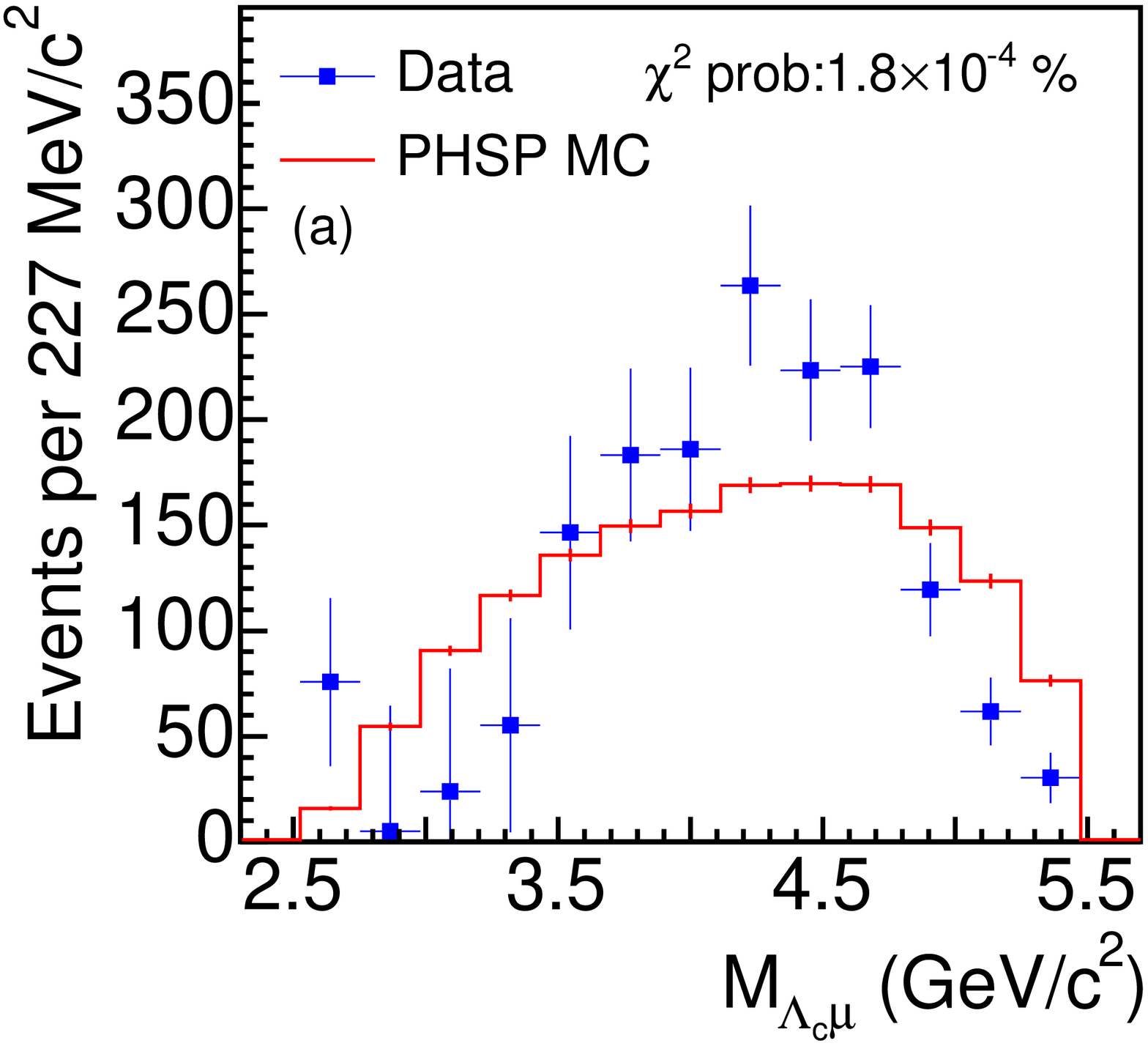}
      \includegraphics[width=250pt, angle=0]
	{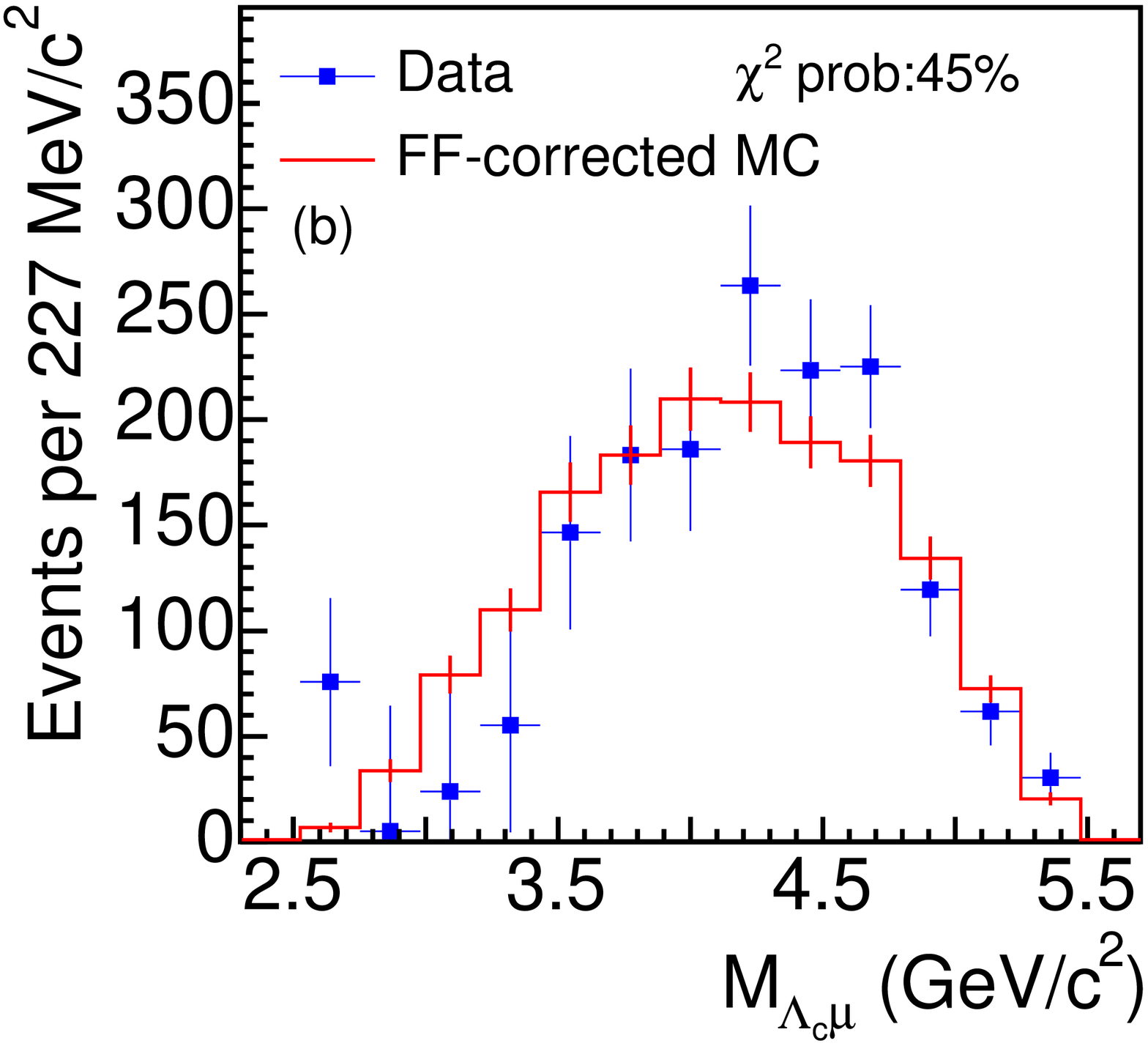}
	\caption
	{ Comparison of reconstructed $M_{\Lamc\mu}$ between the data and the 
	phase space (PHSP) full simulation, (a) before and (b) after  
	the MC histogram is corrected to account for the 
	proper \Lb\ semileptonic decay dynamics~\cite{ratio}. Note that the 
	feed-in backgrounds which are present in the $\Lc\mu^-$ sample are 
	already included in the simulation.}
 	\label{fig:mcdatalcsemi1}
     \end{center}
  \end{figure*}

In order to obtain accurate estimates of the acceptance and efficiency, 
several scale factors are applied to the number of events 
selected in simulation and their values are listed in Table~\ref{t:effscale}. 
As mentioned earlier, \evtgen\ contains only a phase space simulation of 
semileptonic \Lb\ decays. 
In order to estimate the effect of decay models on the signal acceptance, a 
weighting of the flat 
phase space distribution according to a form factor model from Huang~\etal~\cite{Huang:2005me} for the hadronic current of the \Lb\ to \Lc\ transition, and 
a $V-A$ model for the leptonic current, is performed at the generator level. 
The ratio of the generator-level acceptance after weighting relative to that 
before weighting, $C_\mathrm{model}$, is found to be $0.994 \pm 0.025$. 
Since this ratio is consistent with unity, the PHSP full simulation samples are 
used throughout the \Lb\ analysis. The correction factor, $C_\mathrm{model}$, 
which accounts for the \Lb\ semileptonic decay dynamics, is applied to the 
efficiencies for semileptonic decays. 
The shape of the $M_{\Lamc\mu}$ distribution is sensitive to the decay 
dynamics and may be used to cross-check the form factor and $V-A$ models 
(termed FF). 
\figc~\ref{fig:mcdatalcsemi1} shows the reconstructed $M_{\Lamc\mu}$ 
distributions from the data and from the PHSP full simulation, before and after 
multiplying the MC histogram with the bin-by-bin ratios which are derived from the 
same generator-level simulation samples for $C_\mathrm{model}$~\cite{ratio}. 
The corrected distribution has a significantly improved 
agreement with the data, which confirms the procedure for 
deriving $C_\mathrm{model}$.

In addition, the CMU muon reconstruction efficiency is found to be 
over-estimated in the full simulation; the resulting scale factor, 
$C_\mathrm{CMU}$, is measured using a sample of $J/\psi\rightarrow\mu^+\mu^-$ 
decays~\cite{CDF}. The dependence of the XFT trigger efficiency on the 
particle species and \pt\ is not included in the full simulation. 
Using a pure proton sample from the $\Lambda^0\rightarrow p^+\pi^-$ decays, and
 pure kaon and pion samples from the $\dstar\rightarrow \dzero \pi^+$ decays 
where $\dzero\rightarrow K^-\pi^+$~\cite{Acosta:2004ts,yile:lblcpi,Yu:2005hw}, 
the data-to-MC scaling factors, $C_p$, $C_K$, and $C_{\pi}$, are derived and 
applied to the track which passes the trigger requirements in the 
reconstruction program. 

With a reliable simulation for the modeling of mass spectrum shapes, the 
numbers of signal events can be determined by fitting the invariant 
mass spectra in \fig~\ref{fig:allfit} as described in the following section.   

 \begin{table}[ptb]
  \renewcommand{\arraystretch}{1.5}
   \caption{Acceptance and efficiency scale factors applied to the number of 
	events selected in simulation. The \pt\ is the transverse momentum 
	(in \gevc) of the track which passes the trigger requirements.
        The uncertainty on $C_p$ is obtained by taking the difference between 
        the \pt-dependent formula below and a constant from an 
	average over the $\Lambda^0$ sample, 0.905.  
        The uncertainties on $C_K$ and $C_{\pi}$ are below $0.5\%$ and 
        have negligible effect on the final relative branching fractions.
        \label{t:effscale}}
   \begin{center}
   \begin{tabular}{lr@{\,$\pm$\,}l}
    \hline 
    \hline
     \multicolumn{1}{c}{Scale Factor} & \multicolumn{2}{c}{Value} \\
     \hline
     $C_\mathrm{model}$ & 0.994	& 0.025 \\
     $C_\mathrm{CMU}$ & 0.986 & 0.003 \\
     $C_p$ & \multicolumn{2}{l}{$1.06 -\frac{1.3}{\pt} + \frac{3.2}{\pt^2} 
	-\frac{2.2}{\pt^3}$} \\
     $C_K$ & \multicolumn{2}{l}{$0.969- \frac{0.094}{\pt}$} \\      
     $C_{\pi}$ & \multicolumn{2}{l}{ $1.002 - \frac{0.067}{\pt}$} \\
     \hline	
      \hline
      \end{tabular}
      \end{center}
  \end{table}

\section{\label{sec:fitting}Determination of the Signal Yields}
The numbers of hadronic events $\left(\nhad\right)$ and inclusive semileptonic 
events $\left(\nincsemi\right)$ in \eq~(\ref{eq:bigpicture}) are extracted by 
fitting the $M_{\Lamc\pi}$ and $M_{pK\pi}$ spectra in data, respectively. 
The fit to the mass spectra is performed using an unbinned, extended likelihood
 technique~\cite{Cowan:1998ji}, where the fit parameters are adjusted to 
minimize the negative log-likelihood $\left(-\ln \cal L\right)$. 
The general unbinned, extended log-likelihood is expressed as:
\begin{eqnarray}
 \ln {\cal L} &= & \sum_i\ln\left[ N_\mathrm{sig}S\left(m_i\right)
	+ N_\mathrm{bg} B\left(m_i\right) \right] 
	\nonumber \\
	& &  - N_\mathrm{sig}- N_\mathrm{bg} + \sum_j \ln{\cal C}_j, 
\label{eq:generallog}
\end{eqnarray}
where $i$ represents the $i^\mathrm{th}$ candidate and $m$ represents the 
reconstructed mass $M_{\Lamc\pi}$ or $M_{pK\pi}$. The numbers of signal and 
background events are denoted as $N_\mathrm{sig}$ and $N_\mathrm{bg}$; $S\left(m_i\right)$ and $B\left(m_i\right)$ are the 
normalized functions which describe the shapes of signal and background mass 
spectra, respectively. 
Each ${\cal C}_j$ is a Gaussian constraint on a specific 
fit parameter $x_j$: 
\begin{eqnarray} 
 {\cal C}_j = {\cal G}\left(x_j,\mu_j,\sigma_j\right)
	={1\over\sqrt{2\pi}\sigma_j}e^{-{1\over2}\left({\left(x_j-\mu_j\right)\over\sigma_j}\right)^2},
 \label{eq:gconstraint}
 \end{eqnarray}
where the parameter $x_j$ has a central value of $\mu_j$ and an 
uncertainty of $\sigma_j$. 
Because the data sample size is not large enough to determine these parameters 
accurately from the fit, they are constrained to values which are estimated 
from independent measurements and the full simulation. Definitions of the 
constrained parameters ($x_j$) are given in Section~\ref{sec:lcpifit}. 

Correct modeling of the mass spectra is crucial in the determination of $\nhad$
 and $\nincsemi$. 
Two types of background appear in each mass window of interest. 
The first is combinatorial background. Combinations of four random tracks 
contribute to this background in both the hadronic and semileptonic modes. 
Combinations of a real $c$ hadron and a random track contribute only in the 
hadronic mode. The mass spectrum of the combinatorial background is determined 
using data sidebands. The second background is misidentified or partially-reconstructed decays of $b$ hadrons. Their mass spectrum shapes are determined using the simulations as described in Section~\ref{sec:simulation}. 
The dominant contributing decays
 are identified with a generator-level simulation of inclusive 
$b$-hadron decays, and are categorized according to their mass spectrum shapes.
 Decay modes with similar shapes are 
generated together using a full simulation, with the number of 
generated events for each decay mode proportional to the fragmentation fraction
 times the branching ratio, and are parameterized by a single
 function. The functional form for each combined background spectrum is 
determined empirically to match the shape of simulated 
mass distribution. The parameter values of each function are obtained by fitting the 
simulated spectrum. When fitting data, the values of the shape parameters 
are fixed while the normalization is a free parameter. 

%
%
\subsection{The $M_{\Lamc\pi}$ spectrum for the \lbhad\ yield 
	\label{sec:lcpifit}}
\figc~\ref{fig:allbg}(a) shows the $M_{\Lamc\pi}$ spectrum with the fit 
result superimposed. 
The likelihood fit is performed in the mass 
window $M_{\Lamc\pi}=4.6-7.0$~\gevcsq, whereas \fig~\ref{fig:allbg}(a) shows 
a more restricted mass range near the signal peak. 
The \lbhad\ yield returned by the fit is $\nlbhadc\pm\nlbhade$. 
The signal peak at $M_{\Lamc\pi}\approx 5.6$~\gevcsq\ is described by a 
Gaussian function. The width of the Gaussian is constrained in the fit to 
reduce the uncertainty on $N_\mathrm{sig}$. The constrained width is the 
product of a data/MC scale factor and the width of the $M_{\Lamc\pi}$ 
distribution in the full simulation, 
\(\left(\sigma^\mathrm{data}_{D\pi}/\sigma^\mathrm{MC}_{D\pi}\right)
  \times \sigma^\mathrm{MC}_{\Lamc\pi},
\)
which is $0.0231\pm 0.0012$~\gevcsq. The scale factor, 
$\sigma^\mathrm{data}_{D\pi}/\sigma^\mathrm{MC}_{D\pi}$, 
is obtained by comparing the width of the invariant mass distribution in data 
with that in the simulated events, using the \dhad\ decay which has a similar 
topology and a larger data sample size. 
The combinatorial background is parameterized by an exponential 
(light-gray filled region), where the exponential 
slope is determined by the $\Lc\pi^-$ candidates in the mass region above 
5.7~\gevcsq. 
The functions which describe the mass spectra of backgrounds from the misidentified or partially-reconstructed $b$-hadron decays are determined from the 
simulated mass distributions. 

Details of the background from the misidentified or partially-reconstructed 
$b$-hadron decays follow. 
The doubly Cabibbo-suppressed decays $\Lb \rightarrow \Lc K^-$, with a pion mass mistakenly assumed for the kaon, are indicated by 
the black filled region. 
The ratio of the number of doubly Cabibbo-suppressed decays relative to that 
of the signal mode, $N_{\Lamc K}/N_{\Lamc\pi}$, is
 fixed to $8\%$ in the fit. This value is obtained from the world average of 
measurements in the $\Bd$ modes.   
Fully reconstructed $b$-meson decays with misidentified daughters produce a 
distinct peak at $M_{\Lamc\pi}\approx 5.5$~\gevcsq\ (wavy-line region). The 
\dhad\ decays, where $\dplus \rightarrow K^-\pi^+\pi^+$ and one of the pions is 
reconstructed as a proton, contribute about $50\%$ to this background. The 
background from the remaining partially-reconstructed $b$-meson decays has a 
monotonically falling mass distribution (dark-gray filled region) dominated by 
\dsemi, $\bar{B}^0\rightarrow \dplus\rho^-$ where 
$\rho^-\rightarrow \pi^0\pi^-$, and \dstarhad\ where
 $\dstar\rightarrow \dplus\pi^0$ and the $\pi^0$'s are not reconstructed 
in the event. 
The remaining \Lb\ decays also have a falling mass spectrum (hatched region) 
dominated by $\Lb\rightarrow \Lc \ell \bar{\nu}_{\ell}$ 
and $\Lb\rightarrow \Lc\rho^-$ where $\rho^-\rightarrow \pi^0\pi^-$.

\subsection{The $M_{pK\pi}$ spectrum for the \inclbsemi\ yield 
	\label{sec:lcmufit}}
\figc~\ref{fig:allbg}(b) shows the $M_{pK\pi}$ spectrum for events with muons, 
with the fit result superimposed. 
The inclusive \inclbsemi\ yield returned by the fit is 
$\nlbsemic\pm\nlbsemie$. 
The fit for the $M_{pK\pi}$ spectrum is less complex than that for the 
$M_{\Lamc\pi}$ spectrum described above. Note that the signal peak includes the
 backgrounds which also contain $\Lc\mu^-$ in the final state 
(see Section~\ref{sec:slbackgrounds}). The signal peak at $M_{pK\pi}\approx 2.3$~\gevcsq\ is modeled by a Gaussian function. 
Background from the $b$-hadron semileptonic decays with a $c$-hadron daughter 
misidentified as a \Lc, such as \dsemi\ where $\dplus \rightarrow K^-\pi^+\pi^+$ 
and one of the pions is assigned the proton mass, does not produce a peak or 
distinctive structure and is inseparable from the combinatorial background. 
These two backgrounds are combined and modeled by a second-order polynomial 
(light-gray filled region).

%
\subsection{Summary}
Table~\ref{t:yieldsum} summarizes the \Lb\ hadronic and inclusive 
semileptonic yields and the $\chi^2$ probability of corresponding fits. 
Each model describes the data well, as indicated by the $\chi^2$ probability. 
In order to obtain the number of exclusive semileptonic signal events \nsemi, 
the contributions from backgrounds which also 
produce a $\Lc$ and a $\mu^-$ in the final state, $\nbg$, must 
be subtracted from $\nincsemi$. Section~\ref{sec:slbackgrounds} describes the 
estimation of the composition of the \Lb\ inclusive semileptonic data sample.
 Section~\ref{sec:lbbr} details observations of four new \Lb\ semileptonic 
decays and the estimates of \Lb\ semileptonic and hadronic branching ratios 
which are required to determine the 
sample composition in Section~\ref{sec:slbackgrounds}.

 \begin{table}[ptb]
  \renewcommand{\arraystretch}{1.5}
   \caption{Observed number of events in each \Lb\ decay mode determined from 
	the unbinned, extended likelihood fit, $\chi^2/$NDF, 
	and the corresponding probability computed to indicate 
	quality of the fit. \label{t:yieldsum}}
   \begin{center}
   \begin{tabular}{lrcc}
    \hline 
    \hline
     \multicolumn{1}{c}{Mode} & \multicolumn{1}{c}{Yield} &
     \multicolumn{1}{c}{$\chi^2/$NDF} & Prob $\left(\%\right)$\\	
     \hline
      \lbhad & $\nlbhadc \pm \nlbhade$ & 123/111 & 20.7 \\
      \inclbsemi & $\nlbsemic \pm \nlbsemie$ & 48/38 & 13.0 \\
     \hline	
      \hline
      \end{tabular}
      \end{center}
  \end{table}

\begin{figure*}[tbp]
 \begin{center}
  \renewcommand{\arraystretch}{10}
 \begin{tabular}{cc}
 \includegraphics[width=250pt, angle=0]
	{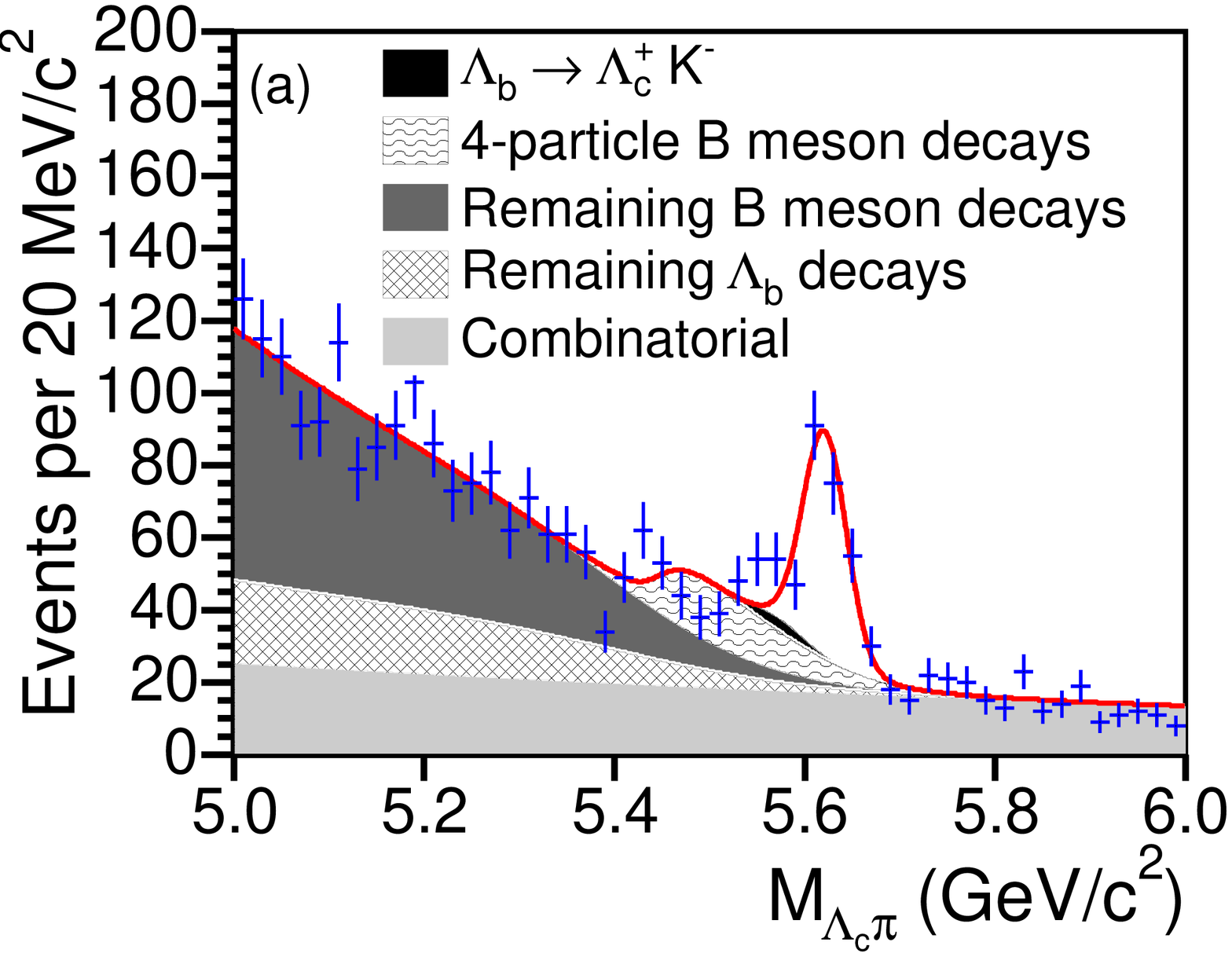} &
 \includegraphics[width=250pt, angle=0]
	{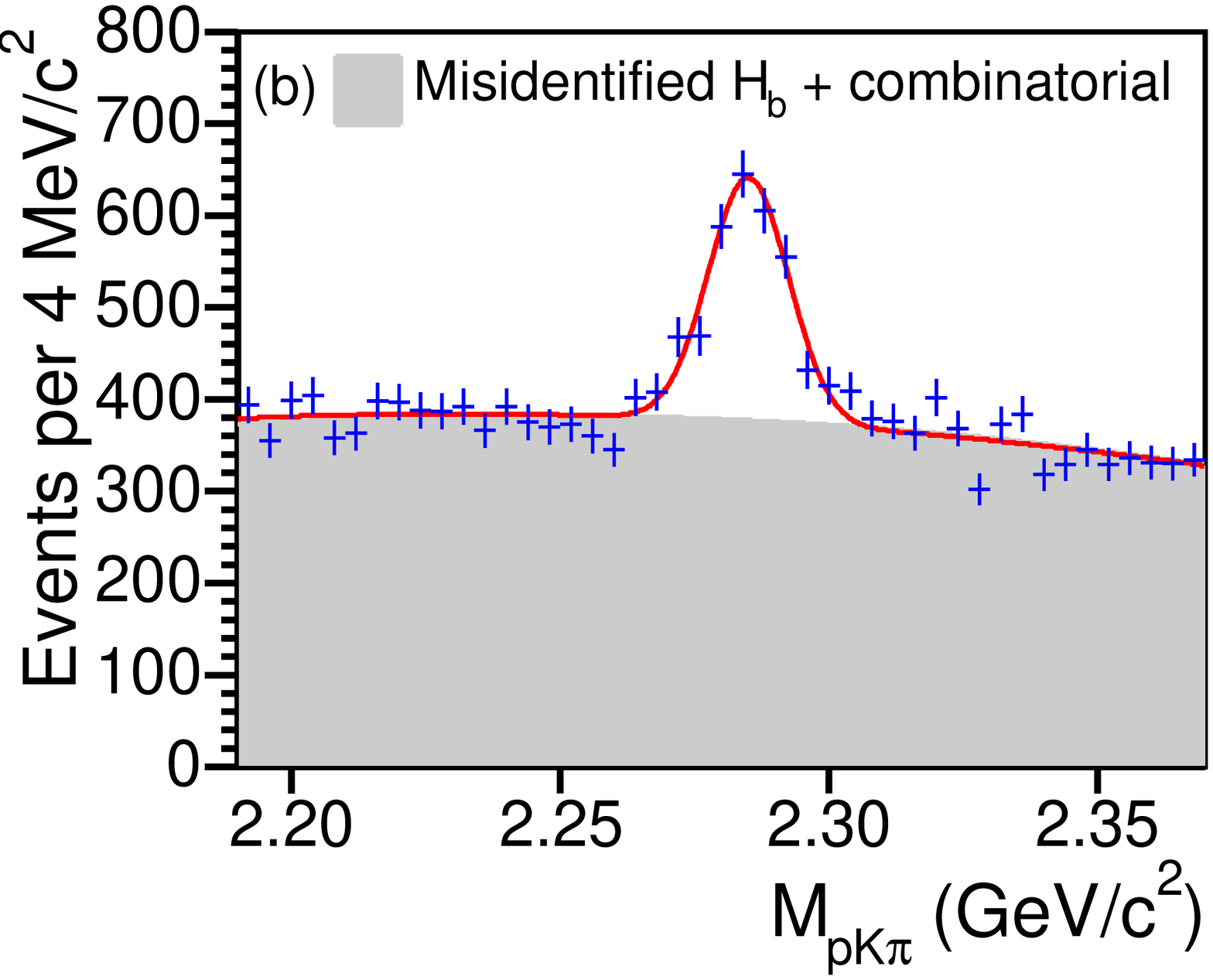} \\
 \end{tabular}
  \caption{Results (curve) of the unbinned, extended likelihood fits for 
determining the numbers of \Lb\ candidates: (a) hadronic and (b) inclusive 
semileptonic. The filled histograms indicate various backgrounds. 
\label{fig:allbg}}
\end{center}
 \end{figure*}
\section{\label{sec:slbackgrounds}Composition of 
	the Inclusive Semileptonic Data}
The $B$ factories~\cite{babar:tdr,belle:tdr,Andrews:1982dp,Albrecht:1988vy} 
produce $b$ hadrons in $e^+e^-$ interactions where the beam 
energy may be used as a constraint when reconstructing events. This feature 
is particularly helpful for reconstructing semileptonic decays where a 
neutrino is missing. At the Tevatron, $b$ hadrons are produced by the 
interactions between quarks and gluons with a broad parton momentum spectrum. 
Therefore, beam energy constraints are not available to aid $b$-hadron 
reconstruction. Backgrounds which contain a \Lc, a $\mu^-$, and other 
particles in the final state can not be separated easily from the exclusive 
semileptonic signal, \lbsemi, and will contribute to the inclusive $\Lc\mu^-$ 
events observed in data. These backgrounds arise from three sources:
 \begin{enumerate}
\item false muon: a \Lc\ and a hadron track 
      ($\bar{p}$, $K^-$ or $\pi^-$) misidentified as a $\mu^-$.
\item $b\bar{b}$/$c\bar{c}$: a \Lc\ from the decay of a heavy-flavor hadron 
$H_b$ ($H_c$) and a $\mu^-$ from the decay of the other heavy-flavor hadron 
$\bar{H_b}$ ($\bar{H_c}$), where the two hadrons are produced by 
the fragmentation of $b\bar{b}$ ($c\bar{c}$) pairs.
\item feed-in: decays of a single $b$ hadron into a \Lc, a $\mu^-$, 
      and particles not reconstructed in data.
\end{enumerate}
The goal is to measure the branching fraction of the exclusive 
semileptonic decay relative to that of the hadronic decay. The backgrounds 
listed above must be subtracted from the observed number of inclusive 
semileptonic events in data. 
\eqc~(\ref{eq:bigpicture}) is then re-written as follows:
\begin{eqnarray}
\label{eq:bigf}
\lefteqn{\frac{\bsemi}{\bhad} =} \nonumber \\
& & \left(\frac{\nincsemi-\nfake -\nbbcc-\nphys}{\nhad}\right) 
 \frac{\effhad}{\effsemi}. 
\end{eqnarray}
The number of false-muon events $\left(\nfake\right)$ is obtained 
from data containing a \Lc\ and a hadron track satisfying reconstruction 
requirements, with the hadron track weighted by an appropriate 
muon-misidentification probability. 
The contributions from the $b\bar{b}$/$c\bar{c}$ $\left(\nbbcc\right)$
 and the feed-in backgrounds $\left(\nphys\right)$ are estimated using 
both data and simulation. 
Instead of the absolute amount, the ratios $\nbbcc/\nhad$ and 
$\nphys/\nhad$ are estimated. Estimating the ratios instead of the absolute 
amount has one advantage: the majority of the background events are decays of 
\Lb, so knowledge of the \Lb\ production cross-section is not necessary. 
The quantities $\nbbcc/\nhad$ and $\nphys/\nhad$ are determined from the ratios
 of the products of branching fractions and efficiencies 
(times production cross-section for non-\Lb\ background). 
The normalization procedure requires measurements or 
estimates of the branching fractions for the \lbhad\ decay and for several 
semileptonic decays which may contribute to the backgrounds; details of 
these measurements and estimates are found in Section~\ref{sec:lbbr}. 
The ratio $\nbbcc/\nhad$ has been estimated to be very small and contributes 
$\le 1\%$ to the \lbsemi\ signal. More information on $b\bar{b}$ and 
$c\bar{c}$ backgrounds may be found in Appendix~\ref{sec:bbcc}. 
The following sections describe the estimation of \nfake\ and \nphys.

\subsection{\label{sec:fakemu}False muons}
\begin{figure}[tbp]
    \begin{center}
  \includegraphics[width=200pt, angle=0]{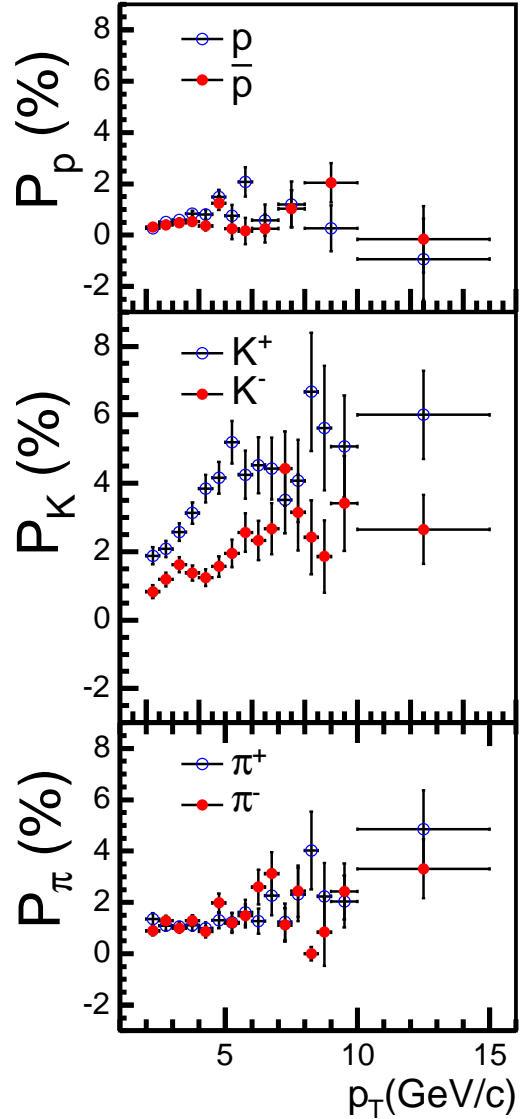} 
   \caption
    {The probability for a proton, kaon or pion to be misidentified as a muon 
	as a function of \pt~\cite{ash:dmm}. Note that for the measurements 
        with negative values, a zero muon-misidentification probability is 
        used to weight data.}
     \label{fig:allfakerate}
     \end{center}
  \end{figure}

 One type of semileptonic background is due to the pairing of a \Lc\ with a 
proton, a kaon, or a pion which is misidentified as a muon. A hadron is 
misidentified as a muon when it passes through the calorimeter into the muon 
detector, or when it decays into a muon in flight. The probabilities for a 
proton, kaon, or pion to be misidentified as a muon (${\cal P}_{p}$, 
${\cal P}_K$, and ${\cal P}_{\pi}$, respectively) are measured using 
a pure proton sample from the $\Lambda^0\rightarrow p^+\pi^-$ decays, and pure 
$K$ and $\pi$ samples from the $\dstar\rightarrow \dzero \pi^+$ decays where 
$\dzero\rightarrow K^-\pi^+$~\cite{ash:dmm}. The muon-misidentification 
probability is defined as the fraction of the CMU-fiducial and SVT-matched 
hadron tracks which satisfy the muon identification requirement (a track 
associated with hits in the CMU and with a matching $\chi^2$ less than 9). 
\figc~\ref{fig:allfakerate} shows the ${\cal P}_{p}$ (measured in twelve 
\pt\ bins) and ${\cal P}_{\pi}$, ${\cal P}_K$ 
(measured in sixteen \pt\ bins) for positively and negatively charged 
tracks, separately. A difference is observed between ${\cal P}_{K^+}$ 
and ${\cal P}_{K^-}$ in the low \pt\ region, which is not seen for 
protons and pions. The larger hadronic cross-section for the $K^-p$ scattering 
relative to that for the $K^+p$ scattering results in a lower rate of 
$K^-$ being misidentified as muons passing through the calorimeter. 

\begin{figure}[tbp]
 \begin{center}
 \includegraphics[width=250pt, angle=0]{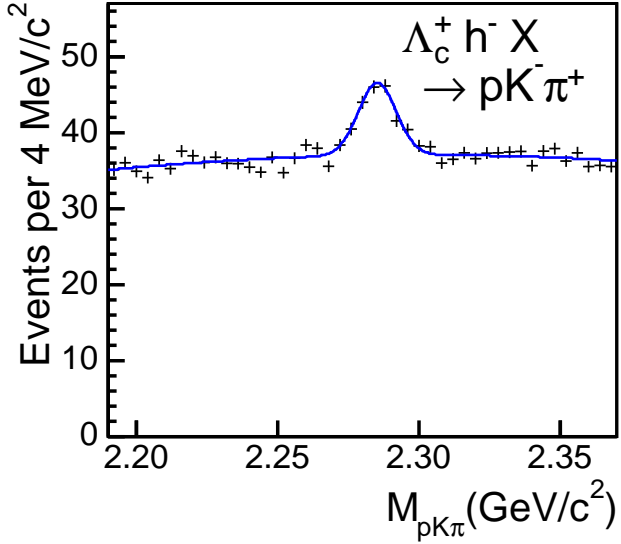} 
  \caption
   { Distribution of $M_{pK\pi}$ produced from data with a hadron track 
	$\left(h^-\right)$ and a \Lc\ candidate in the final state, 
	after weighting the hadron track with an average 
	muon-misidentification probability $\left({\cal P}_\mathrm{avg}\right)$. 
	The curve indicates the result of the $\chi^2$ fit.
     \label{fig:cmfake}}
     \end{center}
  \end{figure}

The contribution of the false-muon background to the \lbsemi\ signal, \nfake, 
is obtained by weighting data containing a \Lc\ and a hadron track 
$\left(h^-\right)$, with 
the muon-misidentification probability $\left({\cal P}_\mathrm{avg}\right)$ 
as a function of the momentum of $h^-$. 
This hadron track must extrapolate to the fiducial region of the CMU and fail 
the muon identification requirements in order to remove real muons. The 
other selection criteria for the \Lc $h^-$ sample are the same as 
those for the \lbsemi\ reconstruction. 
The $\nfake$ is then extracted from a $\chi^2$ fit of the 
$M_{pK\pi}$ distribution produced from the weighted \Lc$h^-$ sample. 
\figc~\ref{fig:cmfake} shows the result of the $\chi^2$ fit. 

Since no particle identification requirement is applied, whether $h^-$ is a 
proton, a kaon, or a pion can not be determined from data. 
The muon-misidentification probability, ${\cal P}_\mathrm{avg}$,
 is, therefore, an average of ${\cal P}_{p}$, ${\cal P}_K$, and ${\cal P}_{\pi}$ weighted by $F_{p}$, $F_K$, and $F_{\pi}$ (the fractions of $p$, 
$K$, $\pi$ in $h^-$):
\begin{eqnarray}
{\cal P}_\mathrm{avg} = F_p{\cal P}_{p}+ F_K{\cal P}_{K} + F_{\pi}{\cal P}_{\pi}.
\end{eqnarray}
In order to determine $F_{p}$, $F_K$, and $F_{\pi}$, physics processes which 
produce these hadrons must be understood. 
The principal sources of these hadrons after analysis requirements are the 
decays  
$H_b \rightarrow \Lc h^- X$, where $h^-$ is a $\bar{p}$, $K^-$, or $\pi^-$ misidentified as a muon and $X$ could be nothing (\eg, \lbhad) or 
any other particles which are not reconstructed 
(\eg, $B^- \rightarrow \Lc \bar{p} \mu^- \bar{\nu}_{\mu}$). 
Other sources include: fragmentation of a primarily produced quark or gluon, 
inelastic collisions of secondary particles with the detector material, 
and decays of $c$ hadrons. 
Hadrons which are not from $b$-hadron decays are suppressed 
by requiring that the transverse impact parameter $\left(d_0\right)$
 of the muon candidate is in the range 120~$\mu$m--1000~$\mu$m, and that the 
\Lc\ and the muon candidates form a vertex significantly displaced 
from the beam line (see Section~\ref{sec:reconstruction}). 
In addition, the \pythia\ simulation indicates that the background 
where a false muon and a \Lc\ signal originate from decays of 
two different $b$ or $c$ hadrons is less than 0.1$\%$ of the inclusive 
semileptonic signal and can be ignored. 
 Therefore, $F_{p}$, $F_K$, and $F_{\pi}$ are obtained from the
 $H_b \rightarrow \Lc h^- X$ full simulation.

Table~\ref{t:bigfake1} shows values obtained for $F_{p}$, $F_K$, $F_{\pi}$, and
 $\nfake$. The uncertainty on $\nfake$ includes: the statistical 
uncertainty from the $\chi^2$ fit, the uncertainties on $F_{p}$, $F_K$, and 
$F_{\pi}$, and the uncertainties on the measured muon-misidentification 
probabilities. The \nfake\ is approximately \fracfakemu\% of the number of 
\inclbsemi\ events.

\renewcommand{\arraystretch}{1.5}
  \begin{table}[h]
   \caption{The fractions of $p$, $K$, and $\pi$ in the 
	$h^-$ ($F_{p}$, $F_K$, and $F_{\pi}$), the estimated 
	number of false-muon events to the \lbsemi\ signal, and for 
	comparison, the number of the inclusive semileptonic events in data.}
  \label{t:bigfake1}
   \begin{center}
   \begin{small}
  \begin{tabular}{lr} 
   \hline
   \hline
    $F_{p}$ & $0.24 \pm 0.16$ \\
    $F_{K}$ & $0.05 \pm 0.08$ \\
    $F_{\pi}$ & $0.71 \pm 0.16$ \\
   \hline
    \nfake  
	& $40 \pm 9$ \\ 
    \nincsemi   
   & $\nlbsemic \pm \nlbsemie$ \\
    \hline 
   \hline
 \end{tabular}
 \end{small}
 \end{center}
\end{table}

\subsection{\label{sec:physicsb}Feed-in backgrounds}
 The feed-in backgrounds to the \lbsemi\ signal fall into three 
categories:
\begin{enumerate}
 \item $\nphys^\mathrm{meson}$: Baryonic, semileptonic decays of $\bar{B}^0$/$B^-$/$\bar{B}_s^0$, 
which decay into \Lc, an anti-nucleon and leptons 
(\eg, $B^- \rightarrow \Lc \bar{p} \mu^- \bar{\nu}_{\mu}$).
 \item $\nphys^\mathrm{b-baryon}$: Semileptonic decays of other $b$ baryons 
(\eg, $\Xi_b^0 \rightarrow \Lc \bar{K}^0 \mu^- \bar{\nu}_{\mu}$). 
 \item $\nphys^\mathrm{other\Lb}$: Other semileptonic decays of \Lb, which 
include either additional 
particles (\eg, $\lblcpim$) or a higher mass $c$ baryon with subsequent 
decay into the \Lc\ signal (\eg, \dmitridecayone, $\Lcstar \rightarrow \Lc\gamma$),
 \end{enumerate}
 and the ratio \(\nphys/\nhad\) is expressed as:
\begin{eqnarray}
\label{eq:feed}
\frac{\nphys}{\nhad} = 
\frac{\nphys^\mathrm{meson}+\nphys^\mathrm{b-baryon}+\nphys^\mathrm{other\Lb}}
{\nhad}.
\end{eqnarray}
The $\nphys^\mathrm{meson}$ and $\nphys^\mathrm{b-baryon}$ have been 
estimated to be very small and contribute $\le 1\%$ to the \lbsemi\ signal. 
Details of these estimates are found in Appendices~\ref{sec:mesonbaryon} 
and \ref{sec:omegab}. 
The contributions from other \Lb\ semileptonic decays are estimated below.

\begin{figure}[tbp]
    \begin{center}
     \includegraphics[width=250pt, angle=0]{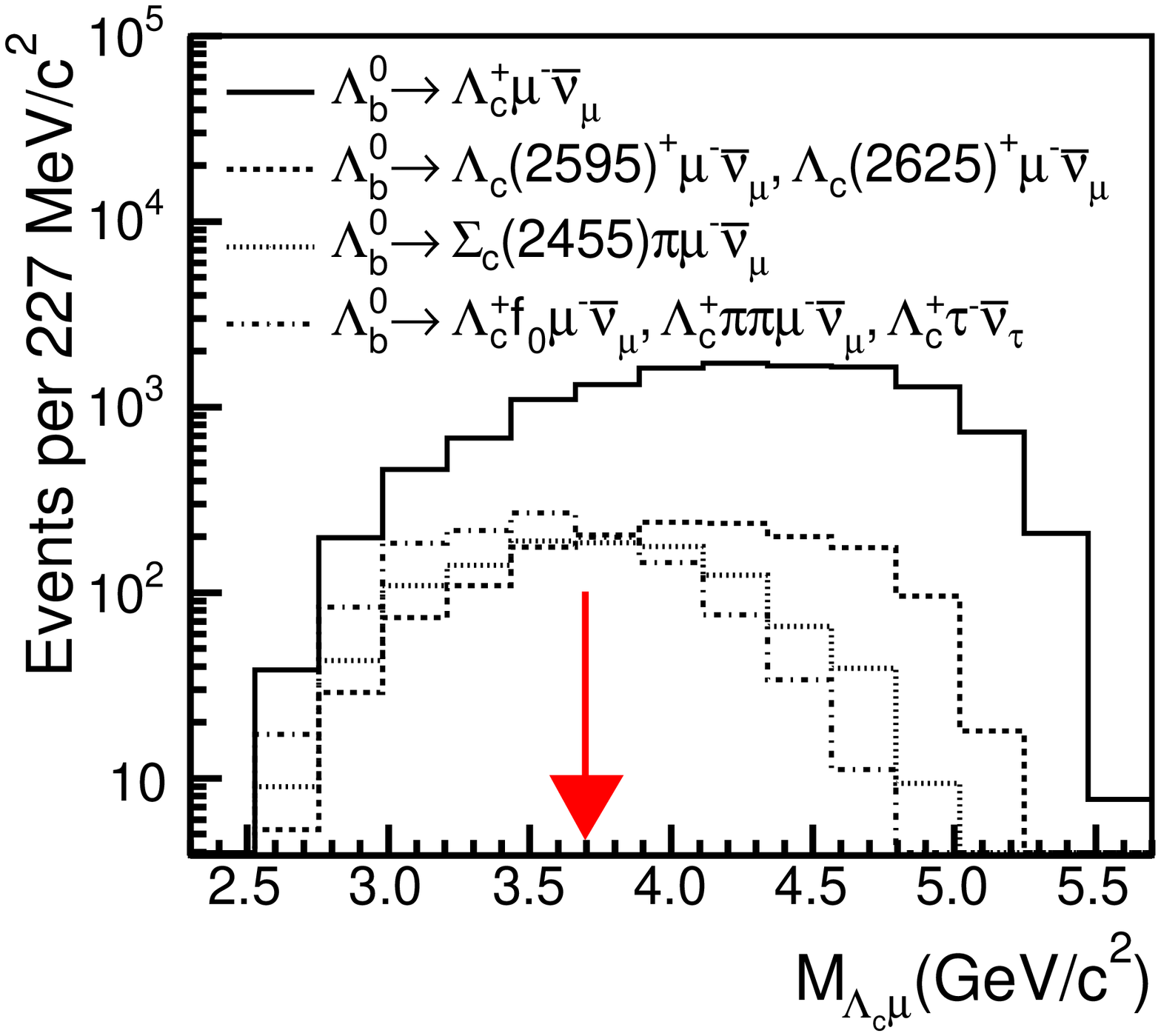}
     \caption{The fully-simulated $M_{\Lamc\mu}$ distributions 
	of the signal and backgrounds from 
	other semileptonic decays of \Lb, on a log scale. 
        The arrow indicates the minimum analysis requirement 
	at 3.7~\gevcsq. Note that the backgrounds tend to be at 
	the lower mass region. }
     \label{fig:m4track}
     \end{center}
  \end{figure}

The ratio $\nphys^\mathrm{other\Lb}/\nhad$ is given by: 
\begin{eqnarray}
\label{eq:bgexample}
\frac{\nphys^\mathrm{other\Lb}}{\nhad} =  
\frac{ \sum_{i} {\cal B}_{i}\eff_{i}}
     { {\cal B}\left(\lbhad\right)\eff_{\lbhad}}, 
\end{eqnarray}
where ${\cal B}_{i}$ and $\eff_{i}$ are the branching fraction of \Lb\ 
semileptonic decay mode $i$ and the efficiency of partially reconstructing 
the decay $i$ as the semileptonic signal, respectively. 
The estimate of $\nphys^\mathrm{other\Lb}$ starts by identifying the dominant 
background decay modes that enter \eq~(\ref{eq:bgexample}). 
The observation of spin-$1/2$ $\Lcstar$ and spin-$3/2$ 
$\Lcsstar$~\cite{Edwards:1994ar,Albrecht:1997qa} indicates 
the existence of \dmitridecayone\ and \dmitridecaytwo\ decays. 
In addition, the following decays may contribute to the \inclbsemi\ final state:
\begin{center}
\begin{tabular}{l}
 \dmitridecaythree, \\
 \lbsigmacp, \\
 \dmitridecayfour, \\
 \lblcfzero, \\
 \lblcpim\ (non-resonant), \\
 \lblcpizero\ (non-resonant).\\ 
\end{tabular}
\end{center}
The decay in the tau channel, \lblctau\ where 
$\tau^-\rightarrow \mu^-\bar{\nu}_{\mu}\nu_{\tau}$, also makes a small 
contribution.
\eqc~(\ref{eq:bgexample}) requires knowledge of the branching fractions of 
\lbhad\ and these background decays. In order to reduce systematic 
uncertainties from theoretical predictions, the dominant background 
decays, \dmitridecayone, \dmitridecaytwo, \dmitridecaythree, and 
\dmitridecayfour, have been reconstructed in the data. Measurements of their 
branching fractions relative to the branching fraction of the \lbsemi\ decay 
and estimates of the branching fractions of \lbhad\ and the other \Lb\ 
semileptonic decays are found in Section~\ref{sec:lbbr}. 
Once the list of background decay modes is established and their branching 
fractions are estimated, the acceptances and efficiencies of these 
backgrounds relative to that of the hadronic mode ($\eff_i/\eff_{\lbhad}$) are 
obtained from the full simulation as described in Section~\ref{sec:simulation}.
 \figc~\ref{fig:m4track} shows that a minimum requirement on $M_{\Lamc\mu}$ 
of 3.7~\gevcsq\ reduces the backgrounds 
from other \Lb\ semileptonic decays which have more particles in the final 
state. 

Table~\ref{t:physicslc} summarizes the feed-in backgrounds from 
the \Lb\ semileptonic decays discussed above and lists the 
hadronic and inclusive semileptonic yields observed in data. 
The two leading backgrounds after all selections are 
\dmitridecayone\ and \dmitridecaytwo. 
The total contribution from feed-in backgrounds has been estimated to be 
\fracphysb\% of the number of \inclbsemi\ events.

\renewcommand{\arraystretch}{1.1}
\begin{normalsize}
 \begin{table*}[tbp]
 \caption{Feed-in backgrounds to \lbsemi\ from other 
	\Lb\ semileptonic decays. 
	The ``$\ast$'' indicates decays which have been reconstructed for this 
	measurement and seen in data for the first time 
	(Section~\ref{sec:lbsemibr}). 	
	The second column lists the estimated branching 
	fractions from Section~\ref{sec:lbbr}. Numbers in parentheses are 
	estimated uncertainties~\cite{cerror,berror}.
	The third column lists $\eff_i/\eff_{\lbhad}$ with statistical uncertainty.
	All efficiencies are determined from the full simulation as described 
	in Section~\ref{sec:simulation}. 
	The fourth and the fifth columns list the normalization for each 
	background relative to the hadronic and the exclusive semileptonic 
	signals, respectively. The last column lists the number of events for 
	each background after multiplying 
	$\left(\nphys^\mathrm{other\Lb}/\nhad\right)^i$ by  
	\nhad, and the uncertainty includes only the statistical uncertainty 
	on \nhad. Note that while the numbers listed in the fourth column are 
	used in the final measurement, the last two columns are shown only for 
	a comparison with the \lbsemi\ signal.  \label{t:physicslc}
}

	 \begin{center}
	 \begin{tabular}{clr@{\,$\pm$\,}lcccr@{\,$\pm$\,}r}
	 \hline
	 \hline
	 \multicolumn{2}{c}{Mode}
	 & \multicolumn{2}{c}{${\cal B}$ $\left(\%\right)$}
	 & \multicolumn{1}{c}{$\frac{\eff_i}{\eff_{\lbhad}}$}
	 & \multicolumn{1}{c}{$\left(\frac{\nphys^\mathrm{other\Lb}}{\nhad}\right)^i$}
	 & \multicolumn{1}{c}{$\left(\frac{\nphys^\mathrm{other\Lb}}{\nsemi}\right)^i$}
	 & \multicolumn{2}{c}{$N_{event}$} \\
	 \hline
	 & \lbhad
		 &  \multicolumn{2}{c}{\brlbhadc \brlbhadecombine}
		 &  \multicolumn{1}{l}{1.000}
		 &  -- 
		 &  -- &  \nlbhadc & \nlbhade\\
	 & \inclbsemi
		 &  \multicolumn{2}{c}{--}
		 &  -- & --
		 &  -- & \nlbsemic & \nlbsemie \\
	 \hline
	 & \lbsemi
		 &   \brlbsemic & (\brlbsemie)
 & $\lbeffratioc \pm \lbeffratioe$
		 &  6.118
		 &  1.000 & \multicolumn{2}{c}{--} \\
	 \hline
	 $\ast$ & \dmitridecayone
		 &  \brlcstarc & (\brlcstare)
		 & $0.198 \pm 0.003$
		 &  0.503
		 &  0.082 & 90 & 10\\
	 & $\hspace{36pt} \hookrightarrow \Sigcpp\pi^{-}$
		 &  24 &  7
		 &
		 & & & \multicolumn{2}{c}{} \\
	 & $\hspace{60pt} \hookrightarrow \Lc \pi^+$ 
			& 100 & (5) 
		 & 
		 & & & \multicolumn{2}{c}{} \\
	 &  $\hspace{36pt} \hookrightarrow \Sigczero\pi^{+}$
		 &  24 &  7
		 & 
		 & & & \multicolumn{2}{c}{} \\
	 & $\hspace{60pt} \hookrightarrow \Lc \pi^-$
			& 100 & (5) 
		 & 
		 & & & \multicolumn{2}{c}{} \\
	 & $\hspace{36pt} \hookrightarrow \Sigcp\pi^{0}$
		 &  24 &  $\left(1.2\right)$ 
		 & 
		 & & & \multicolumn{2}{c}{} \\
	 & $\hspace{60pt} \hookrightarrow \Lc \pi^0$
			& 100 & (5) 
		 & 
		 & & & \multicolumn{2}{c}{} \\
	 & $\hspace{36pt} \hookrightarrow \Lc \pi^+ \pi^-$
		 &  18 &  10 
		 & & & & \multicolumn{2}{c}{}\\
	 & $\hspace{36pt} \hookrightarrow \Lc \pi^0 \pi^0$
		 &  9 &  $\left(0.45\right)$ 
		 & & & & \multicolumn{2}{c}{}\\
	 & $\hspace{36pt} \hookrightarrow \Lc \gamma$
		 &  1 &  $\left(0.05\right)$
		 & & & & \multicolumn{2}{c}{}\\
	 \hline 
	 $\ast$ & \dmitridecaytwo
		 &  \brlcsstarc & (\brlcsstare)
		 &  $0.192 \pm 0.003$
		 &  0.815
		 &  0.133& 146 & 15\\
	 & $\hspace{36pt} \hookrightarrow \Lc \pi^+ \pi^-$
		 &  66 &  $\left(3.3\right)$
		 & & & & \multicolumn{2}{c}{}\\
	 & $\hspace{36pt} \hookrightarrow \Lc \pi^0 \pi^0$
		 &  33 &  $\left(1.7\right)$
		 & & & & \multicolumn{2}{c}{} \\
	 & $\hspace{36pt} \hookrightarrow \Lc \gamma$
		 &  1 &  $\left(0.05\right)$
		 & & & & \multicolumn{2}{c}{}\\
	 \hline
	 $\ast$ & \dmitridecaythree
		 &  \brsigcc & (\brsigce)
		 & $0.082 \pm 0.004$
		 & 0.089 & 0.015 & 16 & 2\\
	 & $\hspace{36pt} \hookrightarrow \Lc \pi^-$
		 & 100 & (5)
		 & 
		 & & & \multicolumn{2}{c}{}\\
	 & $\lbsigmacp$
		 &  \brsigcc & (\brsigce)
		 & $0.073 \pm 0.004$
		 & 0.080
		 & 0.013 & 14 & 2\\
	 & $\hspace{36pt} \hookrightarrow \Lc \pi^0$
		 & 100 & (5)
		 &
		 & & & \multicolumn{2}{c}{}\\
	 $\ast$ & \dmitridecayfour 
		 &  \brsigcc & (\brsigce) 
		 & $0.077 \pm 0.004$
		 & 0.084 & 0.014 & 15 & 2\\
	 & $\hspace{36pt} \hookrightarrow \Lc \pi^+$
		 & 100 & (5) 
		 & 
		 & & & \multicolumn{2}{c}{} \\
	 \hline
	 &	$\lblctau$ 
		 & \brlcsemitau & $\left(\brlcsemitau\right)$
		 & $0.041 \pm 0.003$
		 & 0.040
		 & 0.006 & 7 & 1\\
	 & $ \hspace{48pt}    \hookrightarrow \mu^{-}\bar{\nu}_{\mu}\nu_{\tau}$
		 &  \brtautomuc &  \brtautomue
		 & 
			& & & \multicolumn{2}{c}{}\\
	 \hline
	& $\lblcfzero$ 
		 & 0.00 & $\left(\brfzero\right)$
		 & $0.023 \pm 0.002$
		 & 0.000
		 & 0.000 & \multicolumn{2}{c}{0} \\
	 &  $\lblcpim$ 
		 & 0.00 & $\left(\brlcpipi\right)$
		 & $0.032 \pm 0.002$
		 & 0.000
		 & 0.000 & \multicolumn{2}{c}{0}  \\
	 & $\lblcpizero$
		 & 0.00 & $\left(\brfzero\right)$
		 & $0.033 \pm 0.002$
		 & 0.000
		 & 0.000 & \multicolumn{2}{c}{0} \\
	 \hline 
	 \hline 
	 \end{tabular}
	 \end{center}

	\end{table*}
\end{normalsize}

\subsection{Summary}
Table~\ref{t:lbsemibg} lists the values of all the background variables which 
enter \eq~(\ref{eq:bigf}) and summarizes the composition of the inclusive 
$\Lc\mu^-$ sample. 
The dominant signal contamination is from the feed-in background. The second 
largest background arises from false muons. The smallest background source is 
$b\bar{b}$/$c\bar{c}$. 
 The estimate of $\nphys^\mathrm{other\Lb}/\nhad$ 
requires knowledge of the branching fraction of each feed-in decay and also 
the hadronic decay \lbhad. The next section details the measurements and 
assumptions used to estimate these branching fractions.

\renewcommand{\arraystretch}{1.2}
\renewcommand{\tabcolsep}{0.12in}
  \begin{table*}[tbp]
   \caption{
    The values of background variables in \eq~(\ref{eq:bigf}) and 
    the composition of the \inclbsemi\ sample. 
    Uncertainties on the $b\bar{b}$/$c\bar{c}$ and feed-in 
    backgrounds to the \lbsemi\ decay are statistical only. 
    The values of \nhad, \nincsemi, and $\effhad/\effsemi$ in 
    \eq~(\ref{eq:bigf}) are $\nlbhadc \pm \nlbhade$, 
    $\nlbsemic \pm \nlbsemie$, and $\lbeffratioc \pm \lbeffratioe$, 
	respectively. 
   }

   \label{t:lbsemibg}
   \begin{center}
   \begin{normalsize}
   \begin{tabular}{cccr}
   \hline
   \hline
   Source &  $N$ & $N/\nhad$ & $N/\nincsemi$ $\left(\%\right)$ \\
   \hline
   Signal                 & -- & --  &  \fracsignal \\
   False muon             & $40\pm 9$  &  --       & \fracfakemu \\
   $b\bar{b}$/$c\bar{c}$  & -- & $\lbbbccinputc \pm \lbbbccinpute$ &  \fracbbcc \\
   Feed-in                & -- & $\lbphysinputc \pm \lbphysinpute$ &  \fracphysb \\
  \hline
   \hline
 \end{tabular}
 \end{normalsize}
 \end{center}
 \end{table*}

\section{Observations of Four New \Lb\ Semileptonic Decays and 
	Estimates of \Lb\ Semileptonic and Hadronic Branching Fractions
	\label{sec:lbbr}}
The size of the background contribution from the feed-in of other semileptonic 
decays of \Lb, $\nphys^\mathrm{other\Lb}$, is normalized to the observed 
hadronic signal yield in data, with corrections for the relative acceptance 
times efficiency for each decay mode [see \eq~(\ref{eq:bgexample})]. 
This procedure requires estimates of the 
branching fractions for each background decay and also for the hadronic 
signal. In order to reduce systematic 
uncertainties from these branching fractions, resonant \Lb\ semileptonic decays
 expected to contribute to the $\Lc\mu^-$ sample have been reconstructed in 
data. These reconstructed decays are then used to 
estimate the branching fractions of non-resonant \Lb\ semileptonic decays 
with a constraint of the world average of $ {\cal B}\left(\Lb \rightarrow \Lc \ell^- \bar{\nu}_{\ell}~ \mathrm{anything}\right)$ and 
an estimate of ${\cal B}(\lbsemi)$. 
Note that the estimate of the \lbhad\ and \lbsemi\ branching fractions appear 
in the estimate of the backgrounds (Section~\ref{sec:physicsb}). It must be 
pointed out here that from Table~\ref{t:lbsemibg}, the total contribution of 
feed-in background is \fracphysb\% and contributes at most this amount to the 
total uncertainty for this measurement. Furthermore, the ratio of the estimated
 ${\cal B}\left(\lbsemi\right)$ to ${\cal B}\left(\lbhad\right)$ need not be 
the same as the final measured result. 

Section~\ref{sec:lbsemibr} first presents measurements of the branching 
fractions of four new \Lb\ semileptonic decays relative to that of the 
\lbsemi\ decay: 
\dmitridecayone, \dmitridecaytwo, \dmitridecaythree, and \dmitridecayfour, 
and then describes the estimation of  
${\cal B}\left(\lbsemi\right)$ and branching fractions of 
several non-resonant \Lb\ semileptonic 
decays. Section~\ref{sec:lbxsec} shows that how the \pt\ 
distribution of the \Lb\ baryon produced from $p\bar{p}$ 
collisions is significantly different from that of the $\Bd$ meson and gives 
the corresponding modification to the ratio, \rxsec, with respect to the 
CDF I measurement~\cite{Affolder:1999iq}. The ratio \rxsec\ is then used
 to estimate ${\cal B}\left(\lbhad\right)$.

\subsection{Observations of four new \Lb\ semileptonic decays and 
	estimates of the \Lb\ semileptonic branching fractions 
	\label{sec:lbsemibr}}
The following \Lb\ semileptonic decays are considered in the estimate of 
$\nphys^\mathrm{other\Lb}$ in Section~\ref{sec:physicsb}:
\begin{center}
\begin{tabular}{l}
	\dmitridecayone,\\
 	\dmitridecaytwo,\\ 
        \dmitridecaythree, \\
        \lbsigmacp, \\
        \dmitridecayfour, \\
        \lblcfzero, \\
        \lblcpim\ (non-resonant), \\
        \lblcpizero\ (non-resonant),\\ 
 	\lblctau.
\end{tabular}
\end{center}
Among the nine decays above, none have been observed previously~\cite{DELPHI}; 
only the branching fractions for the 
\dmitridecayone\ and the \dmitridecaytwo\ decays have been predicted, but 
with an uncertainty as large as 100$\%$~\cite{Leibovich:1997az}. 

\begin{figure}[tbp]
 \begin{center}
 \includegraphics[width=250pt, angle=0]{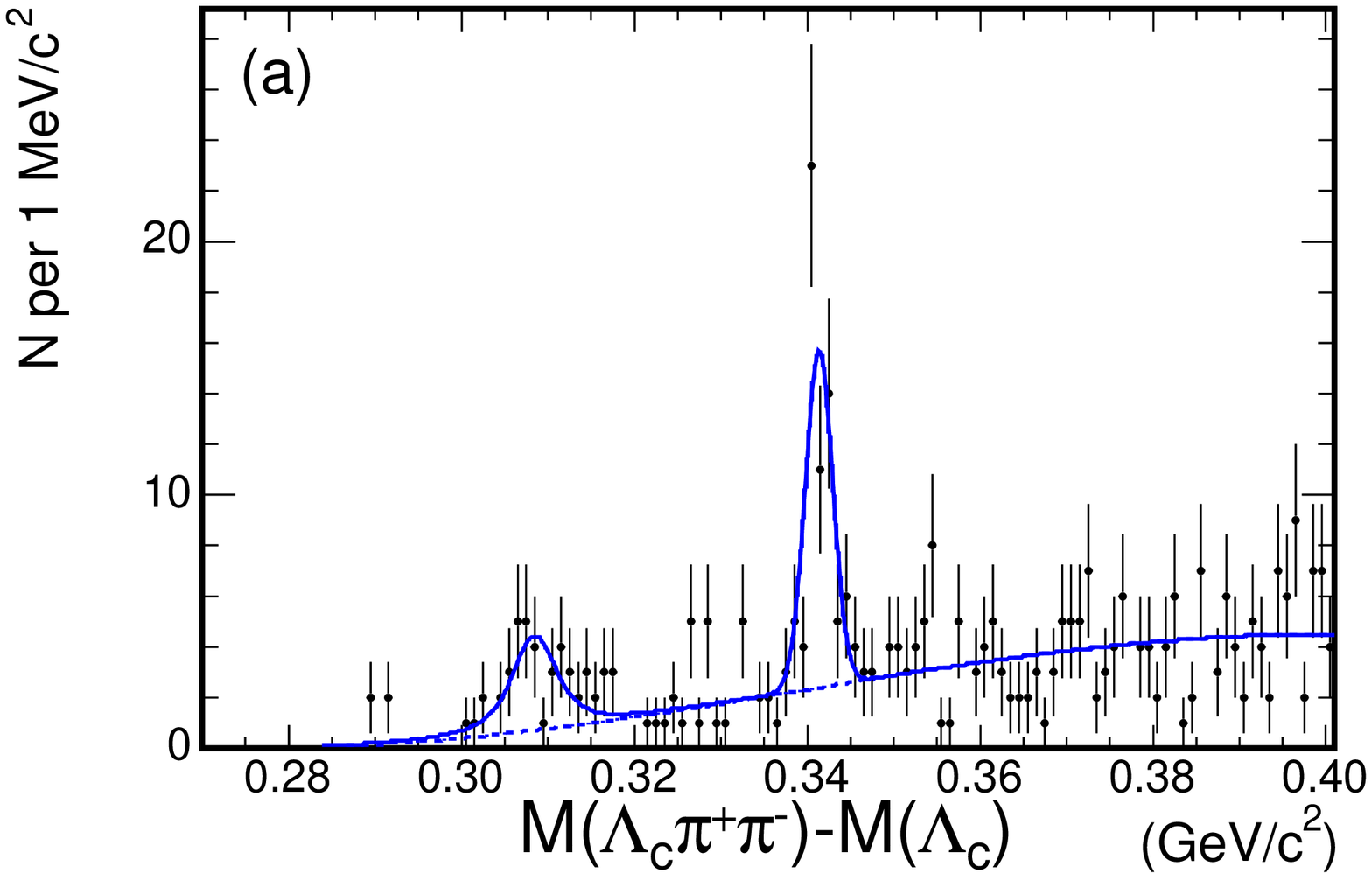}
 \includegraphics[width=250pt, angle=0]{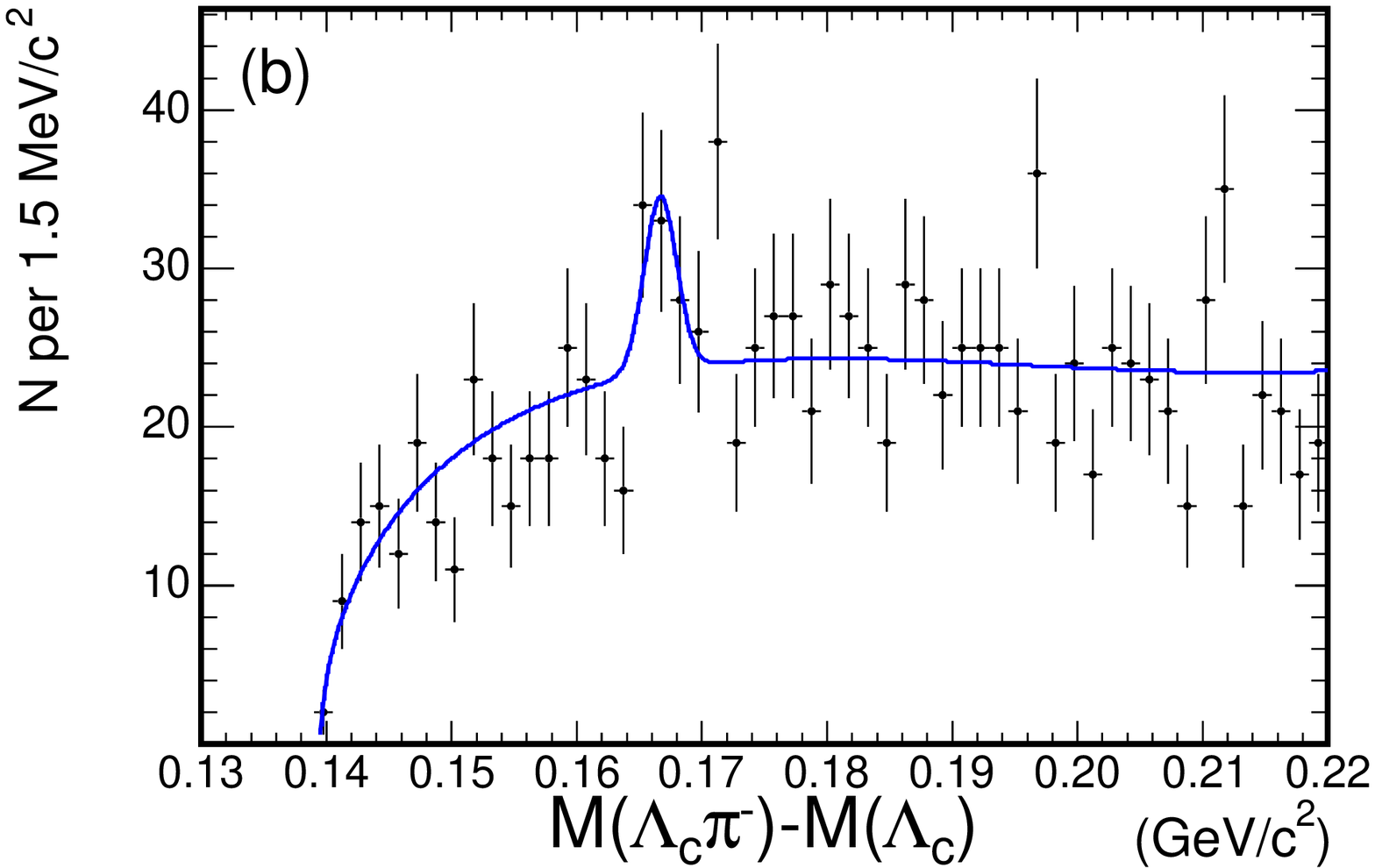}
 \includegraphics[width=250pt, angle=0]{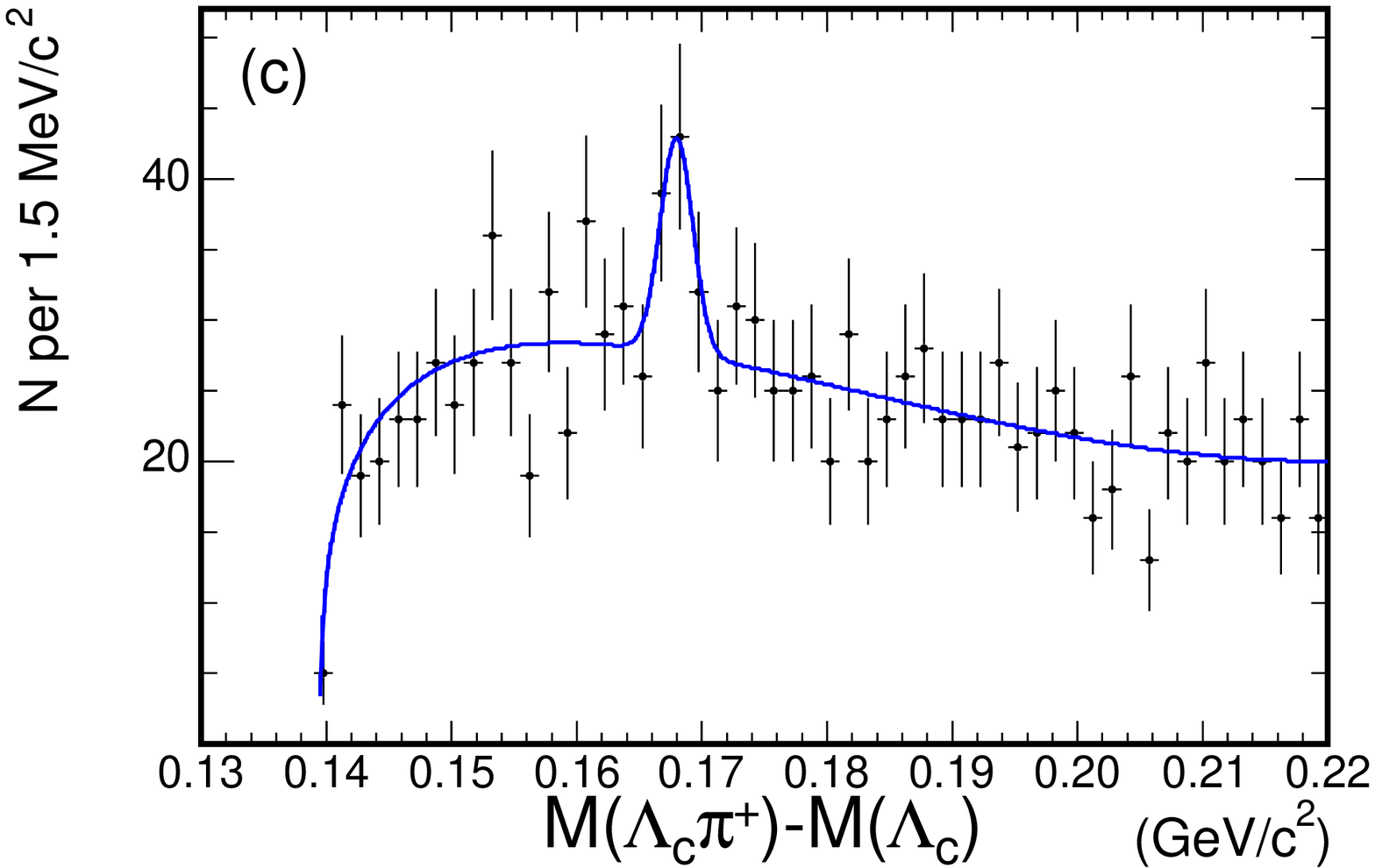}
  \end{center}
  \caption
   { 
     The excited-$c$-baryon candidates which are associated with a $\mu^-$: (a) $\Lcstar$ and $\Lcsstar$, (b) $\Sigczero$, and (c) $\Sigcpp$. 
     The curves indicate fit results to the spectra of mass differences. 
     \label{fig:dmitrilc}}
  \end{figure}

In order to reduce the systematic uncertainty on the final measurement 
coming from the branching ratios of these backgrounds, the following decays 
are searched for in a larger $\Lc\mu X$ data sample (360~pb$^{-1}$):
\begin{enumerate}
\item $\dmitridecayone X$ where $\Lcstar \left(\rightarrow \Sigcpp\pi^{-}, \Sigczero\pi^{+} \right)\rightarrow \Lc \pi^+\pi^-$. 
\item $\dmitridecaytwo X$ where $\Lcsstar \rightarrow \Lc \pi^+\pi^-$. 
\item $\dmitridecaythree X$ where $\Sigczero \rightarrow \Lc \pi^-$.
\item $\dmitridecayfour X$ where $\Sigcpp \rightarrow \Lc \pi^+$.
\end{enumerate}
All four decays modes above contain $\Lc\pi^+\pi^-\mu^-$ in the final state. 
The selection criteria are the same as those for the \inclbsemi\ sample 
(see Section~\ref{sec:reconstruction}), except that two oppositely-charged 
tracks are added to determine the secondary vertex for the $\Lcstar$ and $\Lcsstar$ modes, 
and one track is added for the $\Sigmac$ modes. In all cases, the pion mass 
is assumed for each additional track and each track is required to have \pt\ 
$>$ 0.4~\gevc. 
The available four-momentum transferred to the daughters in the decays of 
these $c$ baryons into \Lc\ is small. Therefore, the 
mass differences $M_{\Lamc\pi^+\pi^-}-M_{\Lamc}$, $M_{\Lamc\pi^-}-M_{\Lamc}$, 
and $M_{\Lamc\pi^+}-M_{\Lamc}$, have a better resolution than the 
masses of the $c$-baryon candidates and are the figure of merit for detecting 
signal peaks. \figc~\ref{fig:dmitrilc} shows the mass difference distributions,
 where the numbers of signal events are determined by fitting the mass 
differences to a Gaussian for the signal and a kinematically-motivated line 
shape for the combinatorial background. Table~\ref{t:dmitriyield} summarizes 
the signal yields, the corresponding significances, and the fitted mass 
differences. In this table, contributions of $\Lcstar$ in 
the $\Sigmac\pi$ modes have been subtracted from the $\Sigmac\pi$ modes and the significances of the 
$\Sigczero$ and $\Sigcpp$ modes are combined. 
Systematic uncertainties on the yields are determined by varying the 
functions for the combinatorial background in the fit. 
This is the first observation of the \dmitridecaytwo\ decay.

\renewcommand{\arraystretch}{1.5}
\begin{normalsize}
 \begin{table*}[tbp]
 \caption{The observed number of signal events, the corresponding  
          significance $\left(S/\sqrt{S+B}\right)$, and the fitted mass 
	  difference in data for each \Lb\ 
 	  semileptonic decay mode.}
 \label{t:dmitriyield}
   \begin{center}
   \begin{tabular}{lccc}
    \hline
    \hline	
     Mode & Yield & Significance ($\sigma$) & $\Delta M$~[\mevcsq] \\
     \hline	
      $\Lcstar\mu^-X$ 
	& $31 \pm 8 (\rm{stat}) \pm 7 (\rm{syst})$ &  2.9 
	& $308.47 \pm 0.99 (\rm{stat}) $ \\
      $\Lcsstar\mu^-X$ 
        & $53 \pm 9 (\rm{stat}) \pm 5 (\rm{syst})$ &  5.2 
	& $341.39 \pm 0.31 (\rm{stat}) $ \\
      $\Sigczero\pi^+\mu^-X$                   
	& $16 \pm 11 (\rm{stat}) \pm 7 (\rm{syst})$ & 
	& $166.72 \pm 0.69 (\rm{stat}) $ 	\\
      $\Sigcpp\pi^-\mu^-X$                
	& $26 \pm 12 (\rm{stat}) \pm 9 (\rm{syst})$ & 
	& $168.01 \pm 0.51 (\rm{stat}) $ \\
      $\Sigmac$ modes combined            &  & 2.1 \\
      \hline
      \hline
     \end{tabular}
    \end{center}
  \end{table*}
\end{normalsize}

After estimating, with simulation, the acceptance times efficiency of these 
reconstructed decays relative to that of the \lbsemi\ decay, and 
taking into account the false-muon background~\cite{physbinnew}, the relative 
branching ratios $\left(R_i\right)$ are extracted:
\begin{eqnarray*}
\label{eq:dmitriratio1}
R_1 & \equiv & \frac{{\cal B}\left(\dmitridecayone\right)}{{\cal B}\left(\lbsemi\right)} \\
&= & \dmitrione\ \pm \dmitrionestat \left(\rm{stat}\right) \dmitrionesyst \left(\rm{syst}\right), \nonumber \\
R_2 & \equiv & \frac{{\cal B}\left(\dmitridecaytwo\right)}{{\cal B}\left(\lbsemi\right)} \\
& = & \dmitritwo\ \pm \dmitritwostat \left(\rm{stat}\right) \dmitritwosyst \left(\rm{syst}\right), \nonumber \\
R_{3,4} &  \equiv & \frac{1}{2}\left[\frac{{\cal B}\left(\dmitridecaythree\right)}{{\cal B}\left(\lbsemi\right)}\right. \\
& + & \left.\frac{{\cal B}\left(\dmitridecayfour\right)}{{\cal B}\left(\lbsemi\right)}  \right]
	\nonumber \\
& = & \dmitrithree\ \pm \dmitrithreestat\left(\rm{stat}\right) \dmitrithreesyst \left(\rm{syst}\right), \nonumber 
\end{eqnarray*}
where the two $\Sigmac\pi$ modes are averaged.
Assuming isospin symmetry leads to the estimate:
\begin{eqnarray*}
\label{eq:dmitriratio5}
R_5 \equiv \frac{{\cal B}\left(\lbsigmacp\right)}{{\cal B}\left(\lbsemi\right)} 
	= R_{3,4} = \dmitrithree. 
\end{eqnarray*}
The systematic uncertainties on the relative branching fractions come from 
variation of background fitting models and uncertainties on the low-momentum 
pion \pt\ spectrum and the correction to the reconstruction efficiency. 

In order to convert the above measurements of the relative branching fractions 
into absolute branching fractions, an estimate of ${\cal B}\left(\lbsemi\right)$ is required. A recent measurement by the DELPHI collaboration reported 
$\left(5.0{+1.1 \atop -0.8}\mathrm{(stat)}{+1.6\atop -1.2}\mathrm{(syst)}\right)\%$ for this branching fraction~\cite{Abdallah:2003gn}. However, from 
heavy quark symmetry, the semileptonic 
decay widths for all $b$ hadrons are expected to be the same. Therefore, 
semileptonic branching fractions of the $b$ hadrons, $\Gamma^\mathrm{semi}/\Gamma^\mathrm{total}$, vary only due to their lifetime differences. Since the 
\Lb\ decays to a spin-$1/2$ \Lc, contributions from both S and P wave 
amplitudes are expected.  
A sum of ${\cal B}\left(\bar{B}^0\rightarrow \dplus \ell^- \bar{\nu}_{\ell}\right) + {\cal B}\left(\bar{B}^0\rightarrow \dstar \ell^- \bar{\nu}_{\ell}\right)$, where the decays to $\dplus$ [$\dstar$] correspond to the S(P) wave 
amplitudes, yield
$\left(\bmesondecaywidthc \pm \bmesondecaywidthe\right)\%$. 
 The number, \bmesondecaywidthc\%, is then scaled by the world average of the 
ratio of lifetimes, $\tau_{\Lb}/\tau_{\Bd}=\lifetimeRc\pm \lifetimeRe$ 
(stat+syst)~\cite{pdg}. The ${\cal B}\left(\lbsemi\right)$ is estimated to be 
$\left(\brlbsemic \pm \brlbsemistate \pm \brlbsemisyste \right)\%$, where the 
first uncertainty arises from the propagation of errors and the second is half 
of the difference between the above estimate and the DELPHI result~\cite{whynotDELPHI}. 
The ${\cal B}\left(\Lb \rightarrow \Lc \tau^- \bar{\nu}_{\tau}\right)$ can be estimated by scaling ${\cal B}\left(\lbsemi\right)$ by the ratio
 of phase space area: $Ph. Sp. \left(\Lb \rightarrow \Lc \tau^- \bar{\nu}_{\tau}\right)/Ph. Sp. \left(\lbsemi\right) =0.277$. 
The middle portion of 
Table~\ref{t:brlc} summarizes the branching fractions of the \Lb\ 
semileptonic decays discussed above. Uncertainties on the observed \Lb\ 
semileptonic decays and the \lbsigmacp\ decay include uncertainties from the 
relative branching fraction measurement and uncertainty from the assumed 
${\cal B}\left(\lbsemi\right)$. A 100$\%$ systematic uncertainty is also 
assigned to ${\cal B}\left(\lblctau\right)$. 

The sum of ${\cal B}\left(\lbsemi\right)$ and the branching fractions in the 
middle portion of Table~\ref{t:brlc} is already larger than the inclusive \Lb\ semileptonic branching fraction in the 2008 PDG summary:
\begin{eqnarray*}
{\cal B}\left(\Lb \rightarrow \Lc \ell^- \bar{\nu}_{\ell}~ \mathrm{anything}\right) = \brinclbsemic \pm \brinclbsemie \%. 
\end{eqnarray*}
The following decays are, therefore, ignored in the central values but will 
be included in the systematic uncertainty:
\begin{center}
\begin{tabular}{lll}
 \lblcfzero, \\
 \lblcpim (non-resonant), \\
 \lblcpizero (non-resonant). \\ 
\end{tabular}
\end{center}
An estimate of these branching fractions is obtained by moving 
 ${\cal B}\left(\Lb \rightarrow \Lc \ell^- \bar{\nu}_{\ell}~ \mathrm{anything}\right)$ upward by 1$\sigma$.
 The remaining branching fraction is calculated to be:
\begin{eqnarray*}
 \lefteqn{(\brinclbsemic+\brinclbsemie)\%}  \\
	&-&  \brlbsemic\%\times\left[1+ R_1 +  R_2 +  3 R_3 + 
	0.277\times{\cal B}\left(\tau^-\rightarrow\mu^{-}\bar{\nu}_{\mu}\nu_{\tau}\right)
	\right] \\
 &\approx&\diffbr \%.
\end{eqnarray*}
The \diffbr\% is then attributed to the above decays which are ignored in the 
central value. The branching fraction of \lblcpim\ is 
estimated to be twice that of \lblcpizero\ based 
on the isospin invariance, and the $f_0(980)$ mode is assumed to have the same 
branching fraction as that of the $\pi^0\pi^0$ mode. 
The bottom portion of Table~\ref{t:brlc} lists zero central values for 
these three decays and uses their estimated branching fractions above 
as the systematic uncertainties on the branching fractions.

\renewcommand{\arraystretch}{1.5}
\begin{normalsize}
 \begin{table}[tbp]
 \caption{The \Lb\ semileptonic branching fractions for decays 
	which are included in the central value (middle portion) and those 
	which are not (bottom portion).
	All the numbers in parentheses are estimated uncertainties.}
 \label{t:brlc}
   \begin{center}
   \begin{tabular}{lc@{\,$\pm$\,}c}
    \hline
    \hline	
     \multicolumn{1}{c}{Mode}  
     & \multicolumn{2}{c}{BR $\left(\%\right)$}	\\
     \hline	
	 \lbsemi
	&   \brlbsemic & (\brlbsemie) \\
     \hline	
        \dmitridecayone
	&  \brlcstarc & (\brlcstare) \\
	\dmitridecaytwo
	&  \brlcsstarc & (\brlcsstare) \\
	\dmitridecaythree
	&  \brsigcc & (\brsigce) \\
	\lbsigmacp
	&  \brsigcc & (\brsigce) \\
	\dmitridecayfour
	&  \brsigcc & (\brsigce) \\
        \lblctau
	& \brlcsemitau & $\left(\brlcsemitau\right)$ \\
	\hline
	\lblcfzero
	&  0 & $\left(\brfzero\right)$ \\
        \lblcpim
	&  0 & $\left(\brlcpipi\right)$ \\
        \lblcpizero
	&  0 & $\left(\brfzero\right)$ \\

	\hline
	\hline
  	\end{tabular}
       \end{center}
	\end{table}
\end{normalsize}

\subsection{\label{sec:lbxsec}Modification of $\sigma_{\Lb}\left(\pt>6.0\right)\left/\sigma_{\Bd}\left(\pt>6.0\right)\right.$ and estimate of ${\cal B}\left(\lbhad\right)$} 
\eqc~(\ref{eq:bgexample}) requires ${\cal B}\left(\lbhad\right)$ to obtain 
the ratio $\nphys^\mathrm{other\Lb}/\nhad$. 
 Combining the CDF measurement 
\(
 G\equiv\yile\)~\cite{yile:lblcpi},
the world average of ${\cal B}\left(\dhad\right)$, and the ratio 
of production cross-sections 
\(
 \rho\equiv\frac{\sigma_{\Lb}\left(\pt>6.0\right)}{\sigma_{\Bd}\left(\pt>6.0\right)},
\)
one may express 
\begin{eqnarray}
\label{eq:lbhadbr}
{\cal B}\left(\lbhad\right) =  \frac{G}{\rho}  {\cal B}\left(\dhad\right). 
\end{eqnarray}
The ratio of cross sections, $\rho$, is calculated from the expression:
\begin{eqnarray}
\lefteqn{\rho  \equiv 
\frac{\sigma_{\Lb}\left(\pt\ > 6~\gevc\right)}
     {\sigma_{\Bd}\left(\pt\ > 6~\gevc\right)}}\nonumber \\
& = &\left(\frac{\sigma_{\Lb}}{\sigma_{\Bd}}\right)^\mathrm{CDF~I} 
	C_{\cal BR} C_{\alpha}  C_{\pt},
\end{eqnarray}
where $\left(\sigma_{\Lb}/\sigma_{\Bd}\right)^\mathrm{CDF~I}$ is the 
CDF~I result: $0.236 \pm 0.084$~\cite{Affolder:1999iq}. The $C_{\cal BR}$, 
$C_{\alpha}$, and $C_{\pt}$ are the correction factors to account for 
differences between the CDF~I result and 
this analysis in: the assumed ${\cal B}(\lbsemil)$, kinematic acceptance, and 
requirements on the minimum \pt\ of \Lb\ and \Bd. 
Each correction factor is explained in the text that follows.

The CDF~I analysis used electron-charm final states, 
such as $H_b \rightarrow \dstar e^- X$, $H_b \rightarrow \dplus e^- X$, 
and $H_b \rightarrow \Lc e^- X$, to measure the ratio of production 
cross-sections. The branching fraction, ${\cal B}\left(\lbsemil\right)$, was 
needed and estimated to be 
$7.94 \pm 0.39\%$, while this analysis estimates the value to be 
$\brlbsemic\pm\brlbsemie\%$. The uncertainty \brlbsemie\% is dominated 
by the difference from the DELPHI result (see Section~\ref{sec:lbsemibr}). 
In order to be consistent within this analysis, a correction to the 
branching fraction, $C_{\cal BR}$, is applied. The value of $C_{\cal BR}$ is 
the ratio of 7.94\% to \brlbsemic\% and found to be 
\(\rwendyRc \pm \rwendyRe\).

In the CDF~I analysis, 
the \Lb\ and $\Bd$ \pt\ spectra measured with fully-reconstructed decays were 
not available. In order to extract the signal acceptance and 
efficiency, the Nason-Dawson-Ellis (NDE) 
$b$-quark spectrum~\cite{Nason:1989zy} followed by the 
Peterson fragmentation model~\cite{Peterson:1982ak} was used at CDF~I to obtain
 the \pt\ distributions of $b$ hadrons in simulation~\cite{Taylor:1999fh}. The 
two-track trigger allows CDF~II to collect large samples of fully-reconstructed
 $b$-hadron decays, such as \lbhad\ and \dhad, and to compare the \pt\ 
distributions in data with those from the NDE+Peterson model. 
The \Lb\ and the $\Bd$ \pt\ spectra from the NDE+Peterson model are found to 
be harder (more $b$ hadrons at higher \pt) than those measured in data, 
which indicates an over-estimate of acceptance in the CDF~I analysis, 
particularly for the \Lb\ decays. 
The acceptance correction factor, $C_{\alpha}$, is the ratio of 
acceptances using generator-level simulations with inputs from the measured
 \pt\ spectra (identical to those described in Section~\ref{sec:production}) 
and from the NDE+Peterson model:\begin{eqnarray}
 C_{\alpha} =  \alpha_R^\mathrm{data-based}\left/\alpha_R^\mathrm{NDE+Peterson}\right., 
\label{eq:cepsilon}
\end{eqnarray}
where $\alpha_R$ is the ratio of the kinematic acceptances of \Lb\ and 
\Bd.
The value of the correction factor is found to be: 
$C_{\alpha} = 1.81 {+0.42 \atop -0.22} $ for the CDF~I kinematic requirements~\cite{alpha}. 
 The uncertainty comes from the uncertainties on 
the measured shapes of the \Lb\ and \Bd\ \pt\ distributions in data.

\begin{figure}[tbp]
  \begin{center}
\includegraphics[width=200pt, angle=0]{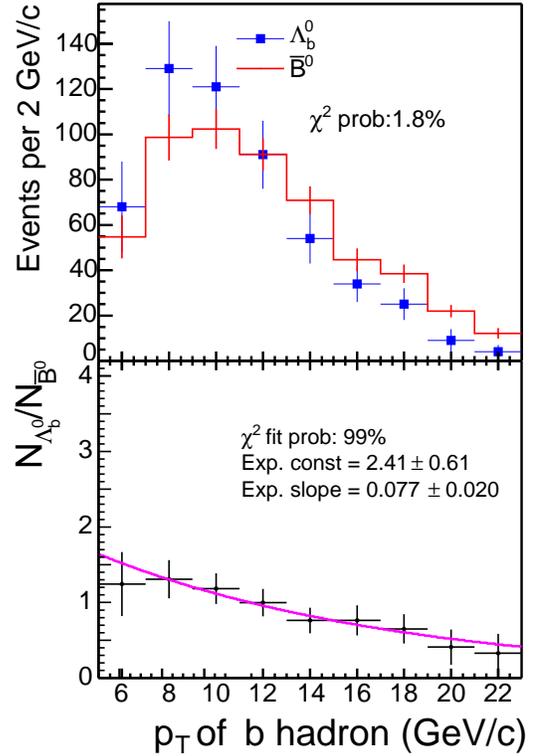}
 \caption[Comparison of the \Lb\ and \Bd\ \pt\ spectra measured in data]
 {Comparison of the reconstructed \Lb\ and \Bd\ \pt\ spectra in data. The 
 negative slope (3--4$\sigma$ away from zero) of the ratio of \Lb\ to $\Bd$ 
 histograms indicates that the $\pt\left(\Lb\right)$ distribution is softer 
 (more $b$ hadrons at lower \pt) than the $\pt\left(\Bd\right)$ distribution. 
 In order to have a fair comparison of \pt\ spectra, the same requirements 
 are applied to the \Bd\ and \Lb\ candidates~\cite{xsecrec}, while 
 \fig~\ref{fig:bptafter} has different selections for the \Bd\ and \Lb. 
 Nevertheless, the \pt\ spectra, used as inputs for the correction 
 factors $C_{\alpha}$ and $C_{\pt}$, have been 
 corrected for acceptance and efficiency and are identical to those 
 described in Section~\ref{sec:production}. }
 \label{fig:bptlbpt} 
 \end{center}
 \end{figure}

\begin{figure*}[tbp]
  \begin{center}
  \begin{tabular}{cc}
  \includegraphics[width=220pt, angle=0]{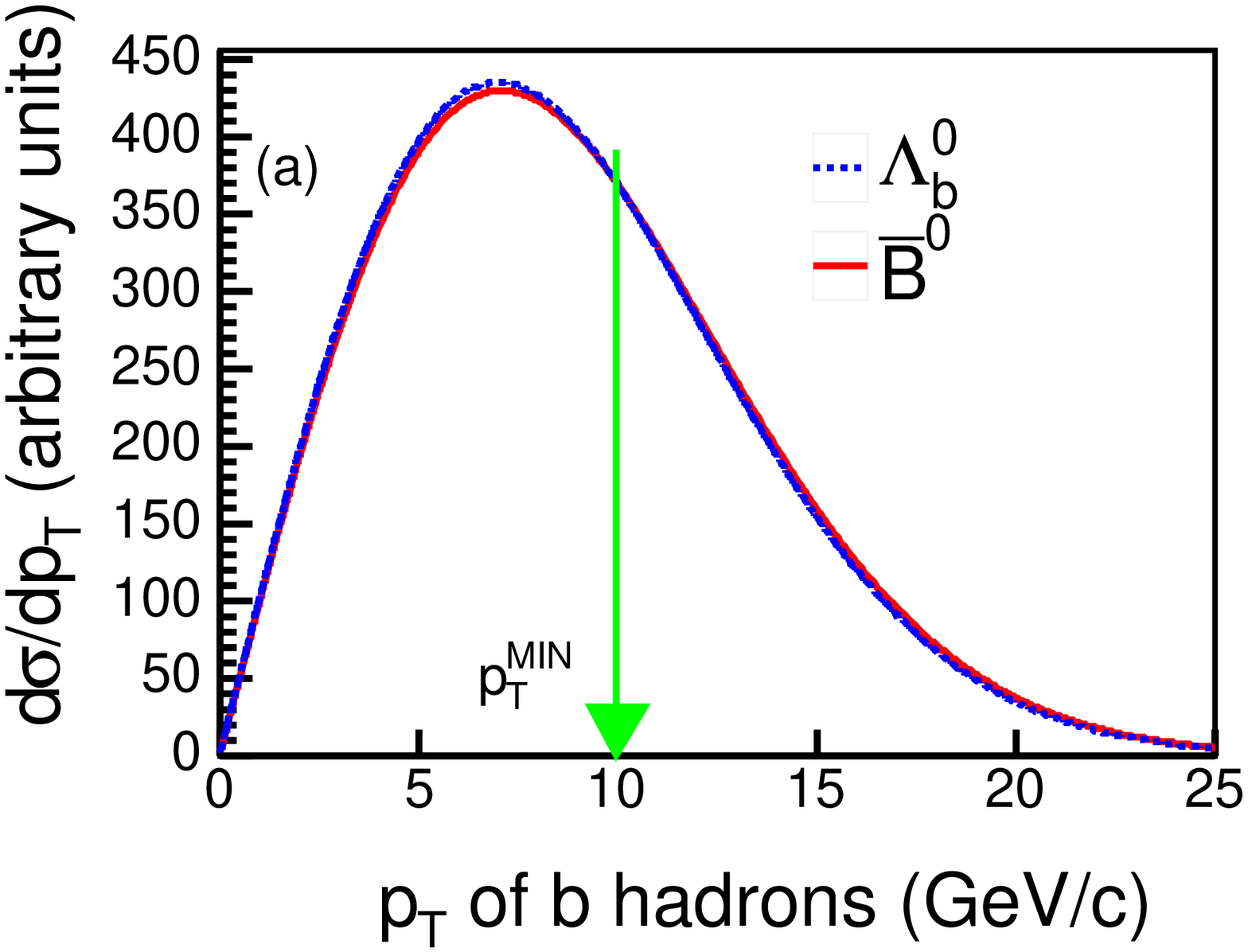} &
  \includegraphics[width=220pt, angle=0]{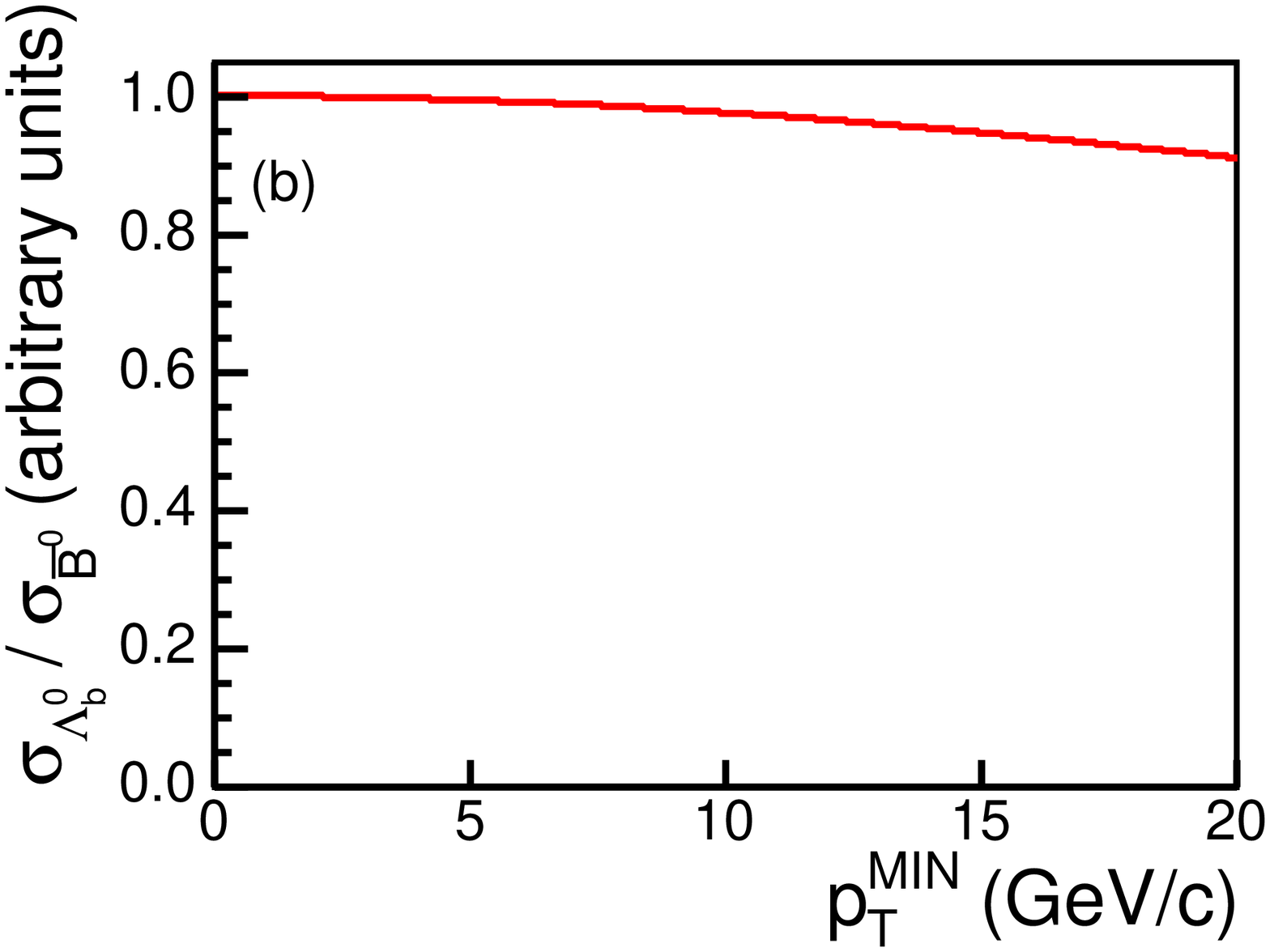} \\
  \includegraphics[width=220pt, angle=0]{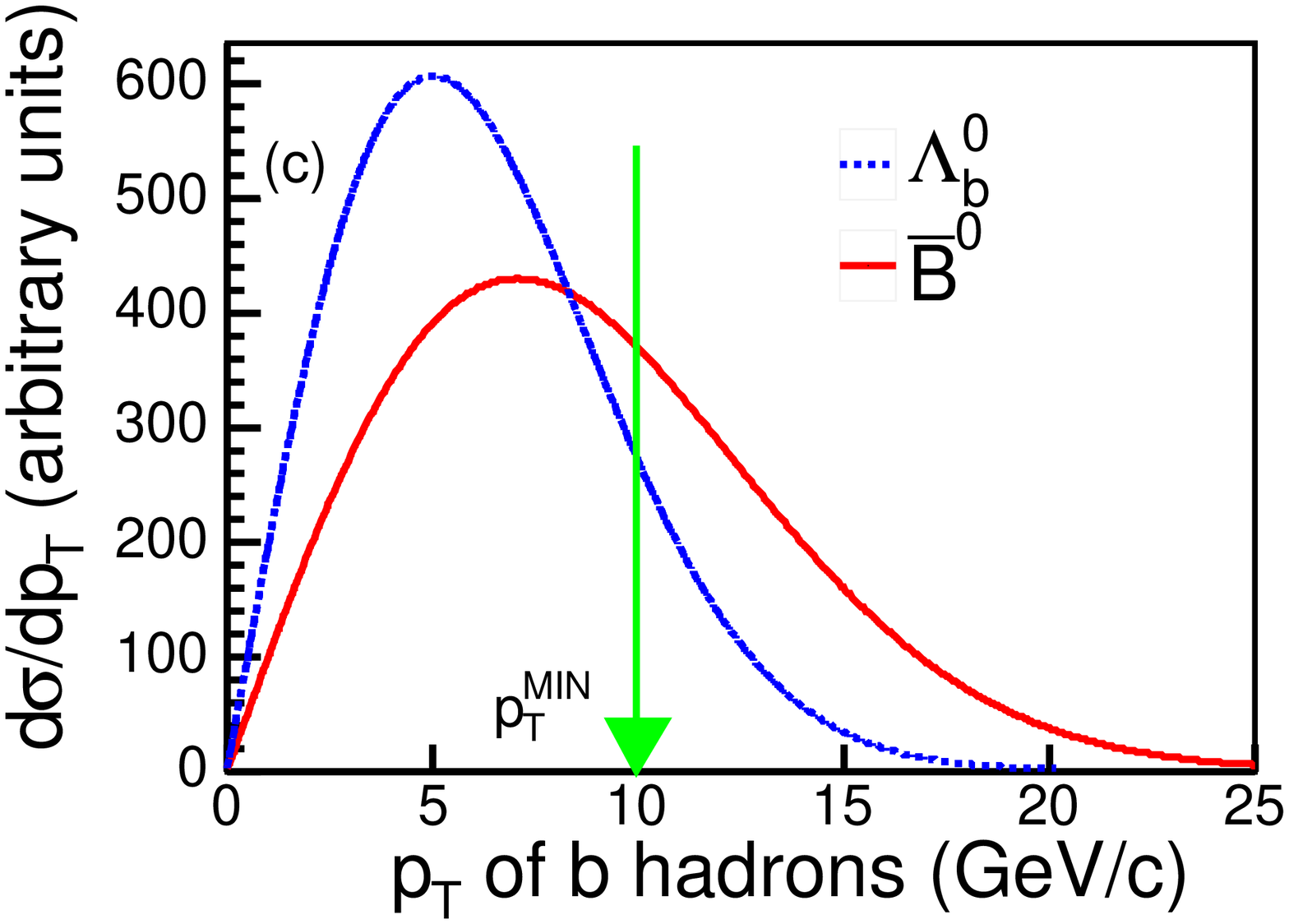} &
  \includegraphics[width=220pt, angle=0]{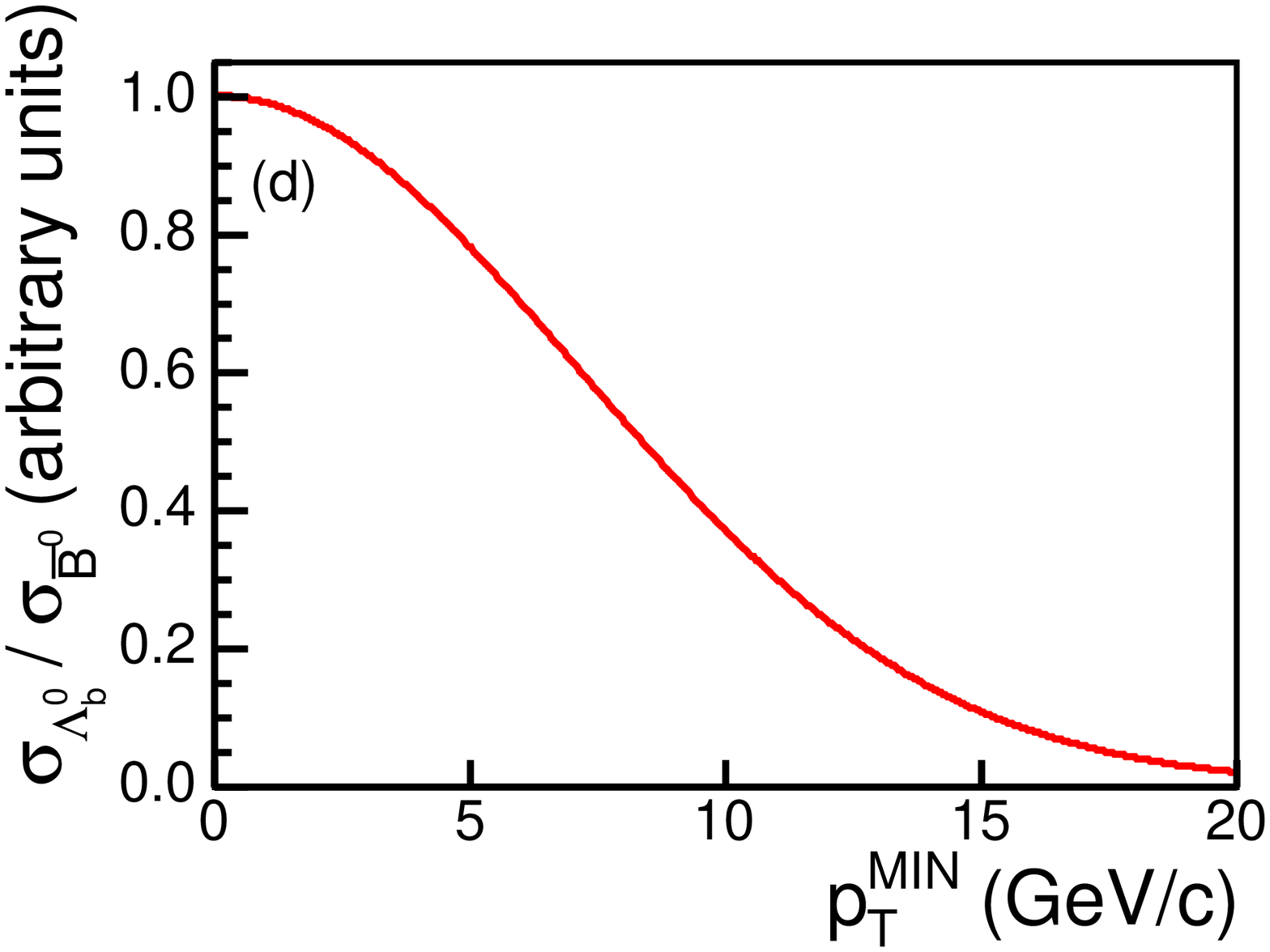} \\
  \end{tabular}
 \caption[Dependence of the production ratio on the minimum \pt\ requirements 
	of $b$-hadrons]
 { Examples of the \Lb\ and \Bd\ \pt\ spectra [(a), (c)] and the dependence of the
 production cross-section ratio on the minimum \pt\ requirements, 
$\pt^{\mathrm{MIN}}$ [(b), (d)]. 
   Figures~\ref{fig:teachxsec}(a) and (b) show the case where both hadrons have similar spectra; the 
   ratio of the integrated areas underneath the spectra, from $\pt^{\mathrm{MIN}}$ and above, depends little on the value of $\pt^{\mathrm{MIN}}$. 
  Figures~\ref{fig:teachxsec}(c) and (d) show that the \Lb\ \pt\ spectrum 
  is significantly softer (more $b$ hadrons at lower \pt) than the \Bd\ \pt\ 
  spectrum; the ratio of the integrated areas depends strongly on the value of 
  $\pt^{\mathrm{MIN}}$. }
 \label{fig:teachxsec} 
 \end{center}
 \end{figure*}

The last correction is due to a difference in the minimum \pt\ requirements 
between the CDF~I analysis [$\pt(H_b) > 10$~\gevc] and this analysis 
[$\pt(H_b)> 6$~\gevc]. By applying the same requirements to the \lbhad\ and 
\dhad\ decays reconstructed in the two-track trigger data, 
\fig~\ref{fig:bptlbpt} shows that the \Lb\ \pt\ spectrum is significantly 
softer (more $b$ hadrons at lower \pt) than that of the $\Bd$~\cite{xsecrec}. 
\figc~\ref{fig:teachxsec} illustrates the dependence of the ratio of 
cross-sections on the minimum \pt\ requirements, for a small [(a)] and a large 
[(c)] difference between the \Lb\ and \Bd\ \pt\ spectra; the scenario 
in \fig~\ref{fig:teachxsec}(c) is what has been observed in data. 
 A correction factor, $C_{\pt}$, is required:
\begin{eqnarray}
  C_{\pt} = \frac{N_{\Lb}\left(\pt> 6\right)}{N_{\Bd}\left(\pt>6\right)}\left/
\frac{N_{\Lb}\left(\pt>10\right)}{N_{\Bd}\left(\pt>10\right)}\right.. 
\end{eqnarray} 
The $C_{\pt}$ is obtained using the generator-level simulation with inputs 
from the measured \pt\ spectra of \Lb\ and $\Bd$ 
(identical to those described in Section~\ref{sec:production}). The value of the correction factor is found to 
be:
\(
C_{\pt} = 1.31 \pm 0.11, 
\)
where the uncertainty also comes from the uncertainties on 
the measured \pt\ distributions in data.

After applying corrections $C_{\cal BR}$, $C_{\alpha}$, and $C_{\pt}$, 
$\rho$ is calculated to be:
\begin{eqnarray}
\rho &= & \rxsecc \pm \rxsececdf \left(\mathrm{CDF~I}\right) \pm \rxsecedelphi \left(\mathrm{DELPHI}\right) \rxsecept \left(\pt\right), \nonumber 
\end{eqnarray}
where the uncertainties are from the uncertainty on the CDF~I measurement 
of $\sigma_{\Lb}/\sigma_{\Bd}$, the difference between the estimated ${\cal B}\left(\lbsemil\right)$ for this analysis and that measured by DELPHI, 
and the uncertainties on the measured shapes of the \Lb\ and \Bd\ \pt\ 
distributions. The value of $\rho$ is also consistent with the result 
from~\cite{Gibson:2006rt}. The ${\cal B}\left(\lbhad\right)$ is then extracted 
following \eq~(\ref{eq:lbhadbr}), 
with the input of the parameters listed in Table~\ref{t:xsecvalue}, and is 
found to be:
\begin{eqnarray*}
\lefteqn{{\cal B}(\lbhad) = }\nonumber \\
&& \left(\brlbhadc \pm \brlbhadhqee \left(\mathrm{DELPHI}\right)
\brlbhadpte \left(\pt\right) 
\brlbhade \left(\mathrm{syst^{other}}\right) 
\right) \%.
\end{eqnarray*} 
The ``syst$^\mathrm{other}$'' uncertainty includes the uncertainty on the 
CDF~I measurement, and the uncertainty on $G$ which is dominated by the world 
average of 
${\cal B}\left(\Lc \rightarrow pK^-\pi^+\right)$~\cite{whylcpkpi}. 
This evaluation of ${\cal B}(\lbhad)$ differs from that of the Particle Data 
Group due to the 
differing production spectrum of the \Lb\ relative to the \Bd\ as described 
in the previous text~\cite{whynotlblcpipdg}. The estimated value is in good 
agreement with the values predicted by 
Leibovich~\etal~\cite{Leibovich:2003tw}, 
$\left(\hqetlbhadc\pm \hqetlbhade\right)\%$, and Cheng~\cite{Cheng:1996cs}, 
$\left(\chenglbhadc\pm \chenglbhade\right)\%$, 
which gives confidence in the procedure described above. 
The estimated ${\cal B}(\lbhad)$ has been used in \eq~(\ref{eq:bgexample}) to 
estimate $\nphys^\mathrm{other\Lb}/\nhad$ (see Section~\ref{sec:physicsb}).

 \begin{table}[tbp]
 \caption{Parameters for calculating ${\cal B}\left(\lbhad\right)$. 
	Text in parentheses indicate the sources of uncertainty: 
        the data sample size, general systematics, shapes of measured \pt\ 
	distributions, and difference from DELPHI's 
	${\cal B}\left(\lbsemil\right)$.}
 \label{t:xsecvalue}
 \begin{normalsize}
  \begin{center}
  \begin{tabular}{cl}  
  \hline \hline 
  Parameter & Value \\
 \hline
  $G$
& $0.82 \pm 0.25 (\mathrm{stat} \oplus \mathrm{syst}) \pm 0.06 (\pt)$ \\
 CDF~I $\frac{\sigma_{\Lb}}{\sigma_\mathrm{\Bd}}$ & $0.236 \pm 0.084 (\mathrm{stat} \oplus \mathrm{syst})$ \\
 $C_{\cal BR}$ & $\rwendyRc \pm \rwendyRe (\mathrm{DELPHI})$ \\
 $C_{\alpha}$ & $1.81 {+0.42 \atop -0.22} (\pt)$ \\
 $C_{\pt}$ & $1.31 \pm 0.11 (\pt)$ \\
 ${\cal B}\left(\dhad\right)$ &  $(\brdhadc \pm \brdhade) \%$ \\
 \hline\hline
\end{tabular}
 \end{center}
 \end{normalsize}
 \end{table}

\section{\label{sec:systematics}Systematic Uncertainties}
\begin{figure}[htbp]
 \begin{center}
 \includegraphics[width=300pt, angle=270]{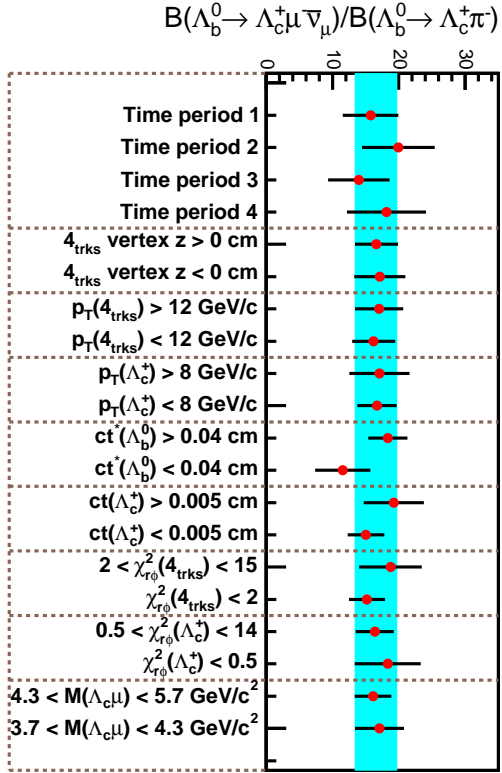}
 \caption
 {Internal-consistency check of the \Lb\ relative branching fractions. 
   The uncertainty on each point is statistical only.  
   Each independent group is separated by a horizontal dashed line.
   The solid bands indicate the relative branching fractions with their 
   statistical uncertainties from the complete, undivided samples.	
\label{fig:xbr}}	
 \end{center}
 \end{figure}

The \Lb\ relative branching fractions $\left(R\right)$, 
with statistical uncertainty only, can now be extracted from 
\eq~(\ref{eq:bigpicture}):
 \begin{eqnarray*}
 R = \frac{{\cal B}\left(\lbsemi\right)}{{\cal B}\left(\lbhad\right)}  & = &
	\rlbc\ \pm \rlbe \left(\rm{stat}\right).
\end{eqnarray*}
A check of internal consistency is performed by dividing the data 
and simulation samples into several groups of independent subsets, according 
to the time period, vertex position, \ctau\ and \pt\ of the \Lc\ 
candidate, $\ctau^*$ and \pt\ of the \Lb\ candidate, {\it etc}. 
\figc~\ref{fig:xbr} shows that the $R$ of 
each subset for each group is consistent with those of the other subsets in 
the same group. The result of this 
check also proves that there is no major problem in the detector, trigger, 
reconstruction, or simulation which produces bias in the measurement.  
 
The systematic uncertainties on $R$ may be classified as internal and 
external. Internal uncertainties are those which affect the final 
measurement through their effects on the observed yields, the numbers of false-muon and $b\bar{b}/c\bar{c}$ background events, and the modeling of acceptance 
times efficiency. External uncertainties are those from production 
fractions and branching ratios which are used in Eqs.~(\ref{eq:bgexample}) 
and (\ref{eq:baryonic}) to determine \nphys. 
The input value for each systematic source is varied by $\pm 1\sigma$ 
where $\sigma$ is the uncertainty on the input value. 
The resulting difference in $R$ from the central value is the systematic 
uncertainty. The following text describes how the uncertainty for each 
systematic source is obtained. 

\subsection{Internal systematic uncertainties \label{sec:internal}}
The signal yields, \nhad\ and \nincsemi, are affected by the 
background functions which describe the mass spectra of misidentified or 
partially-reconstructed decays of $b$ hadrons. The 
systematic uncertainty on the $M_{\Lamc\pi}$ fitting model is estimated by 
changing the relative fraction of the contributing decays in each background 
function. The fragmentation fraction and the branching ratio of every 
contributing decay are varied independently according to their uncertainties by $\pm 1\sigma$~\cite{uncertainty}. 
After combining these contributing decays according to their modified 
fragmentation fractions and branching ratios and producing a new simulated mass
 spectrum, the parameter values for each background function are re-determined.
 The $M_{\Lamc\pi}$ spectrum in data is re-fit again using the new background 
function and the systematic uncertainty is taken from the deviation of the 
\lbhad\ yield from the central value. 
Correlations between different parameters have been taken into account. 
For the fitting of $M_{pK\pi}$, since no branching ratio assumptions are made, 
no systematic uncertainty is assigned.

The uncertainty on the false-muon estimate is driven by: 1) the size of the 
sample used to measure the false-muon probability, 2) the fit to the weighted 
\Lamc\ mass distributions, and 3) the probability of a hadron track being a 
$p$, a $K$, or a $\pi$, which is determined from simulation. 
The resulting changes in the number of false-muon events from the three 
sources above are added in quadrature and already listed in Table~\ref{t:bigfake1}. 
The size of the $b\bar{b}$/$c\bar{c}$ background contribution has a 100$\%$ 
systematic uncertainty, due to a lack of knowledge of the relative 
$b\bar{b}$/$c\bar{c}$ 
production rates between different processes~\cite{Acosta:2004nj} and the 
10--50$\%$ discrepancy of the inclusive hadron production cross-section 
between \pythia\ and data (Appendix \ref{sec:bbcc}). 

The uncertainty on the modeling of acceptance times efficiency for signal 
and background processes arises from: 
the size of simulation samples, the shapes of the measured \Lb\ and \Bd\ \pt\ 
spectra, the efficiency scale factors of the \Lb\ semileptonic decay model/muon reconstruction/XFT trigger (see Section~\ref{sec:simulation}), the 
amount of material in the detector simulation, the \Lb\ lifetime, the \Lc\ 
Dalitz structure, and the \Lb\ and \Lc\ polarizations.  
The \Lb\ and \Bd\ \pt\ distributions used as inputs for \bgen\ 
are varied according to the uncertainties on the exponential slopes 
of data to MC ratios (shown in \fig~\ref{fig:bptafter}). The uncertainties on 
the efficiency scale factors for the \Lb\ semileptonic decay model/muon 
reconstruction/XFT trigger are listed in Table~\ref{t:effscale}.
The uncertainty from the detector material is obtained by switching off 
the hadronic interaction in the detector simulation and multiplying the 
efficiency difference from the central value by 25$\%$. The 25$\%$ is a 
quadrature sum of the 15$\%$ underestimate in the amount of material and 
the 20$\%$ difference between the {\sc GHEISHA} and {\sc FLUKA} models~\cite{geant,numi}. 
The \Lb\ lifetime used as an input for \bgen\ is varied according to 
the uncertainty on the world average~\cite{pdg}. 
 The effect of the \Lc\ Dalitz structure is studied by varying branching 
fractions of the resonant and non-resonant $\Lc\rightarrow pK^-\pi^+$ 
decays measured by E791~\cite{e791} by their uncertainties.  
The unpolarized \Lb\ and \Lc\ simulation samples have been used to obtain 
the central values of acceptance times efficiency of \Lb\ decays. For the
 systematics study, angular distributions in simulation are re-weighted 
according to all combinations of the \Lb\ and \Lc\ polarization 
states: $\pm 1$, assuming the extreme scenario where the \Lb\ and \Lc\ 
baryons are 100$\%$ polarized. The difference in the kinematic acceptances
 between the simulation with re-weighted angular distributions and the 
simulation with unpolarized \Lb\ and \Lc\ is used to assign a systematic 
uncertainty on $R$.

\subsection{External systematic uncertainties}
There are two types of external systematic uncertainties. The first type 
is denoted as the ``PDG'' uncertainty and includes uncertainties on: 
the world average of ${\cal B}\left(\dhad\right)$, the CDF I measurement of 
\rxsec, the CDF measurement of \yile, and the measured branching fractions 
of the four new \Lb\ semileptonic decays relative to that of the \lbsemi\ 
decay (see Section~\ref{sec:lbsemibr}). The second type is denoted as the 
``EBR'' uncertainty and comes from unmeasured 
branching fractions estimated from theory. A 5$\%$ uncertainty is assigned to 
the estimated branching fractions of the excited $c$-hadron decays~\cite{cerror}. 
A 100$\%$ uncertainty is assigned to the other unobserved $b$-hadron decays to 
cover the wide range of theoretical predictions~\cite{berror}. 
Note that the uncertainty on the estimated ${\cal B}\left(\lbsemi\right)$ 
does not affect the final measurement because it affects the 
branching fractions of \lbhad\ and other \Lb\ semileptonic decays in the 
same way so that any change completely cancels.

\subsection{Summary} 
The fractional systematic and statistical uncertainties on the \Lb\ relative 
branching fraction are summarized in Table~\ref{t:sysall}. The leading sources
 of internal systematic uncertainty are the mass fitting model, the shapes 
of the measured \pt\ spectra, and the \Lb\ semileptonic decay model. 
The PDG uncertainty is dominated by the world average of 
${\cal B}\left(\Lc \rightarrow pK^-\pi^+\right)$ and the CDF~I measurement of \rxsec\ which have been used to extract 
${\cal B}\left(\lbhad\right)$~\cite{detail}. 
 The uncertainty on ${\cal B}\left(\Lc \rightarrow pK^-\pi^+\right)$ 
may be reduced in the near future by more precise measurements 
proposed by Dunietz~\cite{Dunietz:1998uz} and 
Migliozzi~\cite{Migliozzi:1999ca}. The EBR uncertainty is dominated by 
the branching fractions of non-resonant 
\lblcpim\ and \lblcpizero\ decays. The \Lb\ relative branching fraction with 
complete uncertainties is found to be: 
\begin{eqnarray*}
\lefteqn{\frac{{\cal B}\left(\lbsemi\right)}{{\cal B}\left(\lbhad\right)} = } \\
&&\rlbc\ \pm \rlbe \left(\rm{stat}\right) \pm \rlbsyste \left(\rm{syst}\right)
	         \rlbbre {\rm \left(PDG\right)} \pm \rlbubre {\rm \left(EBR\right)}.
\end{eqnarray*}
The uncertainties are from statistics (stat), internal systematics 
(syst), world averages of measurements published 
by the Particle Data Group or subsidiary measurements in this analysis
$\left(\rm{PDG}\right)$, and unmeasured branching 
fractions estimated from theory $\left(\rm{EBR}\right)$, respectively. 

      \begin{table}[tbp]
       \caption{Summary of statistical and systematic uncertainties for the 
	\Lb\ mode. The $\sigma_R$ is the uncertainty on the \Lb\ relative 
        branching fraction, $R$.
	\label{t:sysall}}
       \renewcommand{\arraystretch}{1.5}

 	 \begin{center}
 	 \begin{tabular}{lr}
 	 \hline 
 	 \hline 
	 Source  
	 & \multicolumn{1}{c}{$\frac{\sigma_R}{R}$ $\left(\%\right)$} \\
	 \hline
		 Mass fitting &${+ 3.8 \atop - 3.1}$ \\
		 False $\mu$ &1.0 \\
		 $b\bar{b}$/$c\bar{c}$ background &0.3 \\
		 Simulation sample size &2.0 \\
		 $b$-hadron \pt\ spectrum &${+ 0.0 \atop - 2.9}$ \\
		 \Lb\ decay model &3.3 \\
		 XFT/CMU efficiency scale factor &0.4 \\
		 detector material &1.3 \\
		 \Lb\ lifetime &0.3 \\
		 \Lc\ Dalitz &0.4 \\
		 \Lb, \Lc\ polarizations &2.2 \\
	 \cline{2-2}
	 Sum of internal &6.3 \\
	 \hline
	 PDG &${+ 15.6 \atop - 20.4}$ \\
	 Estimated branching fractions &2.1 \\
	 \hline
	 Statistical  &17.8 \\
 	 \hline 
 	 \hline 
 	 \end{tabular}
 	 \end{center}

        \end{table}

\section{Measurements of the \Bd\ Relative Branching Fractions \label{sec:bmeson}}
 The same analysis technique used for the \Lb\ samples is applied to the 
\Bd\ decays. This section only describes the difference in the details of 
event reconstruction, yield determination, and background estimation and 
summarizes the systematic uncertainties. 

%
%

\subsection{\label{sec:b0rec} Reconstruction of the \Bd\ candidates}
The following decay modes are reconstructed in the data collected with the 
two-track trigger:
\begin{enumerate}
  \item \dhad\ and \incdsemi, where $\dplus \rightarrow K^-\pi^+\pi^+$.
  \item \dstarhad\ and \incdstarsemi, where $\dstar \rightarrow \dzero\pi^+, \dzero \rightarrow K^-\pi^+$.
 \end{enumerate}
The requirements on the \Bd\ and \Lb\ candidates are kept as similar as 
possible. 

For the reconstruction of $\dplus \rightarrow K^-\pi^+\pi^+$ decays, the pion mass
 is assigned to the two positively charged tracks and the kaon mass to 
the negatively charged track. The invariant mass of the three tracks 
($M_{K\pi\pi}$), as computed by a three-track kinematic fit, is required to be 
in the range 1.74~--~2.00~\gevcsq. 
The $\dstar$ signals are reconstructed by first looking for 
$\dzero \rightarrow K^-\pi^+$ candidates. 
A two-track kinematic fit determines the $\dzero$ vertex, 
and the invariant mass of the two tracks ($M_{K\pi}$) is required to be 
within the range 1.820~--~1.906~\gevcsq. Then, the pion mass is assigned to 
an additional positively charged track. 
This third track is expected to have a low \pt\ due to the small four-momentum 
transfer in the $\dstar \rightarrow \dzero\pi^+$ decay. However, a minimum 
\pt\ requirement of 0.5~\gevc\ is imposed to ensure a good measurement 
of the pion track. 
For the $\dstar$ candidate, the mass difference, 
$M_{K\pi\pi}- M_{K\pi}$, must be within the range 0.14~--~0.18~\gevcsq. 

In order to form a \Bd\ candidate, the $\dplus$ and $\dstar$ candidates are 
then combined with an additional negatively charged track 
which satisfies the requirements described in Section~\ref{sec:reconstruction}. 
After the four-track kinematic fit, the values of $M_{K\pi\pi}$ for the 
$\dplus$ and $M_{K\pi\pi}-M_{K\pi}$ for the $\dstar$ must be in the range: 
1.8517~--~1.8837~\gevcsq\ and 0.143~--~0.148~\gevcsq\ for the hadronic 
candidates; 1.74~--~2.00~\gevcsq\ and 0.14~--~0.18~\gevcsq\ for the inclusive 
semileptonic candidates. The four-track invariant mass, $M_{D\mu}$ and 
$M_{D^*\mu}$, must be within 3.0~--~5.3~\gevcsq\ for the 
semileptonic decays. Selection criteria for the following variables: 
\pt\ of the fourth \Bd-candidate track [$\pt\left(\pi^-, \mu^-\right)$], 
\pt\ of $\dplus$, \pt\ of $\dstar$, and combined 
\pt\ of the four-track system, 
\chixy\ of the $\dplus$ and $\dzero$ vertex fits, and 
\chixy\ of the four-track kinematic fits, 
$\ctau$ of the $\dplus$ and $\dzero$ candidates, and $\ctau^*$ of the \Bd\ candidate, are also optimized using the simulation and data of hadronic modes, 
as described 
for the \Lb\ sample. Table~\ref{t:fanacutB0} lists the optimized values.

%

 \begin{table*}[tbp]
   \caption{Optimized requirements for reconstructing the 
	\dhad, \incdsemi, \dstarhad, and \incdstarsemi\ 
	decays.	\label{t:fanacutB0}}
   \begin{center}
   \begin{normalsize}
  \renewcommand{\arraystretch}{1.5}
  \renewcommand{\tabcolsep}{0.15in}
  \begin{tabular}{llll} 
   \hline  
   \hline  
    \multicolumn{2}{c}{\dhad} & 
    \multicolumn{2}{c}{\dstarhad}\\
    \multicolumn{2}{c}{\incdsemi} & 
    \multicolumn{2}{c}{\incdstarsemi}\\
   \hline  
   $\pt\left(\pi^-, \mu^-\right)$ & $>$ 2 \gevc & 
   $\pt\left(\pi^-, \mu^-\right)$ & $>$ 2 \gevc \\   
   $\pt\left(\dplus\right)$ & $>$ 5 \gevc 
    & $\pt\left(\dstar\right)$ & $>$ 5 \gevc \\
   $\pt\left(4_\mathrm{trks}\right)$ & $>$ 6 \gevc &  
   $\pt\left(4_\mathrm{trks}\right)$ & $>$ 6 \gevc \\
  
   $\chixy\left(\dplus\right)$ & $<$ 14  &  
   $\chixy\left(\dzero\right)$ & $<$ 16  \\  
   $\chixy\left(4_\mathrm{trks}\right)$ & $<$ 15  &  
   $\chixy\left(4_\mathrm{trks}\right)$ & $<$ 17  \\     
   $\ctau\left(\dplus\right)$ & $>$ $-30$~$\mu$m  &  
   $\ctau\left(\dzero\right)$ & $>$ $-70$~$\mu$m  \\ 
   $\ctau^*\left(\Bd\right)$ & $>$ 200~$\mu$m  &  
   $\ctau^*\left(\Bd\right)$ & $>$ 200~$\mu$m  \\ 
   \hline  
  \hline
 \end{tabular}
 \end{normalsize}
 \end{center}
 \end{table*}

%
%
\subsection{Determination of the $\Bd$ yields \label{sec:controlfit}} 

\figc~\ref{fig:allbgB0}(a) shows the fit result for the 
$M_{D\pi}$ spectrum. 
The \dhad\ yield returned by the fit is $\ndhadc\pm\ndhade$. 
The signal peak at $M_{D\pi}\approx 5.3$~\gevcsq\ and the combinatorial 
background are described by a Gaussian function and an exponential, 
respectively. The ratio of the number of doubly Cabibbo-suppressed decays 
relative to that of the signal mode, $N_{DK}/N_{D\pi}$, is 
Gaussian-constrained to the 
value for the relative branching ratio from the PDG, convoluted with the 
efficiency from the full simulation. The constrained value is 
$0.073 \pm 0.023$. 
Backgrounds from the other $b$-hadron decays consist of the 
following decays. The $\bar{B}_s^0 \rightarrow D_s^{+}\pi^-$ decays, 
where $D_s^{+} \rightarrow \phip\pi^+$, $\phip \rightarrow K^+K^-$ 
and the pion mass is assigned to one of the kaons, appear as a peak at 
around 5.31~\gevcsq. 
Misreconstructed $\lbhad$ decays, where $\Lc \rightarrow pK^-\pi^+$ and
 the pion mass is assigned to the proton, form a broad peak around 5.4 
\gevcsq. The backgrounds from the $\bar{B}^0 \rightarrow \dplus K^-$, 
$\bar{B}_s^0 \rightarrow D_s^{+}\pi^-$, and \lbhad\ decays are combined 
and indicated by the black filled region. 
The $\bar{B}^0\rightarrow \dplus\rho^-$ 
decays, where $\rho^-\rightarrow \pi^0\pi^-$ and the $\pi^0$ is not 
reconstructed in the event, have a triangular mass distribution which
 peaks at $\approx 5.1$~\gevcsq. The 
\dstarhad\ decays, where $\dstar\rightarrow \dplus\pi^0$ and the $\pi^0$ is not
 reconstructed, have a double-peak structure. This structure is 
consistent with the spin-1 $\dstar$ being polarized. 
This polarization results in the $\pi^0$ from the $\dstar$ decay having a 
momentum preferentially parallel or anti-parallel to the momentum of $\dstar$. 
The $\bar{B}^0\rightarrow \dplus\rho^-$ and \dstarhad\ backgrounds are combined 
and indicated by the dark-gray filled region. The
 remaining partially-reconstructed decays of $b$ hadrons, 
$H_b\rightarrow \dplus X$, have a monotonically falling distribution (hatched region). 
The determination 
of the background shapes and the estimation of systematic uncertainty are 
similar to those in the \Lb\ system. 

\figc~\ref{fig:allbgB0}(b) shows the fit result for the $M_{K\pi\pi}$ spectrum 
for events with muons. 
The inclusive \incdsemi\ yield returned by the fit is $\ndsemic\pm\ndsemie$. 
The signal peak at $M_{K\pi\pi}\approx 1.9$~\gevcsq\ is described by a 
Gaussian function. 
The combinatorial background (light-gray filled region) is parameterized by a 
first-order polynomial. Misidentified $H_b\rightarrow D_s^+\mu^-X$ decays 
(black filled region), where the mass of at least 
one $D_s^+$ daughter has been misassigned, appear in the mass window of 
interest. The dominant contributing $D_s^+$ decay modes are 
$D_s^{+} \rightarrow \phip\pi^+$, 
$D_s^+\rightarrow K^{*0} K^+$, and $D_s^+\rightarrow $non-resonant~$K^+K^-\pi^+$.
The function parameters describing the shape of misidentified $D_s^+$ spectrum 
are obtained from the 
$\bar{B}_s^0\rightarrow D_s^+\mu^-\bar{\nu}_{\mu}$ simulation; 
in this simulation, the $D_s^+$'s are forced to decay only to the final 
states which can yield misidentified mass in the 
$M_{K\pi\pi}$ window. The number of the $D_s^+$ background events is 
constrained to the estimated number of 
$H_b\rightarrow D_s^+\mu^-X$ events in the data $\left(N_{D_s\mu}\right)$ as 
described below. First, the following decay mode is reconstructed in the data:
 $H_b\rightarrow D_s^+\mu^-X$, where 
$D_s^{+} \rightarrow \phip\pi^+$ and $\phip \rightarrow K^+K^-$. The narrow $
\phip$ resonance provides a good handle for removing the combinatorial 
background of $D_s^+$. Second, the fraction from the $\phip\pi^+$ mode
 relative to all contributing $D_s^+$ decays $\left(R_{\phi\pi}\right)$ is 
extracted using the world averages of the $D_s^+$ branching ratios, and the 
acceptance times efficiency determined from the full simulation. Then, 
$N_{D_s\mu}$ is simply the yield of the $\phip\pi^+\mu^-$ mode in data divided by 
$R_{\phi\pi}$. The value of $N_{D_s\mu}$ for the constraint is $1812\pm 160$.
The systematic uncertainty is assigned by independently varying the ratio of 
the branching fraction of one specific $D_s^+$ decay relative to that of the 
$D_s^{+} \rightarrow \phip\pi^+$ decay by $\pm 1\sigma$, since the
 branching fractions of all $D_s$ decays have been measured relative to that of
 the $\phip\pi^+$ mode~\cite{pdg}.

\figc~\ref{fig:allbgB0}(c) shows the fit result for the 
$M_{D^{*}\pi}$ spectrum. 
The \dstarhad\ yield returned by the fit is $\ndstarhadc\pm\ndstarhade$. 
The analysis of the $\dstar\pi^-$ signal and backgrounds is similar to 
that in the $\dplus\pi^-$ mode. The only difference is that extra constraints 
are imposed due to the small size of the $\dstar\pi^-$ sample. 
The width of the signal Gaussian $\sigma_{D^*\pi}$, 
the ratio $\frac{N_{D^*K}}{N_{D^*\pi}}$, and the 
ratio of backgrounds $N_{D^*\rho}/\left(N_{D^*\rho} + N_{H_b\rightarrow\mathrm{remaining}\; D^*X}\right)$, are constrained to $0.0259\pm 0.0012$~\gevcsq, 
$0.071\pm 0.019$, and $0.242\pm 0.008$, respectively. 
The systematic uncertainty is assessed in the same way as 
in the \Lb\ and the $\dplus\pi^-$ modes.

\figc~\ref{fig:allbgB0}(d) shows the fit result for the $M_{K\pi\pi}-M_{K\pi}$ 
spectrum for events with muons. 
The likelihood fit for the \incdstarsemi\ mode is performed in the 
mass window $M_{K\pi\pi}-M_{K\pi}=0.14-0.18$~\gevcsq\ whereas 
\fig~\ref{fig:allbgB0}(d) shows a more restricted mass range near the signal 
peaks. 
The inclusive \incdstarsemi\ yield returned by the fit is 
$\ndstarsemic\pm\ndstarsemie$. 
The signal peak at $M_{K\pi\pi}-M_{K\pi}\approx 0.145$~\gevcsq\ is modeled by 
two Gaussian distributions with a common mean and different widths. The 
combinatorial background (light-gray filled region) is parameterized by a 
constant, while the background from other $b$-hadron decays with 
misidentified $c$-hadron daughters is found to be negligible. 
The size of the combinatorial background is very small due to the requirement 
that $M_{K\pi}$ is consistent with the world average $\dzero$ mass, the 
minimum requirement on the mass $M_{D^*\mu}$, and the minimum requirements on 
the \pt\ and the number of SVX hits for the low-momentum pion from the $\dstar$
 decay (Section~\ref{sec:b0rec}). 
The fitting function for this spectrum does not use any branching ratios 
and no systematic uncertainty is assigned.

Table~\ref{t:yieldsumB} summarizes the \Bd\ hadronic and inclusive 
semileptonic yields and the $\chi^2$ probability of corresponding fits. 
Each model describes the data well, as indicated by the $\chi^2$ probability.

 \begin{table*}[ptb]
  \renewcommand{\arraystretch}{1.5}
   \caption{Observed number of events in each decay mode determined from 
	the unbinned, extended likelihood fit, $\chi^2/$NDF, 
	and the corresponding probability computed to indicate 
	quality of the fit. \label{t:yieldsumB}}
   \begin{center}
    \begin{tabular}{lrcc}
    \hline 
    \hline
     \multicolumn{1}{c}{Mode} & \multicolumn{1}{c}{Yield} &
     \multicolumn{1}{c}{$\chi^2/$NDF} & Prob $\left(\%\right)$\\	
     \hline 
      \dhad & $\ndhadc \pm \ndhade$ & 80/91 & 78.9 \\
      \incdsemi & $\ndsemic \pm \ndsemie$ & 47/31 & 3.40 \\ 
      \dstarhad & $\ndstarhadc \pm \ndstarhade$ & 21/12 & 5.40 \\
      \incdstarsemi & $\ndstarsemic \pm \ndstarsemie$ & 108/93 & 14.1 \\
     \hline	
      \hline
      \end{tabular}
      \end{center}
  \end{table*}

\begin{figure*}[tbp]
 \begin{center}
  \renewcommand{\arraystretch}{10}
 \begin{tabular}{cc}
 \includegraphics[width=230pt, angle=0]
	{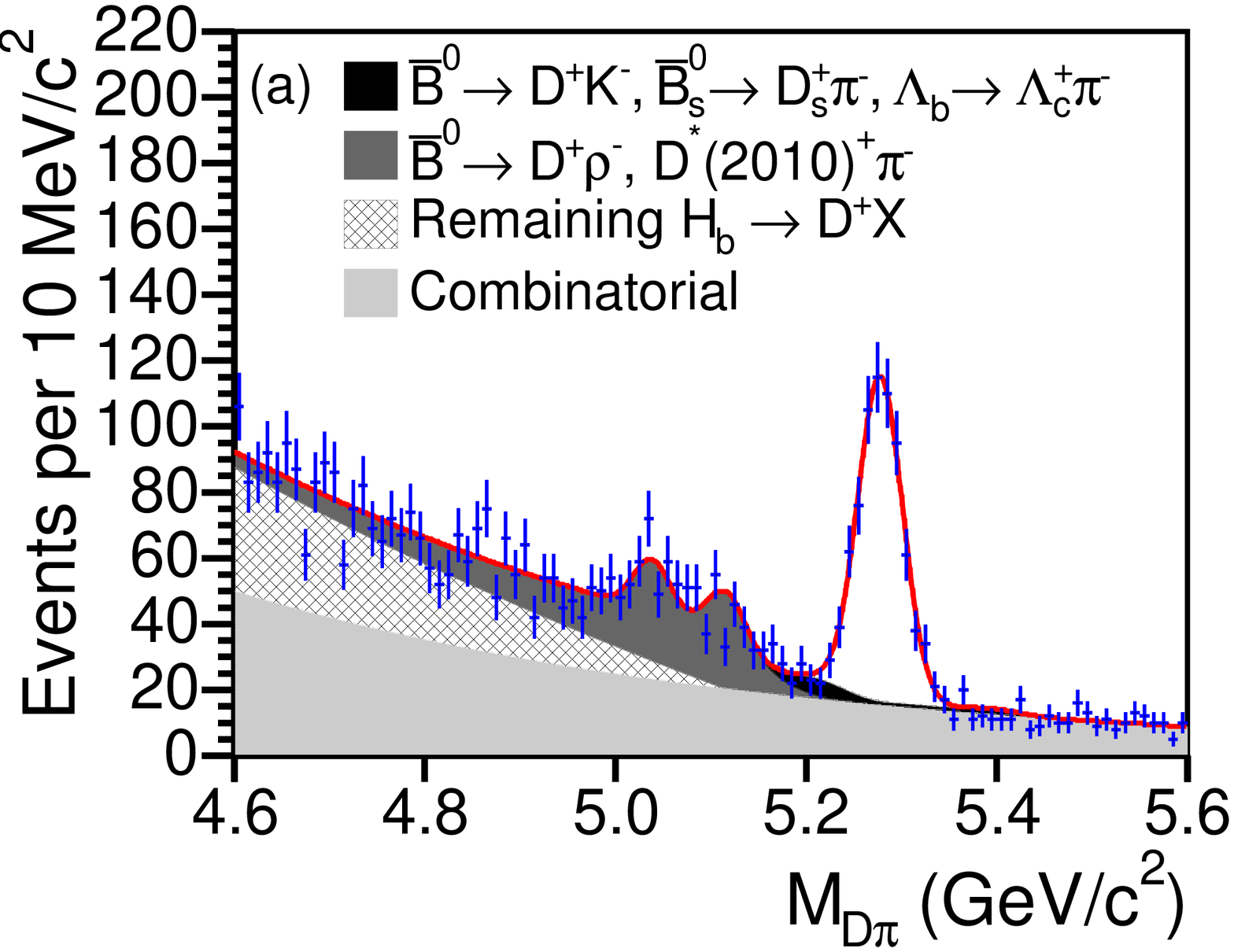}& 
 \includegraphics[width=230pt, angle=0]
	{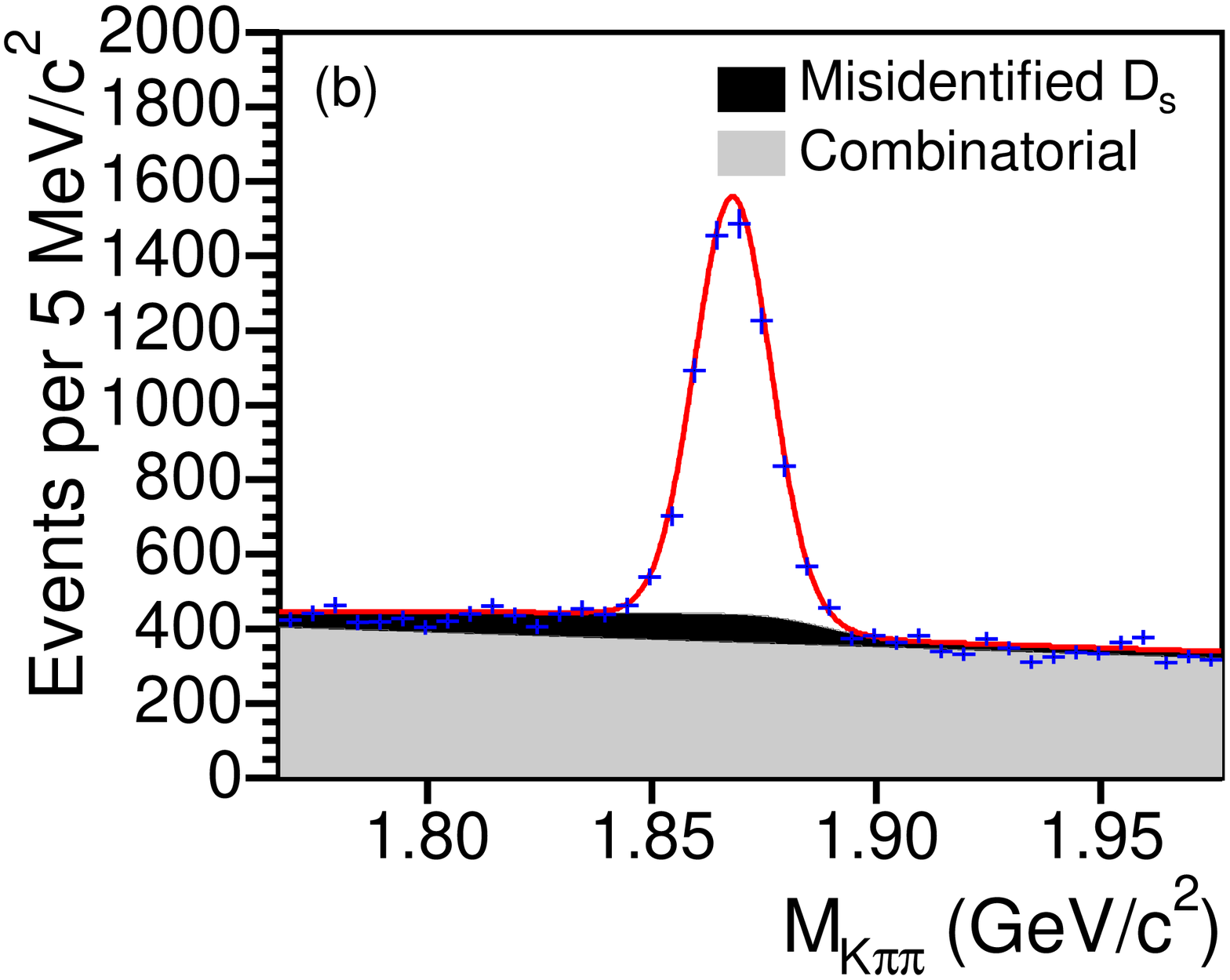}\\
 \includegraphics[width=230pt, angle=0]
	{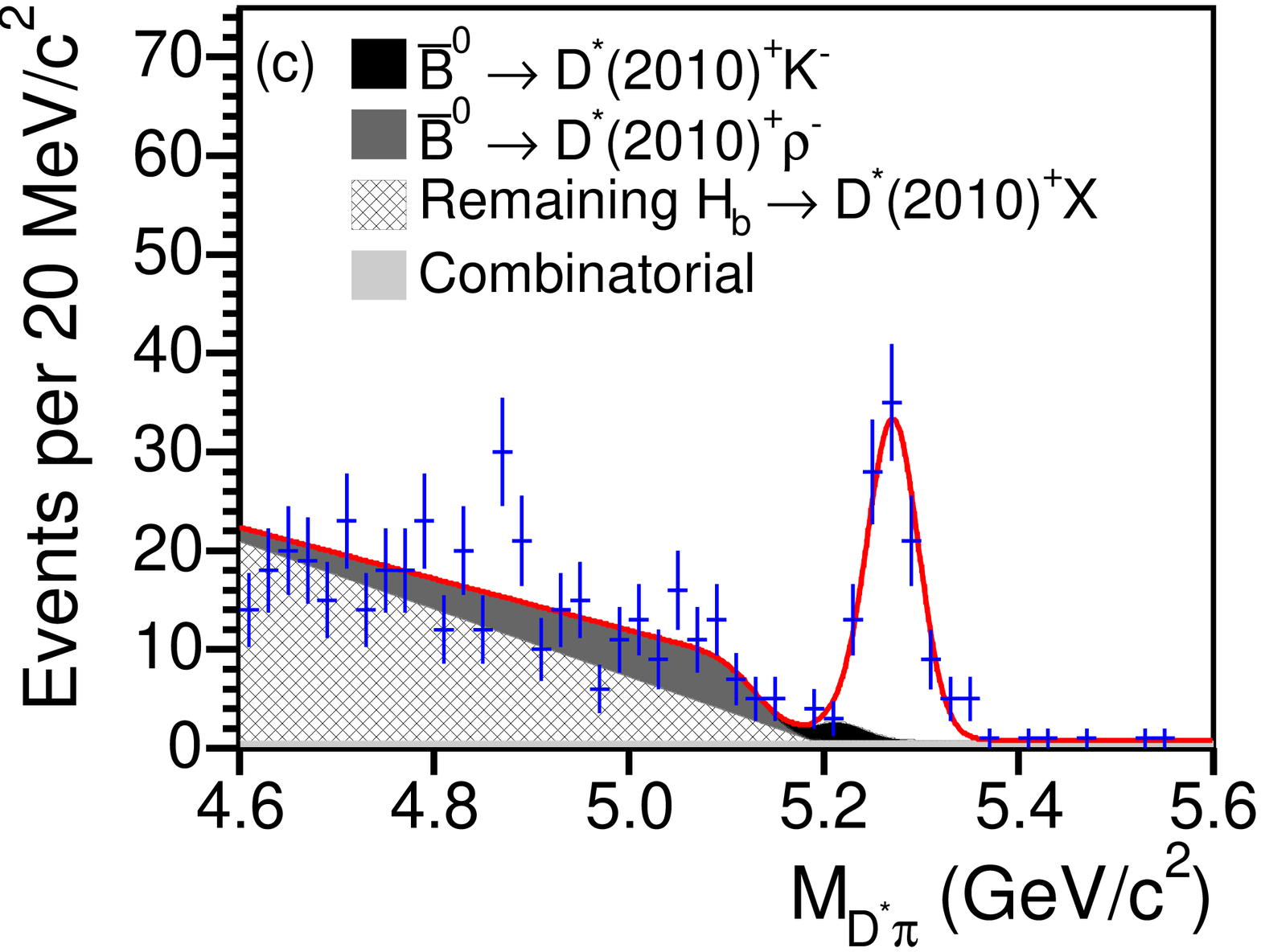} & 
 \includegraphics[width=230pt, angle=0]
	{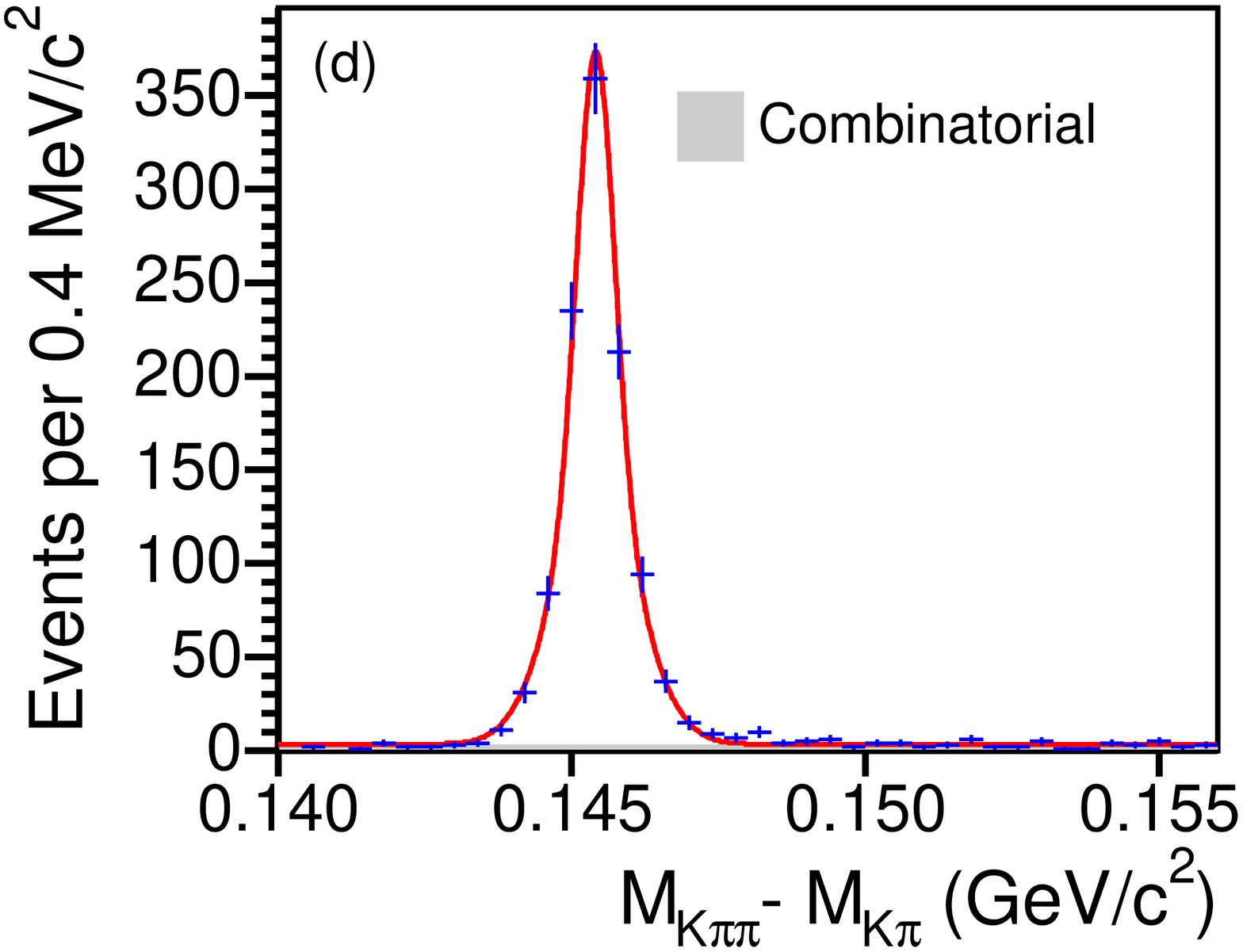}\\
 \end{tabular}
  \caption{Results (curve) of the unbinned, extended likelihood fits for 
determining the numbers of \Bd\ candidates: the hadronic modes (a) 
$M_{D\pi}$ and (c) $M_{D^*\pi}$, and inclusive semileptonic modes (b) 
$M_{K\pi\pi}$ and (d) $M_{K\pi\pi}-M_{K\pi}$. The filled histograms indicate 
various backgrounds. 
\label{fig:allbgB0}
}
\end{center}
 \end{figure*}

\subsection{Compositions of the inclusive semileptonic data 
	\label{sec:b0bg}}
The procedures for estimating the backgrounds to the \dsemi\ 
and \dstarsemi\ decays are similar to those described in 
Section~\ref{sec:slbackgrounds}. The following 
describes the differences when estimating the feed-in background, \nphys, in 
the \Bd\ system. 
Unlike the \Lb\ system, many decays of $b$ and $c$ mesons have been measured 
by other experiments~\cite{babar:tdr,belle:tdr,Andrews:1982dp,Albrecht:1988vy},
 and serve as inputs to the \evtgen\ decay package. In addition, \evtgen\ also 
includes estimates of branching fractions for decay modes which have not yet 
been measured. Therefore, all possible decays which may contribute to the 
\nphys\ in the \Bd\ control samples are studied using the PDG summary and 
the default \evtgen\ decay table~\cite{pdg,evtgen}. 

%
 The feed-in backgrounds to the \dsemi\ and \dstarsemi\ signals fall 
into two categories:
\begin{enumerate}
 \item Semileptonic decays of \Bd/\Bu/\Bs, which include 
either additional particles (\eg, 
$\bar{B}^0\rightarrow \dplus\pi^0\mu^-\bar{\nu}_{\mu}$) or a higher 
mass $c$ meson with subsequent decay into the $c$-meson signal 
(\eg, \dstarsemi, $\dstar \rightarrow \dplus\pi^0$)
  \item Hadronic decays of $b$ mesons into two $c$ mesons: one $c$ meson 
decays hadronically in a reconstructed final state, the other $c$ meson 
decays semileptonically (\eg, $\bar{B}^0\rightarrow \dplus D_s^-$, 
$D_s^- \rightarrow \phip \mu^- \bar{\nu}_{\mu}$). 
 \end{enumerate}
Branching fractions of the $\bar{B}\rightarrow D\bar{D}$ decays relative to the
 signal are all below $3\%$. A generator-level study indicates that they are 
further suppressed after a minimum requirement on the four-track invariant 
mass, $M_{D(D^*)\mu}$, and therefore, contribute less than $1\%$ to the 
signal. Backgrounds from $b$~mesons decaying semileptonically to more particles
 or higher mass $c$ mesons are also reduced or eliminated by the same 
minimum mass requirement. 

\begin{figure*}[tbp]
 \begin{center}
 \includegraphics[width=230pt, angle=0]{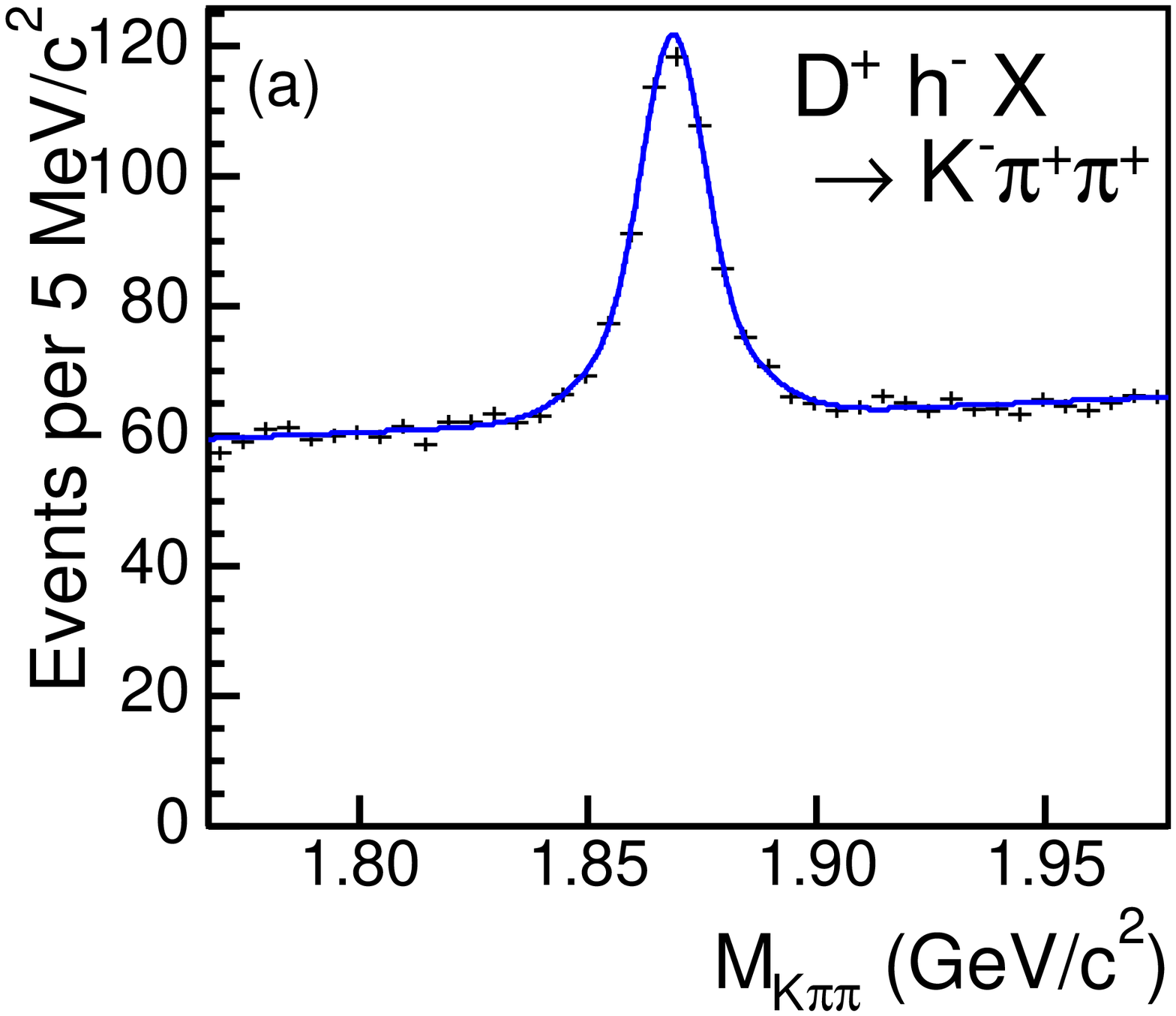} 
 \includegraphics[width=230pt, angle=0]{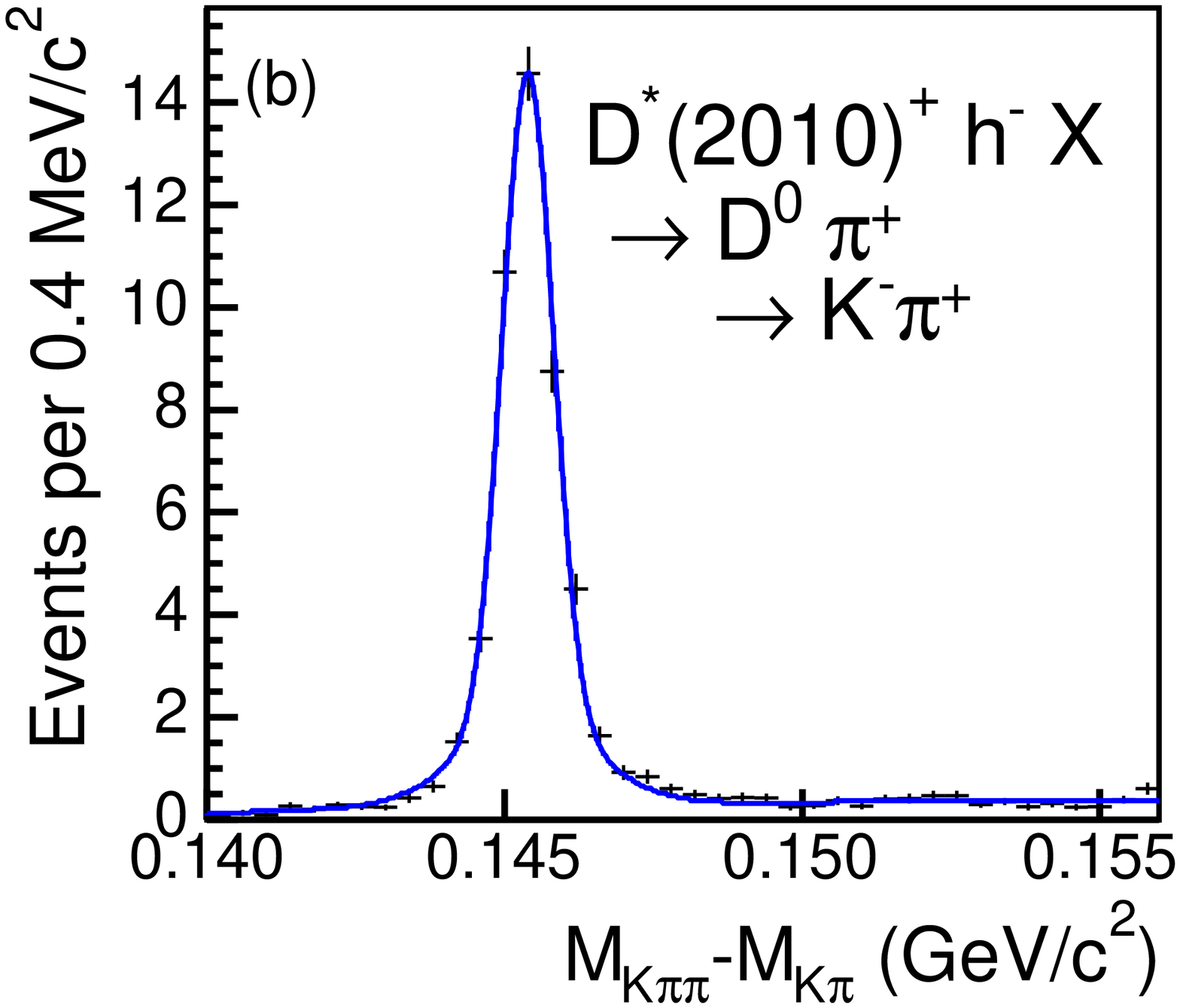}
  \caption
   { The invariant mass distributions produced from data with a hadron track 
	$\left(h^-\right)$ and a $c$-meson candidate in the final 
	state, after weighting the hadron track with an average 
	muon-misidentification probability 
	$\left({\cal P}_\mathrm{avg}\right)$: (a) $M_{K\pi\pi}$ and (b) 
	$M_{K\pi\pi}-M_{K\pi}$. 
	The curves indicate the results of the $\chi^2$ fit.
     \label{fig:cmfake2}}
     \end{center}
  \end{figure*}

Tables~\ref{t:physicsd}--\ref{t:physicsdstar} summarize the feed-in 
backgrounds which contribute $\geq 1\%$ to the \dsemi\ and the \dstarsemi\ 
decays. The definition of quantities listed in each column follows 
Table~\ref{t:physicslc}. 
Only these decays are subtracted from the inclusive semileptonic yield.
The leading background to \dsemi\ is \dstarsemi\ where 
$\dstar \rightarrow \dplus\pi^0$. 
The leading background to \dstarsemi\ is 
$B^-\rightarrow D_1(2420)^{0}\mu^-\bar{\nu}_{\mu}$ where 
$D_1(2420)^{0}\rightarrow \dstar\pi^{-}$. 
Combining information compiled in the PDG, backgrounds from 
$B\rightarrow D^{(*,**)} \mu^-\bar{\nu}_{\mu} X$ which are not considered in 
Tables~\ref{t:physicsd}--\ref{t:physicsdstar} contribute less than 
$2\%$ to the signal. 
The estimates of \nfake\ and \nbbcc\ for the \Bd\ are identical to those for 
the \Lb. Table~\ref{t:b0final} lists the results. 
\figc~\ref{fig:cmfake2} shows 
the $M_{K\pi\pi}$ and $M_{K\pi\pi}-M_{K\pi}$ distributions weighted with 
muon-misidentification probabilities and the results of the $\chi^2$ fit.

 \begin{normalsize}
 \begin{table*}[tbp]
 
\renewcommand{\arraystretch}{1.1}
\renewcommand{\tabcolsep}{0.05in}
 \caption{The feed-in backgrounds to the \dsemi\ signal. 
	For the $\bar{B}_s^0\rightarrow \dplus K^0\mu^-\bar{\nu}_{\mu}$ 
	decay, the $\Gamma(\Bs)/\left[\Gamma(\Bu)+\Gamma(\Bd)\right]$ from 
	the PDG is used to obtain 
	$\nphys/\nhad$~\cite{Bs,support}. 
	Numbers in parentheses are estimated uncertainties for the 
	unmeasured branching fractions~\cite{b0error}.
	The definition of quantities listed in each column follows 
	Table~\ref{t:physicslc}. 
	}
 \label{t:physicsd}

	 \begin{center}
	 \begin{tabular}{lr@{\,$\pm$\,}lcccr@{\,$\pm$\,}r}
	 \hline
	 \hline
	 \multicolumn{1}{c}{Mode}
	 & \multicolumn{2}{c}{${\cal B}$ $\left(\%\right)$}
	 & \multicolumn{1}{c}{$\frac{\eff_i}{\eff_{\dhad}}$}
	 & \multicolumn{1}{c}{$\left(\frac{\nphys}{\nhad}\right)^i$}
	 & \multicolumn{1}{c}{$\left(\frac{\nphys}{\nsemi}\right)^i$}
	 & \multicolumn{2}{c}{$N_{event}$} \\ 
	 \hline
		 \dhad
		 &    \brdhadc &  \brdhade
		 &   \multicolumn{1}{l}{1.000}
		 & -- 
		 & --  & \ndhadc & \ndhade \\ 

		 \incdsemi
		 &  \multicolumn{2}{c}{--}
		 &  -- & -- 
		 & -- & \ndsemic & \ndsemie \\ 
	 \hline
		 \dsemi
		 & 2.17 & 0.12
		 & $0.455 \pm 0.004$
		 & 3.688
		 & 1.000 & \multicolumn{2}{c}{--} \\
	 \hline
		 \dstarsemi 
		 &  5.16 & 0.11
		 &  $0.372 \pm 0.004$
		 & 2.314
		 & 0.627 & 1340 & 69 \\
		 $\hspace{36pt} \hookrightarrow \dplus\pi^{0}/\gamma$
		 & 32.30 & 0.64
		 &  & & & \multicolumn{2}{c}{}\\ 

		 \bddpizeromunu 
		 & 0.30 & (0.30)
		 & $0.165 \pm 0.003$
		 & 0.185
		 & 0.050 & 107 & 6 \\
		 \bddtau
		 &  1.00 &  0.40
		 &  $0.100 \pm 0.004$
		 & 0.065
		 & 0.018 & 37 & 2\\

		 $ \hspace{45pt}    \hookrightarrow \mu^{-}\bar{\nu}_{\mu}\nu_{\tau}$
		 &  \brtautomuc &  \brtautomue
		 & 
		 & & & \multicolumn{2}{c}{} \\

		 \bpdonezeromunu
		 &  0.40 & 0.07
		 & $0.278 \pm 0.005$
		 & 0.134
		 & 0.036 & 78 & 4\\
		 $   \hspace{36pt}   \hookrightarrow \dstar\pi^{-}$
		 &  \multicolumn{2}{c}{}
		 & 
		 & & & \multicolumn{2}{c}{} \\
		 $\hspace{60pt} \hookrightarrow \dplus\pi^{0}/\gamma$
		 &  32.30 & 0.64
		 & 
		 & & & \multicolumn{2}{c}{} \\

		 \bpdpronezeromunu
		 &  0.37 & (0.37)
		 & $0.273 \pm 0.005$
		 & 0.081
		 & 0.022 & 47 & 2 \\
		 $\hspace{36pt} \hookrightarrow \dstar\pi^{-}$
		 &  66.67 & (3.33)
		 & 
		 & & & \multicolumn{2}{c}{} \\
		 $\hspace{60pt} \hookrightarrow \dplus\pi^{0}/\gamma$
		 &  32.30 &  0.64
		 & 
		 & & & \multicolumn{2}{c}{} \\
		 \bpdpimunu 
		 &  0.42 & 0.05
		 & $0.165 \pm 0.003 $
		 & 0.259
		 & 0.070 & 150 & 8\\

		 \bsdkzero	
		 & 0.30 &  (0.30)
		 &  $0.137 \pm 0.004$
		 & 0.064
		 & 0.017 & 37 & 2\\
	 \hline 
	 \hline 
	 \end{tabular}
	 \end{center}

 \caption{The feed-in backgrounds to the \dstarsemi\ signal. 
	}
 \label{t:physicsdstar}
	 \begin{center}
	 \begin{tabular}{lr@{\,$\pm$\,}lcccr@{\,$\pm$\,}r}
	 \hline
	 \hline
	 \multicolumn{1}{c}{Mode}
	 & \multicolumn{2}{c}{${\cal B}$ $\left(\%\right)$}
	 & \multicolumn{1}{c}{$\frac{\eff_i}{\eff_{\dstarhad}}$}
	 & \multicolumn{1}{c}{$\left(\frac{\nphys}{\nhad}\right)^i$}
	 & \multicolumn{1}{c}{$\left(\frac{\nphys}{\nsemi}\right)^i$}
	 & \multicolumn{2}{c}{$N_{event}$} \\ 
	 \hline
		  \dstarhad
		  &  0.276 & 0.013
		  &   \multicolumn{1}{l}{1.000}
		  & --
		  & -- &  \ndstarhadc &  \ndstarhade \\

		  \incdstarsemi
		  &  \multicolumn{2}{c}{--}
		  &  -- & --
		  &  -- &  \ndstarsemic & \ndstarsemie \\
	 \hline
		  \dstarsemi 
		  &  5.16 & 0.11
		  &  $0.447 \pm 0.006$
		  &  8.361
		  & 1.000 & \multicolumn{2}{c}{--} \\
	 \hline
		  \bddonemunu
		  &  0.81 & (0.32)
		  &  $0.349 \pm 0.008$
		  &  0.341
		  &  0.041 & 36 & 4 \\
		   $\hspace{36pt} \hookrightarrow \dstar\pi^{0}$
		  & 33.33 & (1.67)
		  & 
		  & & & \multicolumn{2}{c}{} \\
		  \bddpronemunu
		  &  0.37 & (0.37)
		  &  $0.336 \pm 0.008$
		  &  0.150
		  &  0.018 & 16 & 2 \\
		  $\hspace{36pt} \hookrightarrow \dstar\pi^{0}$
		  & 33.33 & (1.67)
		  & 
		  & & & \multicolumn{2}{c}{} \\
		  \bddstarpizeromunu
		  &  0.10 & (0.10)
		  &  $0.239 \pm 0.006$
		  &  0.086
		  &  0.010 & 9 & 1 \\
		 \bddstartau 
		  &  1.60 & 0.50
		  &  $0.136 \pm 0.005$
		  &  0.137
		  &  0.016 & 14 & 2 \\
		  $ \hspace{76pt}    \hookrightarrow \mu^{-}\bar{\nu}_{\mu}\nu_{\tau}$
		 &  \brtautomuc &  \brtautomue
		 & 
		 & & & \multicolumn{2}{c}{} \\

		  \bpdonezeromunu
		  &  0.40 & 0.07
		  &  $0.356 \pm 0.008$
		  &  0.516
		  &  0.062 & 55 & 6 \\
		 $   \hspace{36pt}   \hookrightarrow \dstar\pi^{-}$
		  &  \multicolumn{2}{c}{} 
		 & & & & \multicolumn{2}{c}{} \\
		  \bpdpronezeromunu	
		  &  0.37 & (0.37)
		  &  $0.351 \pm 0.008$
		  &  0.314
		  &  0.038 & 33 & 3 \\
		  $\hspace{36pt} \hookrightarrow \dstar\pi^{-}$
		  & 66.67 & (3.33)
		  & 
		  & & & \multicolumn{2}{c}{} \\
		  \bpdstarpimunu
		  &  0.61 & 0.06
		  &  $0.242 \pm 0.006$
		  &  0.534
		  &  0.064 & 57 & 6 \\
	 \hline 
	 \hline 
	 \end{tabular}
	 \end{center}

 	\end{table*}
 \end{normalsize}

  \begin{table}[tbp]
   \caption{
    The estimated sizes of the false-muon, $b\bar{b}$, $c\bar{c}$, and feed-in 
    background contribution to the \dsemi\ and \dstarsemi\ signals and the 
    observed yields in data. Uncertainties on the $b\bar{b}$, $c\bar{c}$, 
    and feed-in backgrounds are statistical only.}
   \label{t:b0final}
   \begin{normalsize}
   \begin{center}
   \begin{tabular}{lr@{\,$\pm$\,}lr@{\,$\pm$\,}l}
   \hline
   \hline
   & \multicolumn{2}{c}{$\dplus\mu^-$} 
   & \multicolumn{2}{c}{$\dstar\mu^-$} \\
   \hline
    \nfake  & 230 & 19 & 44 & 3 \\ 
   $N_{b\bar{b}}/\nhad$ & \bdbbrc & \bdbbre & \bdstarbbrc & \bdstarbbre \\
   $N_{c\bar{c}}/\nhad$ & \bdccrc & \bdccre & \bdstarccrc & \bdstarccre \\
   $\nphys/\nhad$  &  \bdphysinputc & \bdphysinpute & \bdstarphysinputc & \bdstarphysinpute \\
   \nhad\ 
   &  \ndhadc &  \ndhade &   \ndstarhadc &  \ndstarhade \\
   \nincsemi\ 
   &  \ndsemic &  \ndsemie  
   &  \ndstarsemic &  \ndstarsemie \\
  \hline
   \hline
 \end{tabular}
 \end{center}
 \end{normalsize}
 \end{table}

 The compositions of the inclusive $\dplus\mu^-$ and $\dstar\mu^-$ 
samples are summarized in Table~\ref{t:bgsumallB0}. 
The dominant signal contamination is from the feed-in background. The second 
largest background arises from false muons. The smallest background source is 
from $b\bar{b}$/$c\bar{c}$. 

\begin{table}[h]
\caption{The composition of the inclusive $\dplus\mu^-$ and $\dstar\mu^-$
         data sample.}
\label{t:bgsumallB0}
\begin{center}
\begin{tabular}{crr}
   \hline
   \hline
  & \multicolumn{2}{c}{$N/N_\mathrm{inc\;semi}$ $\left(\%\right)$} \\
 & \multicolumn{1}{c}{$\dplus\mu^-$} & \multicolumn{1}{c}{$\dstar\mu^-$} \\
\hline
  Signal                  &  \bdfracsignal &  \bdstarfracsignal \\
  False muon              &  \bdfracfakemu &  \bdstarfracfakemu \\
  $b\bar{b}$/$c\bar{c}$   &  \bdfracbbcc   &  \bdstarfracbbcc   \\
  Feed-in                 &  \bdfracphysb  &  \bdstarfracphysb   \\
\hline\hline
\end{tabular}
\end{center}
\end{table}

\subsection{Systematic uncertainties}
\begin{figure}[htbp]
 \begin{center}
 \includegraphics[width=300pt, angle=270]{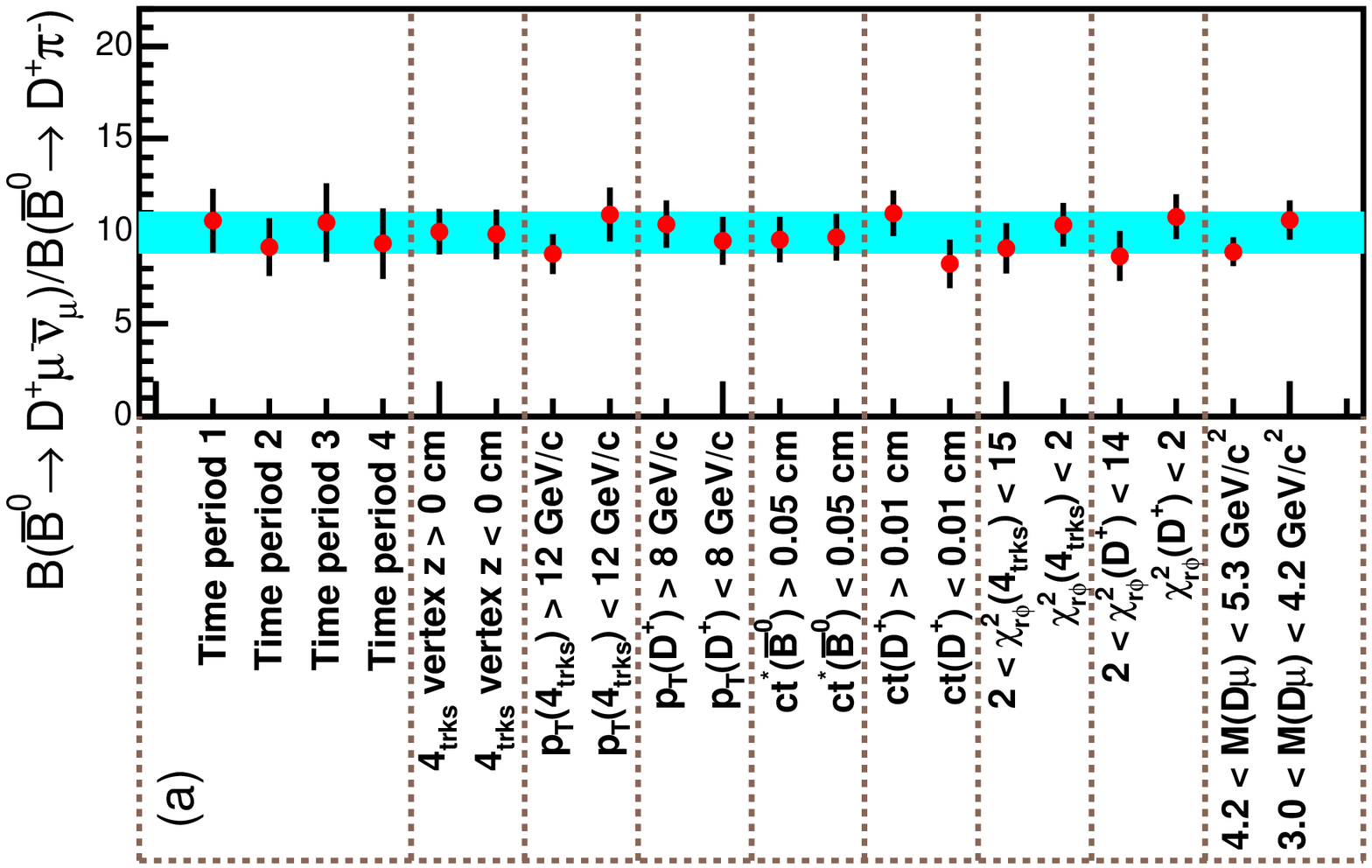} \\
 \includegraphics[width=300pt, angle=270]{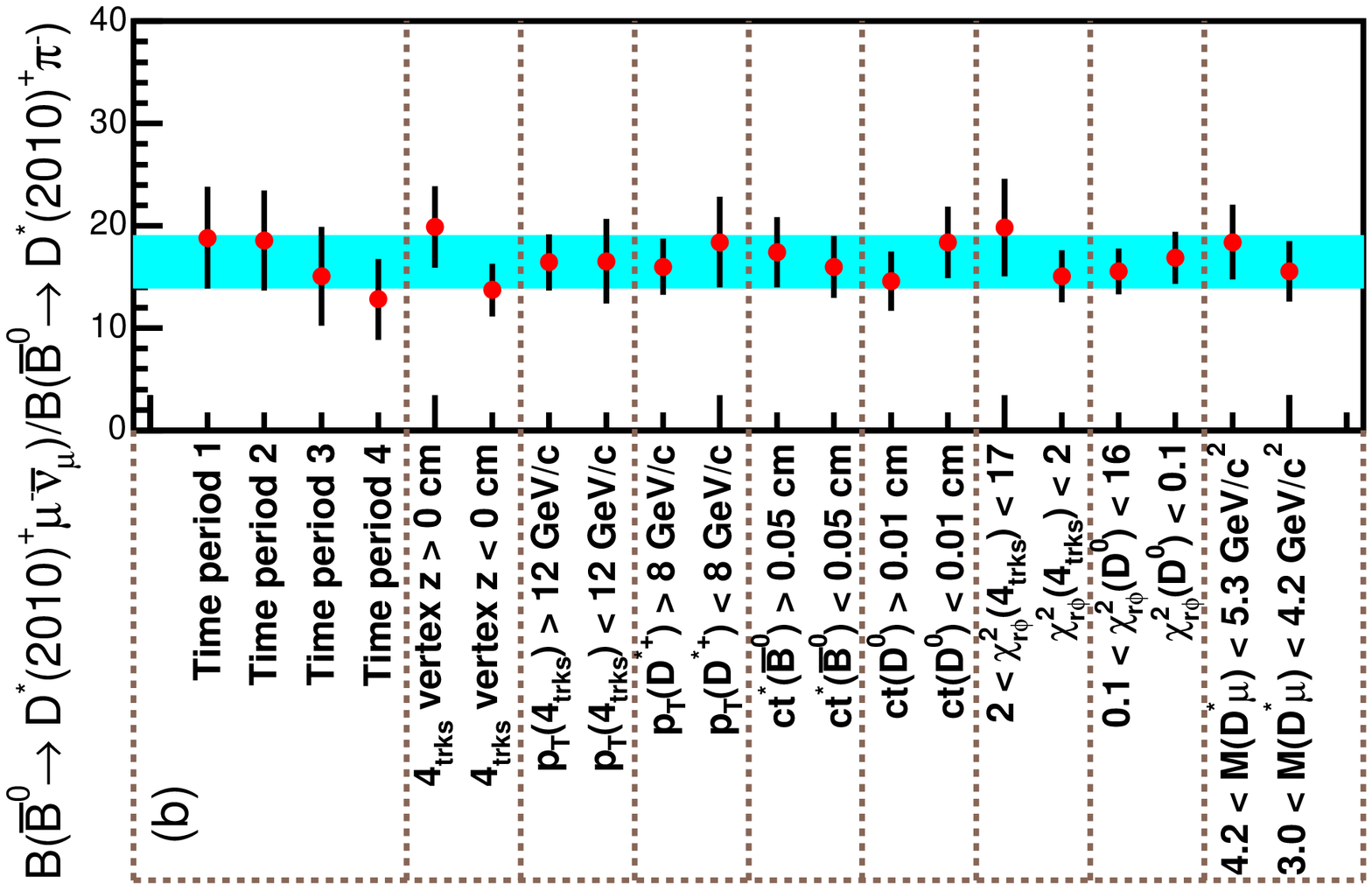} 
 \caption
 {Internal-consistency checks of the relative branching 
fractions measured in the two \Bd\ control samples: 
(a) $\Bd\rightarrow \dplus$ decays and (b) $\Bd\rightarrow \dstar$ decays.
   The uncertainty on each point is statistical only.  
   Each independent group is separated by a horizontal dashed line.
   The solid bands indicate the relative branching fractions with their 
   statistical uncertainties from the complete, undivided samples.	
\label{fig:xbrB0}}	
 \end{center}
 \end{figure}

\figc~\ref{fig:xbrB0} shows a summary of internal consistency checks. 
 The fractional systematic and statistical uncertainties on the \Bd\ relative 
branching fractions are summarized in Table~\ref{t:sysB0}. The leading sources
 of internal systematic uncertainties are the mass fitting models and the 
shape of the measured \Bd\ \pt\ spectrum. 
The dominant PDG uncertainties come from ${\cal B}\left(\dhad\right)$ for the $\dplus$ mode, and ${\cal B}\left(\bddonemunu\right)$ for the $\dstar$ mode. 
The dominant uncertainties on the estimated branching fractions come from 
${\cal B}\left(\bddpizeromunu\right)$
for the $\dplus$ mode and ${\cal B}\left(\bpdpronezeromunu\right)$ for the $\dstar$ mode. 
 The \Bd\ relative branching fractions with complete uncertainties 
are found to be: 
\begin{eqnarray*}
\lefteqn{\frac{{\cal B}\left(\dsemi\right)}{{\cal B}\left(\dhad\right)} = }\\
&& \rdc\ \pm \rde \left(\rm{stat}\right) \pm \rdsyste \left(\rm{syst}\right)
	         \pm \rdbre {\rm \left(PDG\right)} \pm \rdubre {\rm \left(EBR\right)},\\
\lefteqn{\frac{{\cal B}\left(\dstarsemi\right)}{{\cal B}\left(\dstarhad\right)}   = }\\	
&& \rdstarc\ \pm \rdstare \left(\rm{stat}\right) \pm \rdstarsyste \left(\rm{syst}\right)
	         \pm \rdstarbre {\rm \left(PDG\right)} \pm \rdstarubre {\rm \left(EBR\right)}.
\end{eqnarray*}
The uncertainties are from statistics (stat), internal systematics 
(syst), world averages of measurements published 
by the Particle Data Group or subsidiary measurements in this analysis
$\left(\rm{PDG}\right)$, and unmeasured branching 
fractions estimated from theory $\left(\rm{EBR}\right)$, respectively. 

      \begin{table*}[tbp]
       \caption{Summary of statistical and systematic uncertainties for the 
	\Bd\ modes. The $\sigma_R$ is uncertainty on the \Bd\ relative 
	branching fraction, $R$.
	\label{t:sysB0}}
	  \renewcommand{\arraystretch}{1.5}

 \begin{center}
 \begin{tabular}{lrr}
 \hline
 \hline
 & \multicolumn{2}{c}{$\frac{\sigma_R}{R}$ $\left(\%\right)$} \\ 
 Source & 
 $\frac{{\cal B}\left(\dsemi\right)}{{\cal B}\left(\dhad\right)}$ &
 $\frac{{\cal B}\left(\dstarsemi\right)}{{\cal B}\left(\dstarhad\right)}$ \\
 \hline
  Mass fitting                        & 4.1 & $<$ 0.1 \\   
  False $\mu$                         & 0.7 & 0.4 \\	     
  $b\bar{b}$/$c\bar{c}$ background    & 2.9 & 1.7 \\	     
  Simulation sample size              & 1.6 & 1.7 \\	     
  \Bd\ \pt\ spectrum                  & 3.0 &  2.4 \\	     
  XFT/CMU efficiency scale factor     & 0.5 & 0.4 \\	     
  detector material                   & 1.7 & 1.3 \\	     
 \cline{2-3}				      	     
  Sum of internal                     & 6.3 & 3.7 \\	     
 \hline					      	     
  PDG                                 & 4.1 & 2.8 \\	     
  Estimated branching fractions       & 4.7 & 4.9 \\	     
 \hline					      	     
  Statistical                         & 9.7 & 14.1 \\      
 	 \hline 
 	 \hline 
 	 \end{tabular}
 	 \end{center}

\end{table*}


\section{\label{sec:results}Results} 
\begin{figure}[tbp]
     \begin{center}
\includegraphics[width=200pt, angle=0]{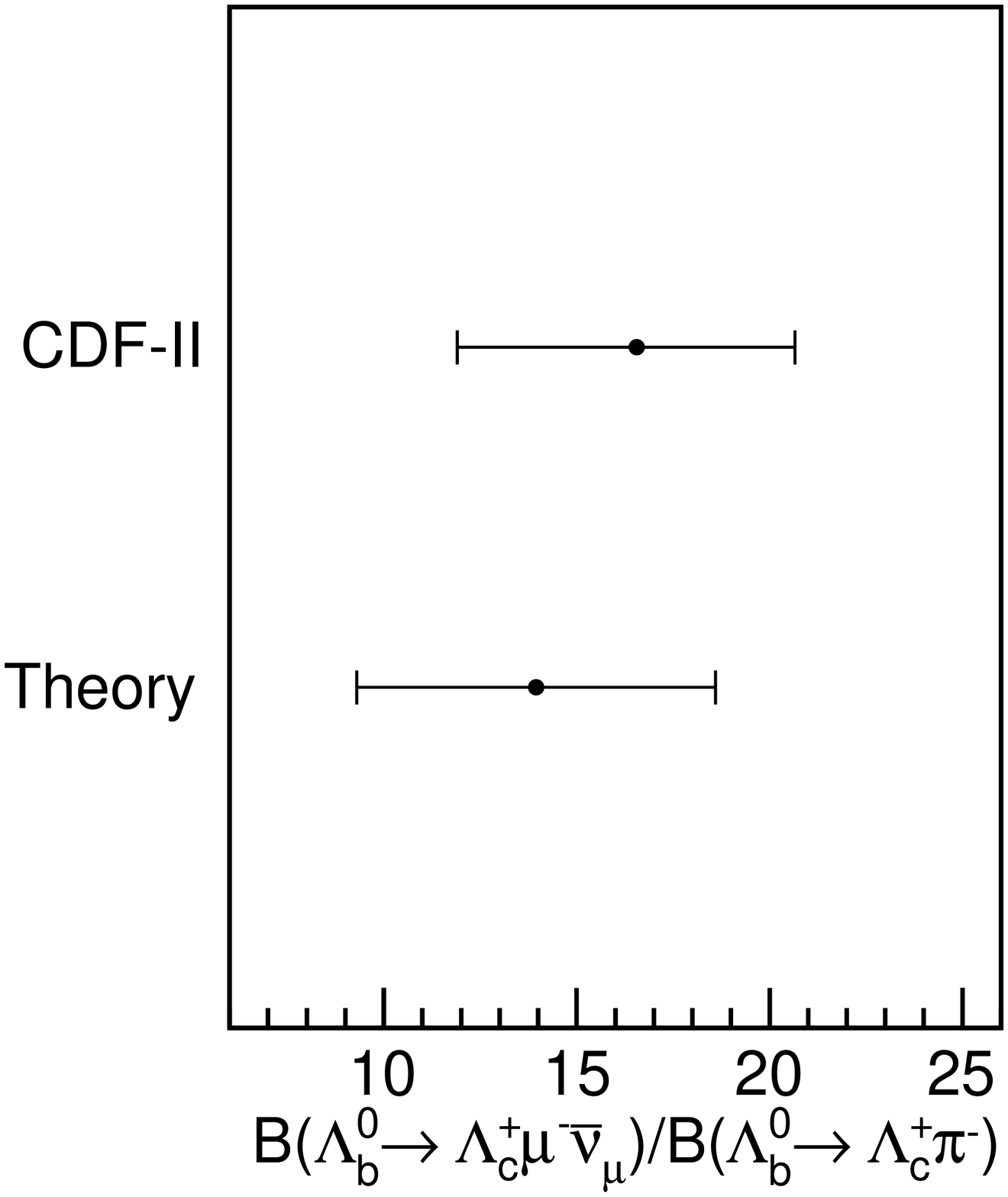}
     \end{center}
        \caption  		 
	{\label{fig:theoryexp}
	Comparison of the measured \({\cal B}\left(\lbsemi\right)/{\cal B}\left(\lbhad\right)\)  
	with the theoretical prediction based on HQET 
	by Huang~\etal~\cite{Huang:2005me} and 
	Leibovich~\etal~\cite{Leibovich:2003tw}.
        The measurement includes both the statistical 
	and the systematic uncertainties.
	} 	
  \end{figure}

The \Lb\ and \Bd\ relative branching fractions are measured to be:
\begin{eqnarray*}
 \lefteqn{\frac{{\cal B}\left(\lbsemi\right)}{{\cal B}\left(\lbhad\right)} =}\\
&& \rlbc\ \pm \rlbe \left(\rm{stat}\right) \pm \rlbsyste \left(\rm{syst}\right)
	         \rlbbre {\rm \left(PDG\right)} \pm \rlbubre {\rm \left(EBR\right)},
 \end{eqnarray*}
\begin{eqnarray*}
\lefteqn{\frac{{\cal B}\left(\dsemi\right)}{{\cal B}\left(\dhad\right)} =}\\
&& \rdc\ \pm \rde \left(\rm{stat}\right) \pm \rdsyste \left(\rm{syst}\right)
	         \pm \rdbre {\rm \left(PDG\right)} \pm \rdubre {\rm \left(EBR\right)},
 \end{eqnarray*}
\begin{eqnarray*}
\lefteqn{\frac{{\cal B}\left(\dstarsemi\right)}{{\cal B}\left(\dstarhad\right)} =}\\	
&& \rdstarc\ \pm \rdstare \left(\rm{stat}\right) \pm \rdstarsyste \left(\rm{syst}\right)
	         \pm \rdstarbre {\rm \left(PDG\right)} \pm \rdstarubre {\rm \left(EBR\right)}.
\end{eqnarray*}
The uncertainties are from statistics (stat), internal systematics 
(syst), world averages of measurements published 
by the Particle Data Group or subsidiary measurements in this analysis
$\left(\rm{PDG}\right)$, and unmeasured branching 
fractions estimated from theory $\left(\rm{EBR}\right)$, respectively. 
The control sample results are consistent with the ratios published by the 2008
 PDG~\cite{pdg} at the \agreebd$\sigma$ and \agreebdstar$\sigma$ level, respectively (see Table~\ref{t:b0pdg}). 
The measured ratio of \Lb\ branching fractions is compared with 
the predicted value based on HQET. 
The prediction has a $\approx30\%$ uncertainty and is obtained by combining 
the results of Huang~\etal~\cite{Huang:2005me} and 
Leibovich~\etal~\cite{Leibovich:2003tw}\cite{whynotcheng}. 
\figc~\ref{fig:theoryexp} shows the consistency between this measurement 
and the theoretical prediction. 

 The branching fractions of the four new \Lb\ semileptonic decays relative to 
that of \lbsemi\ are measured to be:
\begin{eqnarray*}
\lefteqn{\frac{{\cal B}\left(\dmitridecayone\right)}{{\cal B}\left(\lbsemi\right)}= }\\
&& \dmitrione\ \pm \dmitrionestat \left(\rm{stat}\right) \dmitrionesyst \left(\rm{syst}\right), \\
\lefteqn{\frac{{\cal B}\left(\dmitridecaytwo\right)}{{\cal B}\left(\lbsemi\right)} =}\\
 && \dmitritwo\ \pm \dmitritwostat \left(\rm{stat}\right) \dmitritwosyst \left(\rm{syst}\right),
\end{eqnarray*}
\begin{eqnarray*}
 \lefteqn{\frac{1}{2}\left[\frac{{\cal B}\left(\dmitridecaythree\right)}{{\cal B}\left(\lbsemi\right)}\right.}
\\
& + &
\left. \frac{{\cal B}\left(\dmitridecayfour\right)}{{\cal B}\left(\lbsemi\right)} \right]\\
& = &\dmitrithree\ \pm \dmitrithreestat\left(\rm{stat}\right) \dmitrithreesyst \left(\rm{syst}\right).\
\end{eqnarray*}

%

     \begin{table*}[tbp]
       \caption{The \Bd\ relative branching fractions 
	measured in this analysis and those published in 
	the 2008 PDG~\cite{pdg}. 	
	The measurements of this analysis include both the statistical 
	and the systematic uncertainties.
	\label{t:b0pdg}}
	\renewcommand{\arraystretch}{1.5}
       \begin{center}
	\begin{tabular}{cc@{\,$\pm$\,}cc@{\,$\pm$\,}l}
        \hline
	 \hline
	Mode & \multicolumn{2}{c}{PDG} & \multicolumn{2}{c}{This Analysis} \\
        \hline
	$\frac{{\cal B}\left(\dsemi\right)}{{\cal B}\left(\dhad\right)}$ 
	&  \rdpdgc & \rdpdge &  \rdc &  \rdcombinee \\
	$\frac{{\cal B}\left(\dstarsemi\right)}{{\cal B}\left(\dstarhad\right)}$ 
	&  \rdstarpdgc & \rdstarpdge &  \rdstarc & \rdstarcombinee \\
        \hline 
	   \hline
       \end{tabular}
        \end{center}
        \end{table*}

\section{\label{sec:conclusion}Conclusion} 
Using data from an integrated luminosity of $\approx 172~\rm{pb}^{-1}$ 
collected with the CDF II detector, 
 \nlbsemi\ \inclbsemi\ and \nlbhad\ \lbhad\ signal events are reconstructed. 
The large \Lb\ sample enables the measurement of ${\cal B}\left(\lbsemi\right)/{\cal B}\left(\lbhad\right)$ 
and the comparison to the predictions of heavy quark effective theory. 
The uncertainty is dominated by the size of the data sample, the world average 
of ${\cal B}\left(\Lc \rightarrow pK^-\pi^+\right)$, and the CDF~I 
measurement of \rxsec. 
Ratios for the control modes, ${\cal B}\left(\dsemi\right)/{\cal B}\left(\dhad\right)$ 
and ${\cal B}\left(\dstarsemi\right)/{\cal B}\left(\dstarhad\right)$, are found to be in good agreement with the world averages~\cite{pdg}. 
For the first time, 
the semileptonic decay \dmitridecaytwo\ has been observed and three 
other semileptonic decays, \dmitridecayone, \dmitridecaythree, and 
\dmitridecayfour, have been reconstructed, using data from an integrated luminosity of $\approx 360~\mathrm{pb^{-1}}$. Measurements of the ratios
 of their branching fractions to the branching fraction of \lbsemi\ have been 
performed. Finally, the transverse-momentum distribution of the \Lb\ baryon 
produced in $p\bar{p}$ collisions is found to be softer (more 
$b$ hadrons at lower \pt) than that of the $\Bd$ meson; this results in a 
new estimate for ${\cal B}\left(\lbhad\right)$ in better agreement with the 
theory than the PDG evaluation.


\section*{Acknowledgments} 
We thank the Fermilab staff and the technical staffs of the participating institutions for their vital contributions. This work was supported by the U.S. Department of Energy and National Science Foundation; the Italian Istituto Nazionale di Fisica Nucleare; the Ministry of Education, Culture, Sports, Science and Technology of Japan; the Natural Sciences and Engineering Research Council of Canada; the National Science Council of the Republic of China; the Swiss National Science Foundation; the A.P. Sloan Foundation; the Bundesministerium f\"ur Bildung und Forschung, Germany; the Korean Science and Engineering Foundation and the Korean Research Foundation; the Science and Technology Facilities Council and the Royal Society, UK; the Institut National de Physique Nucleaire et Physique des Particules/CNRS; the Russian Foundation for Basic Research; the Ministerio de Ciencia e Innovaci\'{o}n, and Programa Consolider-Ingenio 2010, Spain; the Slovak R\&D Agency; and the Academy of Finland. 



\appendix
\section{Semileptonic Decays of $b$ Mesons to Baryons \label{sec:mesonbaryon}}
The number of feed-in events from semileptonic decays of $b$ mesons to 
baryons 
$\left(\nphys^\mathrm{meson}\right)$ is also normalized to the \lbhad\ yield 
in data $\left(\nhad\right)$ and has an 
expression similar to that of \eq~(\ref{eq:bgexample}):
\begin{eqnarray}
\label{eq:baryonic}
\lefteqn{\frac{\nphys^\mathrm{meson}}{\nhad} =} \nonumber \\
 && \frac{\sigma_{B_{u,d,s}}\left(\pt>6.0\right)}{\sigma_{\Lb}\left(\pt>6.0\right)} 
\frac{ \sum_{i} {\cal B}_{i}\eff_{i}}
     { {\cal B}\left(\lbhad\right)\eff_{\lbhad}}, 
\end{eqnarray}
where $\sigma_{B_{u,d,s}}\left(\pt>6.0\right)$ and $\sigma_{\Lb}\left(\pt>6.0\right)$ are the production 
cross-sections of $b$ mesons and \Lb\ baryons for \pt\ greater than 6~\gevc.

A list of $b$-meson decays that may contribute to the 
\inclbsemi\ sample is obtained from an inclusive sample of $b$-meson 
semi-muonic decays generated using \pythia. After applying the trigger and 
analysis requirements to the \pythia\ generated events, the maximum 
contributing decays are found to be 
$\bar{B}^0 \rightarrow \Lc \bar{n} \mu^- \bar{\nu}_{\mu}$ and
$B^- \rightarrow \Lc \bar{p} \mu^- \bar{\nu}_{\mu}$. 
While there are measurements of branching ratios of the $b$-meson hadronic 
decays to baryons, \eg, $\bar{B}^0 \rightarrow \Lc \bar{p} \pi^+ \pi^-$,
there is only an upper limit for the semileptonic decay of $B^-$:
\begin{eqnarray*}
 {\cal B}\left(B^- \rightarrow \Lc \bar{p} e^- \bar{\nu}_{e}\right) < 0.15 \%.
\end{eqnarray*}
Assuming the branching fractions of the muon-neutron and muon-proton final 
states are the same as that of the proton-electron final state, 
the value of this upper limit is then taken for the branching fraction of the 
 $\bar{B}^0 \rightarrow \Lc \bar{n} \mu^- \bar{\nu}_{\mu}$ and 
the $B^- \rightarrow \Lc \bar{p} \mu^- \bar{\nu}_{\mu}$ decays.
The ratio $N_{\bar{B}^0 \rightarrow \Lc \bar{n} \mu^- \bar{\nu}_{\mu}}/
N_{\lbhad}$, for example, is then given by: 
 \begin{eqnarray}
 \label{eq:bmesonlc}
\lefteqn{\frac{N_{\bar{B}^0 \rightarrow \Lc \bar{n} \mu^- \bar{\nu}_{\mu}}}
{N_{\lbhad}}} \nonumber \\
 & = &
 \frac{\sigma_{\Bd}\left(\pt>6.0\right)}{\sigma_{\Lb}\left(\pt>6.0\right)} 
  \frac{{\cal B}\left(\bar{B}^0 \rightarrow \Lc \bar{n} \mu^- \bar{\nu}_{\mu}\right) \eff_{\bar{B}^0 \rightarrow \Lc \bar{n} \mu^- \bar{\nu}_{\mu}}}
     { {\cal B}\left(\lbhad\right)\eff_{\lbhad}}   \nonumber \\
  & = &
  \frac{1}{G}
 \times
  \frac{0.15\%~~    
  \eff_{\bar{B}^0 \rightarrow \Lc \bar{n} \mu^- \bar{\nu}_{\mu}}}{{\cal B}\left(\dhad\right)  \eff_{\lbhad}},
 \end{eqnarray}
where $G$ is the CDF measurement~\cite{yile:lblcpi}:
\begin{eqnarray}
 \label{eq:yileeq}
 G\equiv\yile. 
\end{eqnarray}
The ratio $N_{B^- \rightarrow \Lc \bar{p} \mu^- \bar{\nu}_{\mu}}/N_{\lbhad}$ follows \eq~(\ref{eq:bmesonlc}), assuming the production fractions are 
the same for the \Bd\ and $B^-$ mesons~\cite{support}. 
Table~\ref{t:physicslc2} lists the estimated size of the feed-in background 
contribution from semileptonic decays of $b$ mesons to baryons.

\renewcommand{\arraystretch}{1.4}
\renewcommand{\tabcolsep}{0.04in}
\begin{normalsize}
 \begin{table*}[tbp]
 \caption{Feed-in backgrounds to \lbsemi\ from $b$ mesons. All the numbers in 
        parentheses are estimated uncertainties. 
	The definition of quantities listed in each column follows 
	Table~\ref{t:physicslc}. 
	}
 \label{t:physicslc2}

	 \begin{center}
	 \begin{tabular}{lr@{\,$\pm$\,}lcccr@{\,$\pm$\,}r}
	 \hline
	 \hline
	 \multicolumn{1}{c}{Mode}
	 & \multicolumn{2}{c}{${\cal B}$ $\left(\%\right)$}
	 & \multicolumn{1}{c}{$\frac{\eff_i}{\eff_{\lbhad}}$}
	 & \multicolumn{1}{c}{$\left(\frac{\nphys^\mathrm{meson}}{\nhad}\right)^i$}
	 & \multicolumn{1}{c}{$\left(\frac{\nphys^\mathrm{meson}}{\nsemi}\right)^i$}
	 & \multicolumn{2}{c}{$N_{event}$}\\
	 \hline
		 \lbhad 
		 & \multicolumn{2}{c}{\brlbhadc \brlbhadecombine} 
	         &   \multicolumn{1}{l}{1.000} & -- & --
		 &   \nlbhadc  &  \nlbhade
		 \\
		 \inclbsemi\
		 &  \multicolumn{2}{r}{--}
		 &  -- & -- & --
		 &   \nlbsemic &  \nlbsemie \\ 
	 \hline
		  \lbsemi\
		 &   \brlbsemic & (\brlbsemie)
		 & $\lbeffratioc \pm \lbeffratioe$
		 & 6.118
		 & 1.000
 & \multicolumn{2}{c}{--} \\
	 \hline
		 $B^- \rightarrow \Lc \bar{p} \mu^- \bar{\nu}_{\mu}$
		 &   0.15 & $\left(0.15\right)$
		 & $0.035 \pm 0.002$ 
		 & 0.024
		 & 0.004
		 & 4.3 & 0.5\\

		 $\bar{B}^0 \rightarrow \Lc \bar{n} \mu^- \bar{\nu}_{\mu}$
		 &   0.15 & $\left(0.15\right)$
		 &  $0.037 \pm 0.002$ 
		 & 0.025
		 & 0.004
		 & 4.5 & 0.5\\
		 \hline 
		 \hline 
		 \end{tabular}
		 \end{center}

	\end{table*}
\end{normalsize}

\section{Semileptonic Decays of Other $b$ Baryons \label{sec:omegab}}
In addition to the feed-in backgrounds from the semileptonic decays of $b$ 
mesons to baryons, contributions are also expected from the semileptonic
decays of other $b$ baryons. Until recently~\cite{:2007rw,:2007un}, the \Lb\ 
was the only $b$ baryon which had been observed unambiguously. Therefore, in 
order to estimate the number of feed-in background events, the production cross
 section of the other $b$ baryons and the branching ratio of the feed-in 
channel must be estimated. The first step in the estimation is to identify 
possible contributions to the feed-in.

Of the lowest lying $b$ baryons, the members of $\Sigma_b$ triplet are expected
 to decay to $\Lb\pi$ via the strong interaction and contribute to the \Lb\ 
signal. 
This leaves $\Xi_b^-$, $\Xi_b^0$, and $\Omega_b^-$ and they are expected 
to decay predominantly to $\Xi_c^+$ and $\Omega_c^0$. However, by vacuum 
production of one or more $q\bar{q}$ pairs, these $b$ baryons can decay 
into a \Lc, specifically,
\begin{center}
\begin{tabular}{l}
 $\Xi_b^0 \rightarrow \Lc \bar{K}^0 \mu^- \bar{\nu}_{\mu}$, \\
 $\Xi_b^- \rightarrow \Lc K^- \mu^- \bar{\nu}_{\mu}$, \\
 $\Omega_b^- \rightarrow \Lc K^-\bar{K}^0 \mu^- \bar{\nu}_{\mu}$. \\
\end{tabular}
\end{center}
Since the $\Xi_b^0$ decays have a decay topology similar to 
\dmitridecaythreefour\ and the $\Omega_b^-$ decay topology is similar to that 
of \lblcpim, a branching fraction of \brsigcc$\%$, \brsigcc$\%$, and 
\brlcpipi$\%$ (given in Table~\ref{t:brlc}) are assigned to these three decays,
 respectively.  CDF reports observing 17 $\Xi_b^- \rightarrow J/\psi \Xi^-$
events in data from an integrated luminosity of approximately 
$1900\;{\rm pb^{-1}}$~\cite{:2007un}. 
Assuming that the branching fraction for $\Xi_b^- \rightarrow J/\psi \Xi^-$ is 
similar to that reported by the Particle Data Group for 
$\Lb \rightarrow J/\psi \Lambda$~\cite{pdg}, using a generator-level 
simulation to estimate the ratio of acceptances of 
$\Xi_b \rightarrow \Lc K \mu^- \bar{\nu}_{\mu}$ relative to that of 
$\Xi_b^- \rightarrow J/\psi \Xi^-$ (0.2), and scaling by the 
ratio of luminosities for this analysis and the $\Xi_b^-$ analysis (172/1900), 
each of the $\Xi_b$ decays is found to contribute approximately \fracxib\%\ to
 the signal.  For the $\Omega_b^-$ decays, a similar calculation is performed
assuming the same production rate as for the $\Xi_b$.  However, because of 
the larger number of particles in the $\Omega_b^-$ decay, the acceptance 
for the $\Omega_b^-$ is an order of magnitude smaller than that of the 
$\Xi_b$.  For the $\Omega_b^-$, the contribution is \fracomegab\%.  These 
three decays are found to contribute $\le\;1\%$ to the signal and may be 
ignored~\cite{Dzero}.

\section{\label{sec:bbcc}The $b\bar{b}$/$c\bar{c}$ Background}
\begin{figure*}[tbp]
  \begin{center}
 \resizebox{350pt}{!}{\includegraphics*[46pt,370pt][550pt,680pt]
{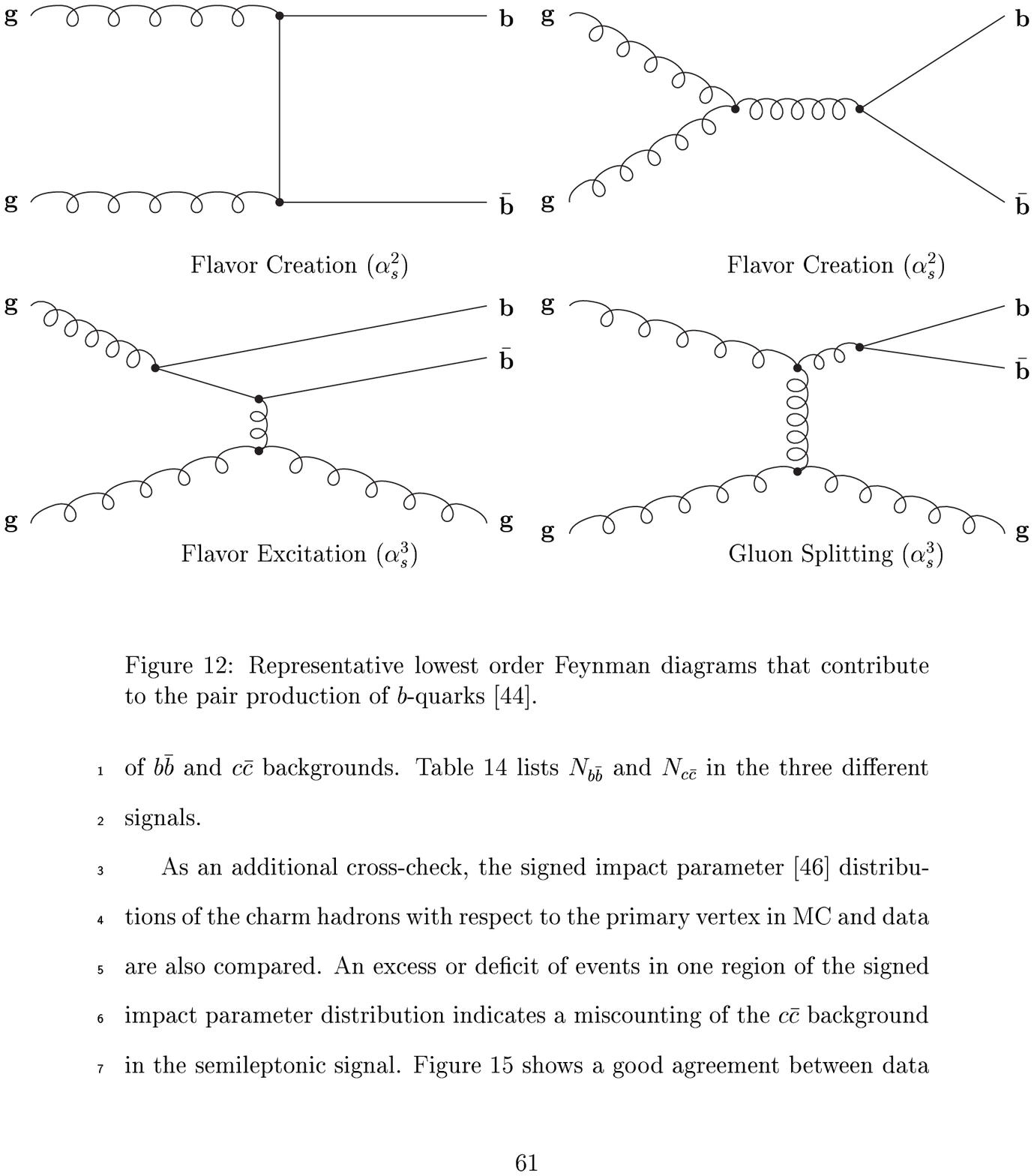}}
  \end{center}
    \caption{Representative lowest-order Feynman diagrams that contribute to 
	the pair production of $b$ quarks~\cite{Mangano:1991jk,Halzen:1982iq}. 
	\label{fig:cbproc}}
  \end{figure*}

\begin{figure*}[tbp]
  \begin{center}
\includegraphics[width=180pt, angle=0]
	{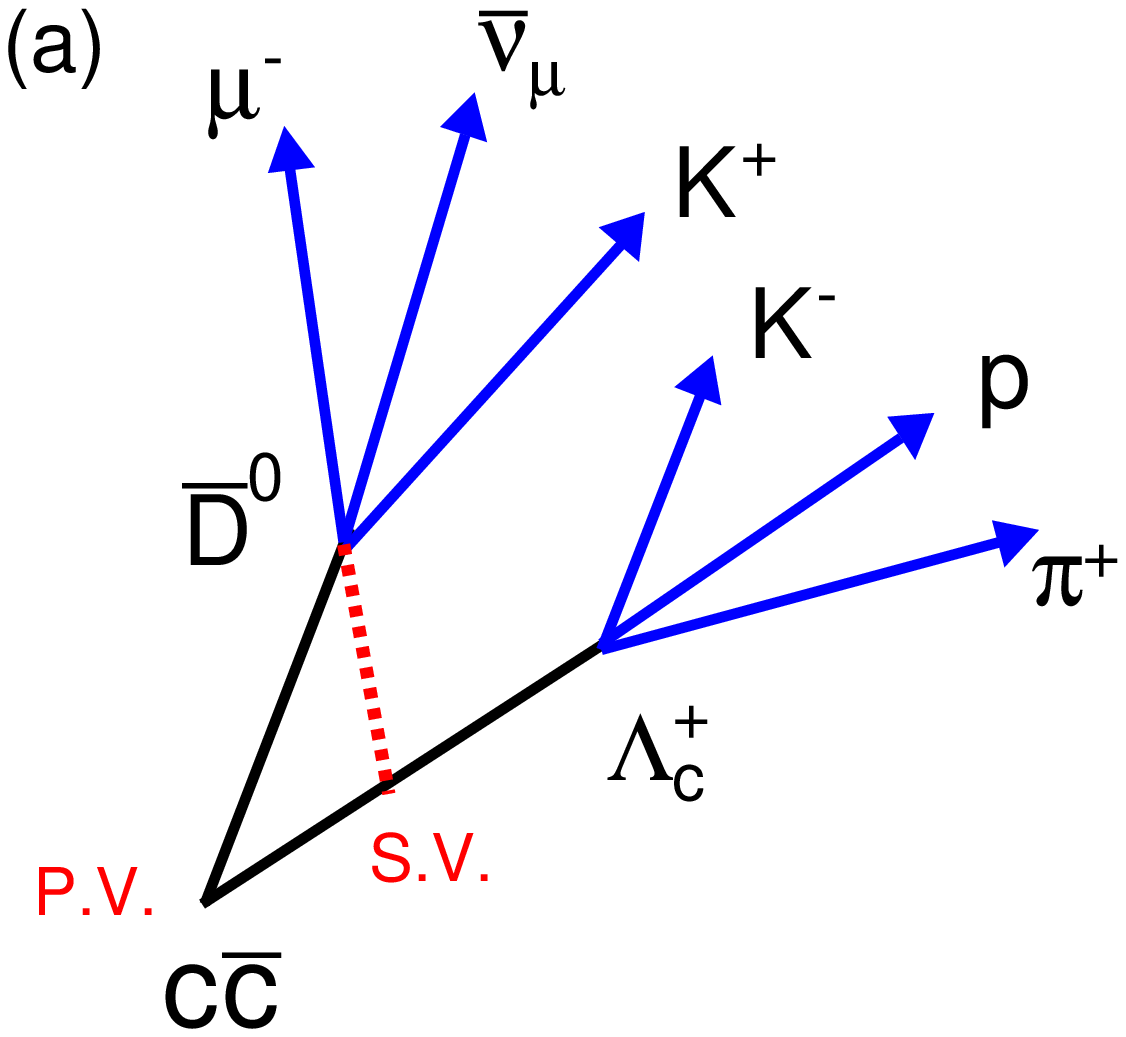}
\includegraphics[width=250pt, angle=0]
	{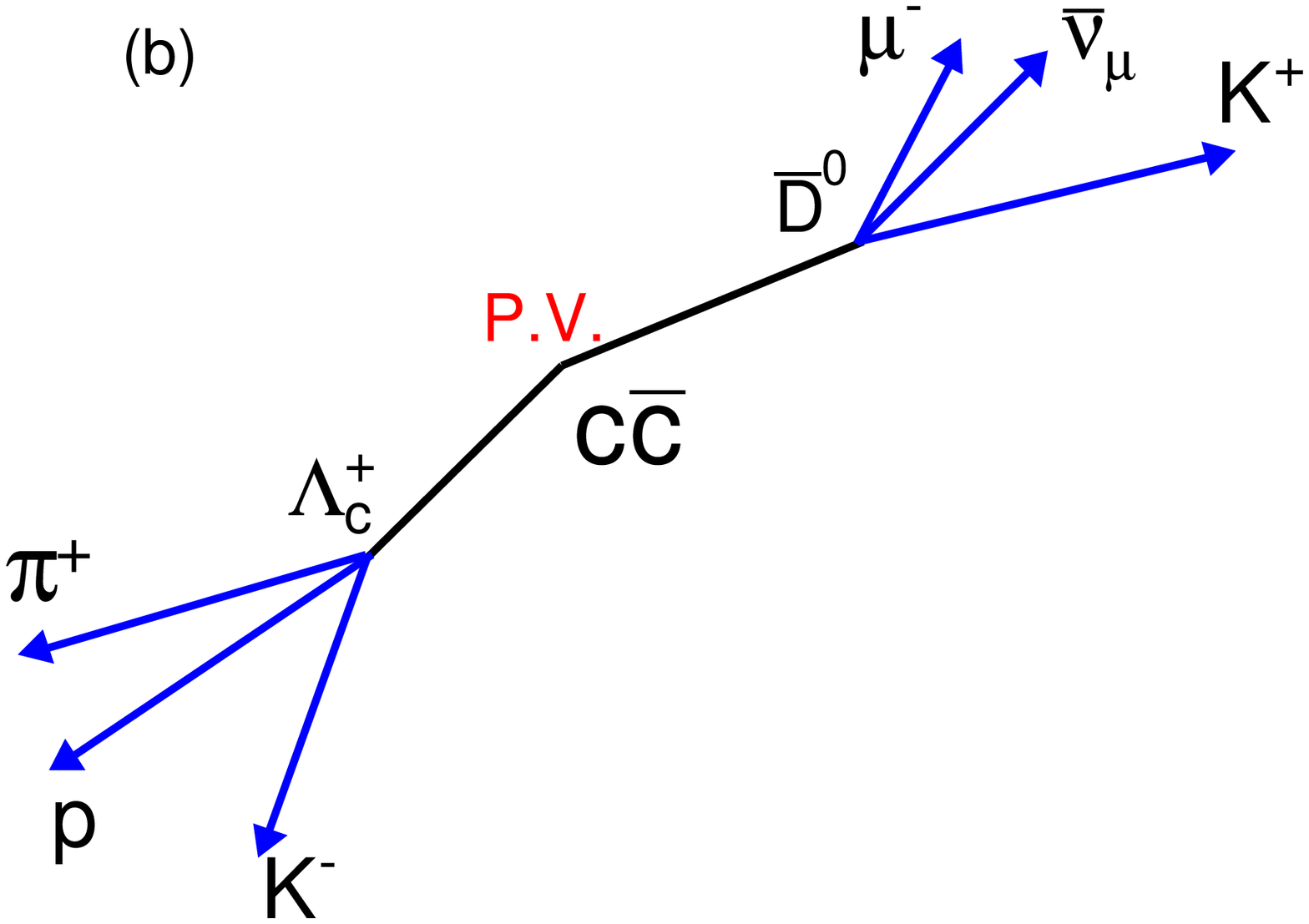}
 \caption{The $c$ hadrons from $c\bar{c}$ with (a) small and (b) large 
	$\Delta \phi$. Unlike \fig~\ref{fig:cbexample}(b), 
	\fig~\ref{fig:cbexample}(a) shows that the \Lc\ and 
	the muon from the semileptonic decay of $\bar{D}^0$ form 
	a secondary vertex and are misidentified as the \lbsemi\ signal.}
 \label{fig:cbexample} 
 \end{center}
 \end{figure*}

The $b\bar{b}$/$c\bar{c}$ background refers to the pairing of a \Lc\ and a 
real muon from the decays of two different heavy-flavor hadrons produced by 
the fragmentation of $b\bar{b}$ or $c\bar{c}$ pairs. 
In $p\bar{p}$ collisions, the $b$ and $c$ quarks are primarily pair-produced 
via the strong interaction; the single-quark production cross-section via the 
electroweak process, $p\bar{p}\rightarrow W^+ \mathrm{anything}
\rightarrow \bar{b}~\mathrm{anything}$ or $c~\mathrm{anything}$, is 
more than 20,000 times smaller~\cite{Acosta:2004uq,wxsec}. 
\figc~\ref{fig:cbproc} shows the Feynman diagrams up to $\alpha_s^3$ for 
the three processes which contribute to the $b\bar{b}$/$c\bar{c}$ 
production~\cite{Mangano:1991jk,Halzen:1982iq}: flavor creation, flavor 
excitation, and gluon splitting. Flavor creation, 
referring to gluon fusion and quark anti-quark annihilation, tends to produce 
$b\bar{b}$/$c\bar{c}$ pairs with an azimuthal angle distribution 
$\left(\Delta \phi\right)$ between the two quarks which peaks at $180^{\circ}$. 
In contrast, the $\Delta \phi$ distribution is more evenly distributed for the 
flavor excitation and the low-momentum gluon splitting and peaks at small 
angles for the 
high-momentum gluon splitting. When $\Delta \phi$ is small, daughters of the 
two heavy-flavor hadrons from the fragmentation of $b\bar{b}$/$c\bar{c}$ may 
appear to come from the same decay vertex, as shown in \fig~\ref{fig:cbexample}. If one hadron decays semileptonically, and the other hadron decays into a 
final state including a $\Lc \rightarrow pK^-\pi^+$ decay, the 
muon from the semileptonic decay together with the \Lc\ may be misidentified 
as the exclusive semileptonic signal, \lbsemi. An estimate using \pythia\ has 
shown that this measurement is most sensitive to the $b\bar{b}$/$c\bar{c}$ 
background from high-momentum gluon splitting. 

In the following, the determination of the $b\bar{b}$ background contribution 
is described. The same procedure is followed for the $c\bar{c}$ background. 
The ratio $N_{b\bar{b}}/\nhad$ is given by:
\begin{eqnarray}
\label{eq:bbcc1}
\lefteqn{\frac{N_{b\bar{b}}}{\nhad} = } \nonumber \\
 & & \frac{ \sigma_{b\bar{b}} {\cal P}\left(b\rightarrow \Lc X\right)
{\cal P}\left(\bar{b}\rightarrow \mu^- X\right)  \eff_{b\bar{b}\rightarrow \Lc \mu^- X}} 
 {\sigma_{\Lb}{\cal B}\left(\lbhad\right)\eff_{\lbhad}}.
\end{eqnarray}
The $\sigma_{b\bar{b}}$ is the production cross-section of $b\bar{b}$ pairs; 
${\cal P}\left(b\rightarrow \Lc X\right)$ and ${\cal P}\left(\bar{b}\rightarrow \mu^- X\right)$ are 
the probabilities for a $b$ and a $\bar{b}$ quark to fragment into a $b$ 
hadron and a $\bar{b}$ hadron and to decay to a final state including 
a \Lc\ and a $\mu^-$, respectively. 
The $\eff_{b\bar{b}\rightarrow \Lc\mu^- X}$ is the 
acceptance times efficiency for reconstructing the background as the \lbsemi\ 
signal. The denominator of \eq~(\ref{eq:bbcc1}) can be re-written using 
the CDF measurement~\cite{yile:lblcpi} defined in \eq~(\ref{eq:yileeq}), 
 the CDF measurement of $\sigma_{B^+}$~\cite{Abulencia:2006ps} assuming $\sigma_{B^+}=\sigma_{\Bd}$~\cite{support}, 
and the world average of ${\cal B}\left(\dhad\right)$:
\begin{eqnarray}
\label{eq:bbcc2}
\lefteqn{\frac{N_{b\bar{b}}}{\nhad} = \frac{1}{G}} \nonumber \\
&& \times\frac{ \sigma_{b\bar{b}}{\cal P}\left(b\rightarrow \Lc X\right)
{\cal P}\left(\bar{b}\rightarrow \mu^- X\right) \eff_{b\bar{b}\rightarrow \Lc\mu^- X}} 
 {\sigma_{\Bd}{\cal B}\left(\dhad\right)\eff_{\lbhad}}.
\end{eqnarray}
%
%
The $\eff_{\lbhad}$ is determined from a signal simulation generated with 
the \bgen\ program~\cite{whygen,restrict} and 
Table~\ref{t:cbhad} lists the parameters for calculating the denominator of 
\eq~(\ref{eq:bbcc2}). 
In order to determine the numerator of \eq~(\ref{eq:bbcc2}), 
inclusive $b\bar{b}$ events are first generated with \pythia. 
The \pt\ of the hard scattering, i.e. the part of the interaction with the 
largest momentum scale, is required to be greater than 5 \gevc. At least one 
$b$ quark must have a \pt\ greater than 4 \gevc\ and $\left|\eta\right|$ less 
than 1.5. 
The value of $\sigma_{b\bar{b}}$, after applying the kinematic 
requirements above, is obtained from \pythia, 
since the status of the $\sigma_{b\bar{b}}$ measurements at the Tevatron 
is still inconclusive~\cite{Happacher:2005gx,Tevatron_bb,giromini}. 
Then, the gluon-splitting events are filtered and the decays are simulated with
 \evtgen. Only events with a $\mu^-$ and a \Lc\ which pass the 
generator-level trigger and analysis requirements are considered further. 
Ancestors of the $\mu^-$ and the \Lc\ determine whether they originate from 
$b\bar{b}$ pairs or single $b$ hadrons, and are retrieved by tracing the 
information from the generator. The number of events satisfying these 
criteria divided by 
the number of generated events gives the product 
${\cal P}\left(b\rightarrow \Lc X\right){\cal P}\left(\bar{b}\rightarrow \mu^- X\right)  \eff_{b\bar{b}\rightarrow \Lc\mu^- X}$.
Table~\ref{t:cbbb} lists the parameters for the determination of the numerator 
of \eq~(\ref{eq:bbcc2}). 

Table~\ref{t:cbfinal} 
lists the estimated ratios, $N_{b\bar{b}}/\nhad$ and $N_{c\bar{c}}/\nhad$, 
based on the values in Tables~\ref{t:cbhad}--\ref{t:cbbb}. 
The \nbbcc\ is found to be only \fracbbcc\% of the number of inclusive \inclbsemi\ 
events.
%
%
%
The production of $b\bar{b}$ and $c\bar{c}$ pairs in $p\bar{p}$ collisions 
has not yet been completely 
understood~\cite{Acosta:2004nj,Happacher:2005gx,Tevatron_bb,giromini}. 
In order to understand how well \pythia\ predicts $\sigma_{b\bar{b}}$ and 
$\sigma_{c\bar{c}}$, an indirect cross-check was performed by comparing the 
differential cross-sections of inclusive $b$ hadrons, $B^+$, and $\dzero$ 
in \pythia\ with the 
CDF measurements~\cite{CDF,Abulencia:2006ps,cchen:dxec} 
(see Appendix~\ref{sec:forbbcc1}). The discrepancy between \pythia\
 and the data cross-sections is generally within $10\%$ for $c$ 
hadrons and $50\%$ for $b$ hadrons, which will be included in the systematic 
uncertainty in Section~\ref{sec:systematics}. Another cross-check using 
the signed impact parameter distributions of the \Lc\ baryons 
(see Appendix~\ref{sec:forbbcc2}) 
indicates a negligible contribution of promptly produced \Lc\ from $c\bar{c}$, 
which is consistent with the above estimate using \pythia.

\renewcommand{\arraystretch}{1.5}
  \begin{table*}[tbp]
   \caption{Parameters used to calculate the denominator of 
	\eq~(\ref{eq:bbcc2}).}
  \label{t:cbhad}
   \begin{center}
   \begin{normalsize}
 \renewcommand{\doublerulesep}{0.03in}
  \begin{tabular}{lr} 
   \hline
   \hline
    CDF $\sigma_{B^+}$ ($\mu$b) & $\bpxsecc \pm \bpxsece$ \\
    \yile\ $\left(G\right)$ & $0.82 \pm 0.26$ \\    
    ${\cal B}\left(\dhad\right)$ &  $(\brdhadc \pm \brdhade)\%$ \\
    $\eff_{\lbhad}$ & $(2.109 \pm 0.002) \times 10^{-2}$  \\
    \hline
    $\sigma_{\Lb}{\cal B}\left(\lbhad\right)\eff_{\lbhad}$ 
	(10$^{-4}$ $\mu$b)     
    & $\lbhadbbccc \pm \lbhadbbcce$ \\
   \hline
   \hline
 \end{tabular}
 \end{normalsize}
 \end{center}
 \end{table*}

\renewcommand{\arraystretch}{1.3}
  \begin{table*}[tbp]
   \caption
	{Parameters used to determine the numerator of \eq~(\ref{eq:bbcc2}). 
	The uncertainties are statistical only.}
  \label{t:cbbb}
   \begin{center}
   \begin{normalsize}
  \begin{tabular}{lrrr} 
   \hline
   \hline
    \pythia\ $\sigma_{b\bar{b}}$ ($\mu$b) & 49.6 \\
${\cal P}\left(b\rightarrow \Lc X\right)
{\cal P}\left(\bar{b}\rightarrow \mu^- X\right)  \eff_{b\bar{b}\rightarrow \Lc\mu^- X}$
    &  $(4.1 \pm 1.4)\times 10^{-8}$  \\
  \pythia\ $\sigma_{c\bar{c}}$ ($\mu$b) & 198.4  \\
${\cal P}\left(c\rightarrow \Lc X\right)
{\cal P}\left(\bar{c}\rightarrow \mu^- X\right)  \eff_{c\bar{c}\rightarrow \Lc\mu^- X}$
     & $(1.2 \pm 0.5)\times 10^{-9}$  \\
   \hline
   \hline
 \end{tabular}
 \end{normalsize}
 \end{center}
 \end{table*}

\renewcommand{\arraystretch}{1.2}
  \begin{table}[tbp]
   \caption{
    The estimated size of the $b\bar{b}$ and $c\bar{c}$ background 
    contribution to the \lbsemi\ signal and the observed yields in data.}
   \label{t:cbfinal}
   \begin{center}
   \begin{normalsize}
   \begin{tabular}{lr@{\,$\pm$\,}l}
   \hline
   \hline
   $N_{b\bar{b}}/\nhad$ & \bbrc & \bbre \\
   $N_{c\bar{c}}/\nhad$ & \ccrc & \ccre \\
   $\nhad$ 
   & \nlbhadc & \nlbhade \\ 
   $N_\mathrm{inc \; semi}$ 
   & \nlbsemic & \nlbsemie \\ 
  \hline
   \hline
 \end{tabular}
 \end{normalsize}
 \end{center}
 \end{table}

\subsection{Cross-check of the inclusive $b$ hadron, $B^+$, and $\dzero$ cross-sections 
	\label{sec:forbbcc1}}
\begin{figure}[tbp]
  \begin{center}
	\includegraphics[width=160pt, angle=0]{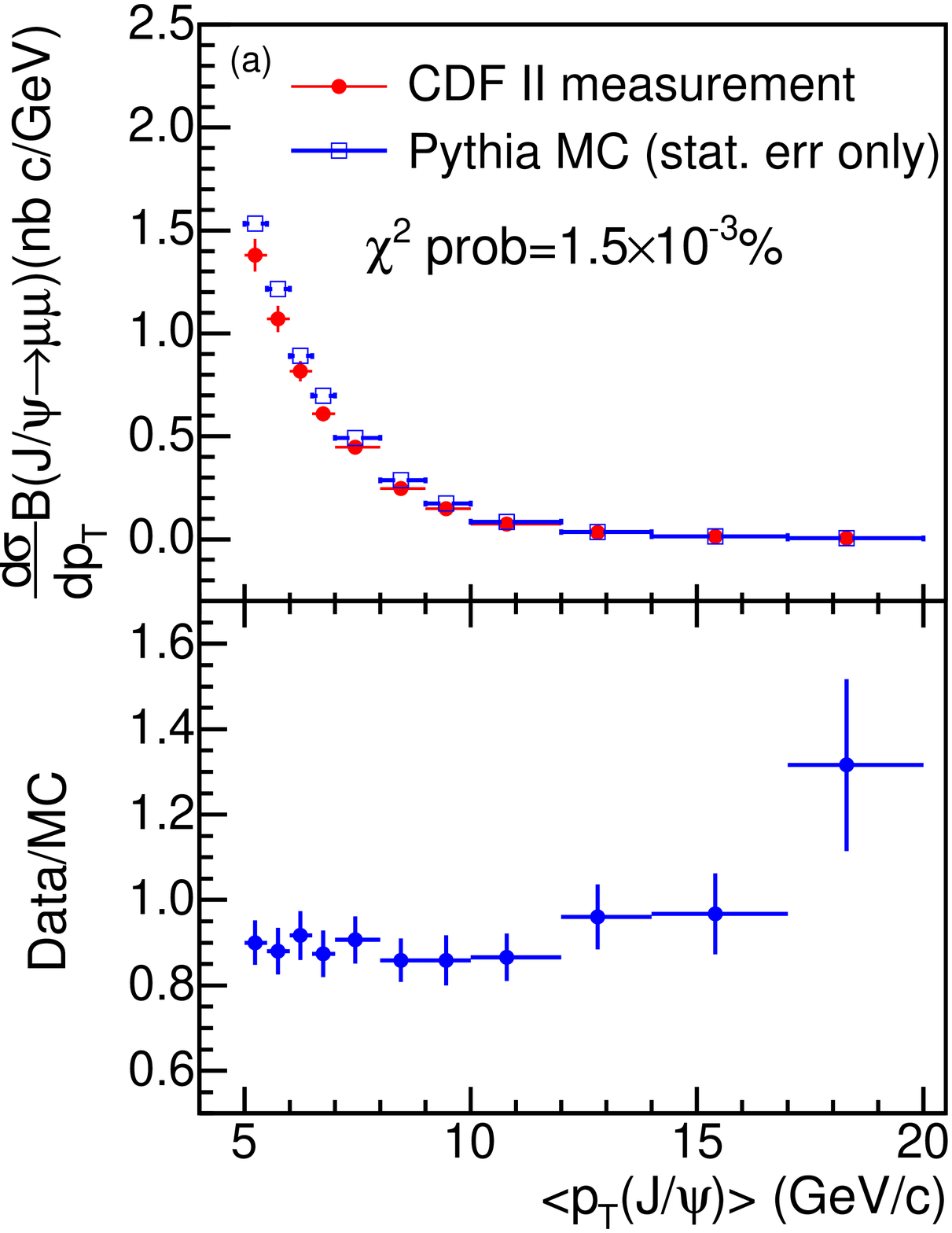}\\
	\includegraphics[width=160pt, angle=0]{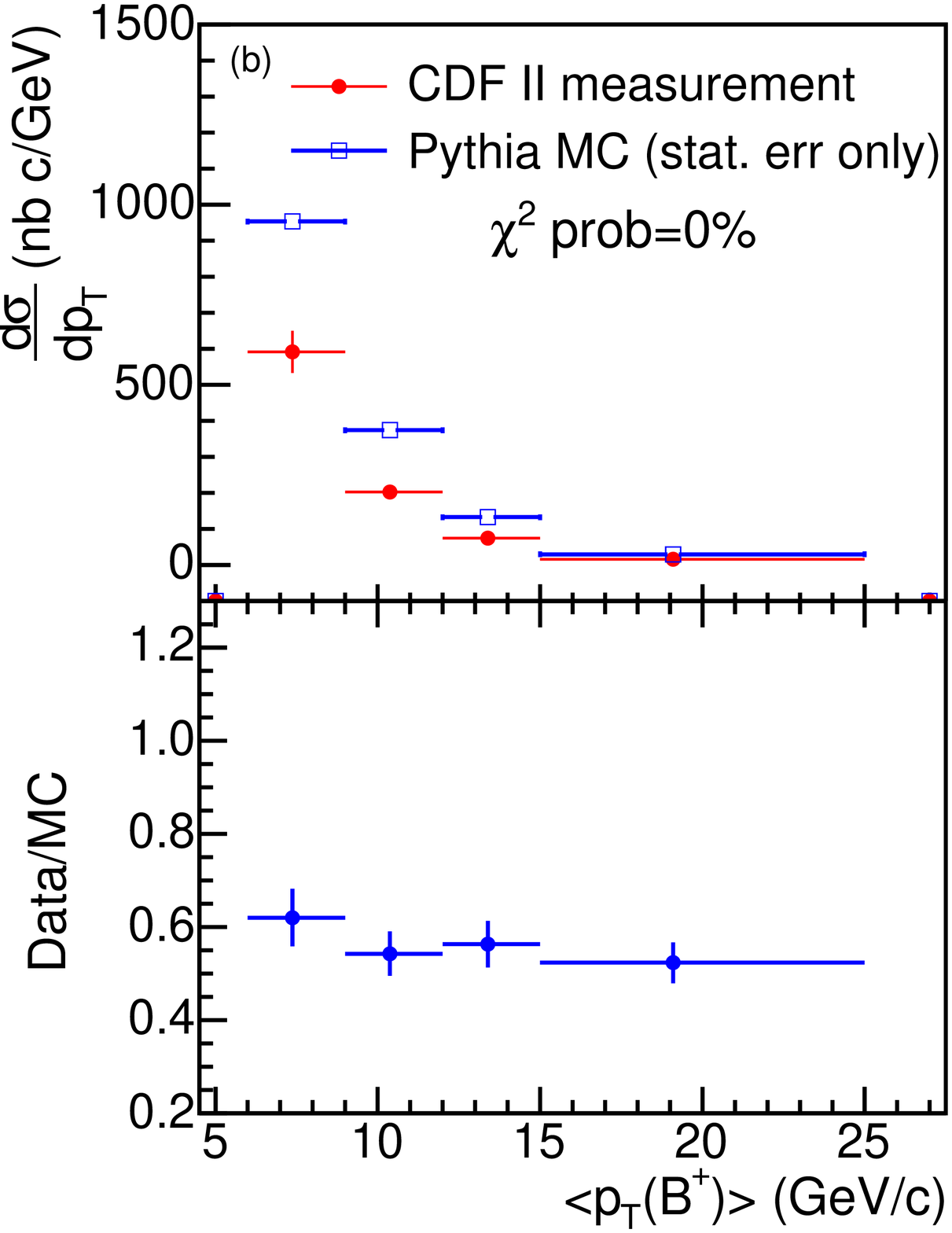}\\
	\includegraphics[width=160pt, angle=0]{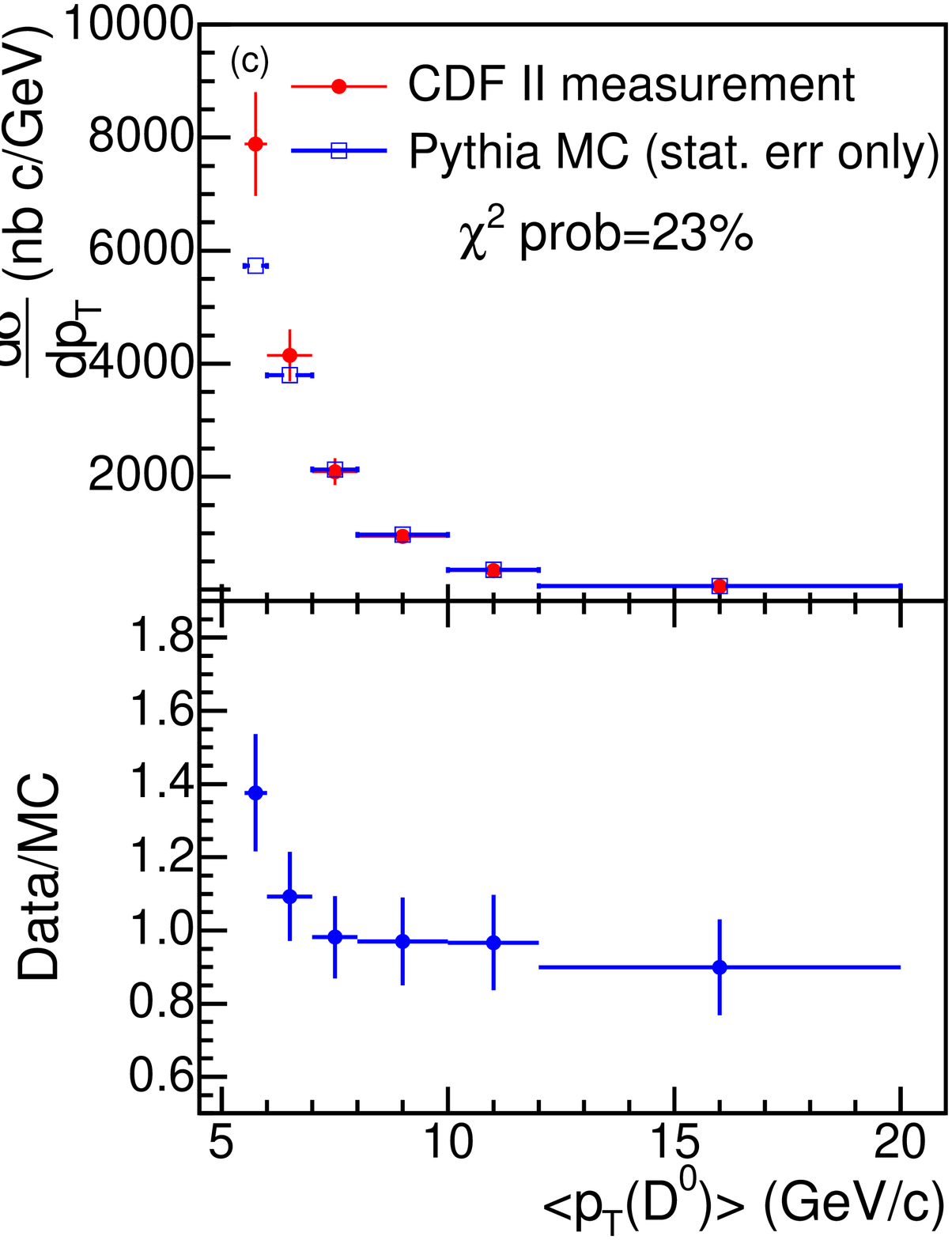}
 \caption
  { The differential cross-sections of (a) inclusive $b$ hadrons, (b) $B^+$, 
	and (c) $\dzero$. The upper plot in each figure shows the 
	differential cross-section for data 
	(closed circles)~\cite{CDF,Abulencia:2006ps,cchen:dxec} 
	and \pythia\ (open squares).
   The lower plot in each figure shows the data to \pythia\ ratio.}
 \label{fig:xseccomp} 
 \end{center}
 \end{figure}

In order to understand how well \pythia\ predicts $\sigma_{b\bar{b}}$ and 
$\sigma_{c\bar{c}}$, a cross-check was performed indirectly by comparing the 
differential cross-sections of inclusive $b$ hadrons, $B^+$, and $\dzero$ 
$\left(d\sigma\left(p\bar{p}\rightarrow \dzero X\right)/d\pt~\it{etc.}\right)$ in 
\pythia\ with the CDF measurements~\cite{CDF,Abulencia:2006ps,cchen:dxec}. 
The differential cross-section of $\dzero$ in \pythia\ (see \fig~\ref{fig:xseccomp}), for instance, is defined as:
\(
   d \sigma\left(p\bar{p}\rightarrow \dzero X\right)^\mathrm{\pythia}/d\pt
	\equiv \sigma_{c\bar{c}}  
	\left(N_{\dzero}/N_\mathrm{gen}\right)/\Delta \pt, 
\)
where $N_{\dzero}$ is the number of $\dzero$ in each \pt\ bin and $N_\mathrm{gen}$ 
is the total number of generated $c\bar{c}$ events. The $\Delta \pt$ 
corresponds to the bin width of each \pt\ bin, which is the same as that 
in~\cite{CDF,Abulencia:2006ps,cchen:dxec}. The discrepancy between the 
\pythia\ and the data cross-sections is generally within 10$\%$ 
for $c$ hadrons and 50$\%$ for $b$ hadrons, which is included in the 
systematic uncertainty in Section~\ref{sec:systematics}.

\subsection{Cross-check using the signed impact parameter distributions \label{sec:forbbcc2}} 
\begin{figure*}[tbp]
\begin{center}
      \includegraphics[width=150pt, angle=0]{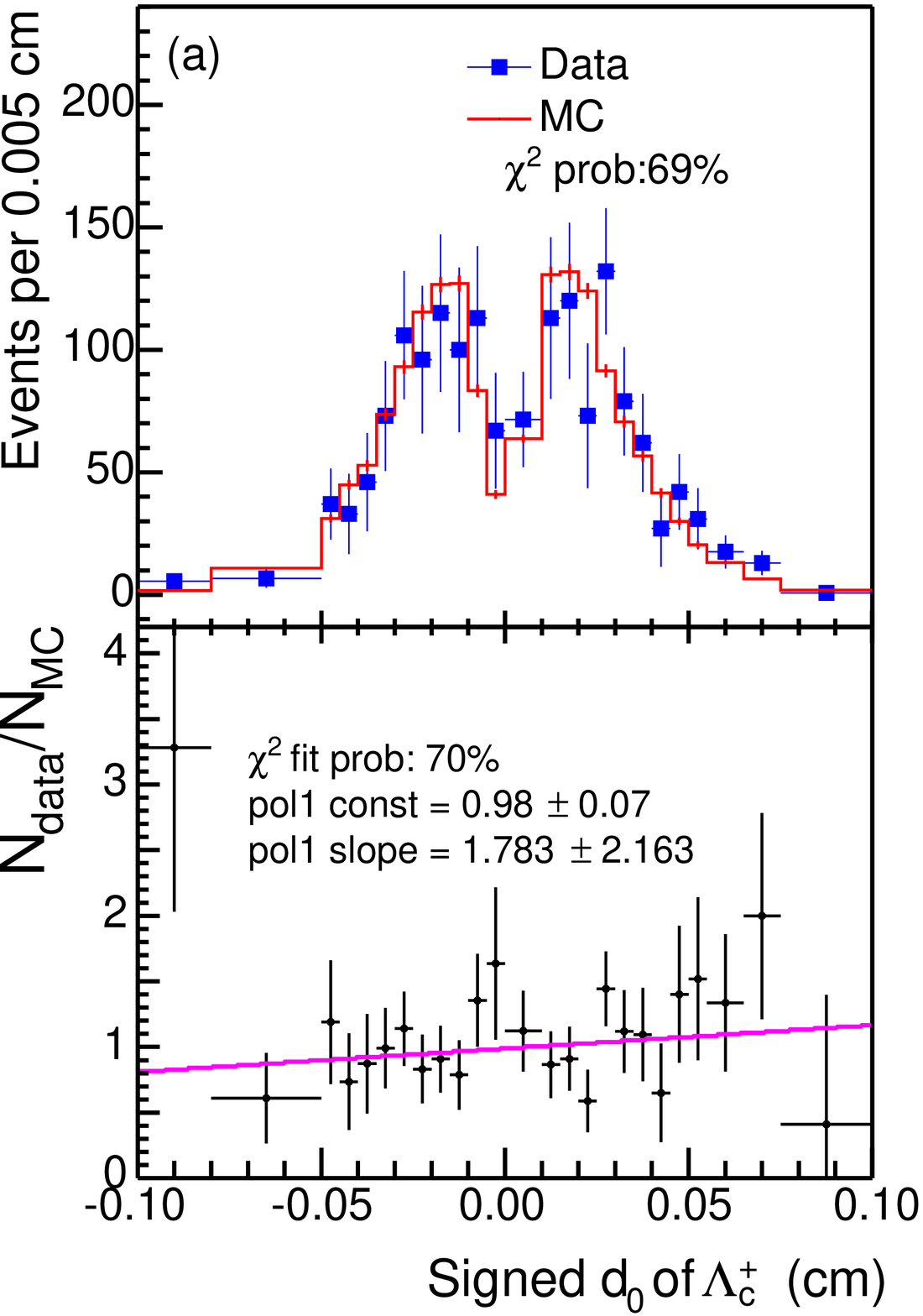}
      \includegraphics[width=150pt, angle=0]{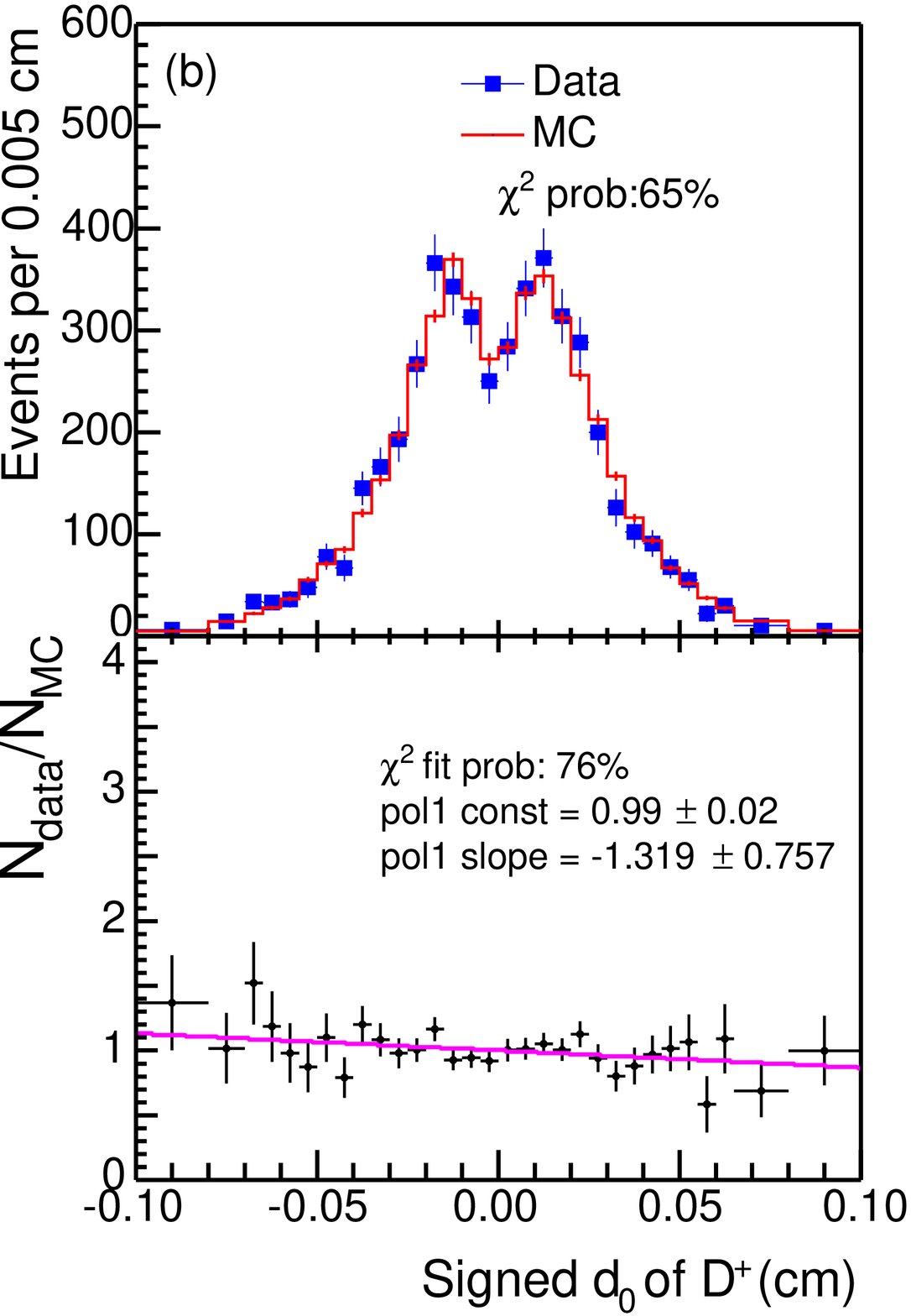}
      \includegraphics[width=150pt, angle=0]{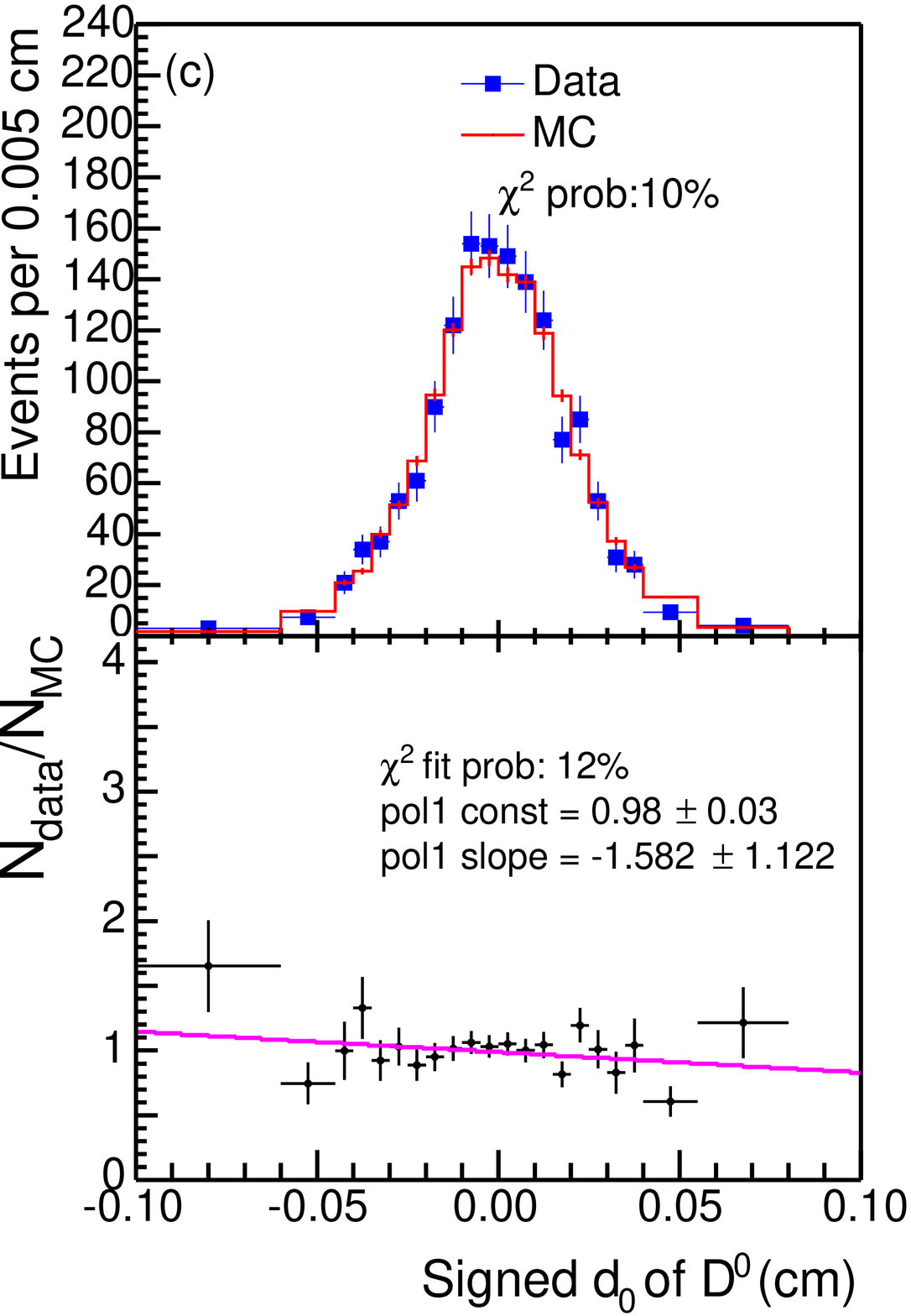}
     \caption
 {Comparison of the signed impact parameters between the full simulation and 
	data for $c$ hadrons which are associated with a $\mu^-$: 
	(a) \Lc, (b) $\dplus$, and (c) $\dzero$ from the $\dstar$. 
	The good agreement of the full simulation with data indicates that 
  backgrounds from the promptly produced \Lc, $\dplus$, and $\dzero$ 
	$\left(c\bar{c}\right)$ are negligible.}
  \label{fig:dimpact}
     \end{center}
  \end{figure*}

As an additional cross-check, the signed impact parameter distributions 
(signed $d_0$) of the \Lc\ baryons with respect to the primary 
vertex, in data and the full simulation, are also compared. 
The signed impact parameter is defined as \(d_0 = Q(r_0-\rho)\),
where $Q$ is the charge of the particle and $r_0$ is the distance between the
beam line and the center of the helix describing the track in the transverse 
plane. The parameter $\rho$ is the radius of the track helix.
The full simulation includes the \lbsemi\ signal and feed-in backgrounds, with 
relative fractions following the estimates in Section~\ref{sec:physicsb}. 
An excess of the signed $d_0$ distribution in the region close to zero would 
indicate a significant contribution of the $c\bar{c}$ background
 in the \inclbsemi\ sample. 
\figc~\ref{fig:dimpact} shows good agreement between data and simulation, 
proving that the promptly produced \Lc\ from $c\bar{c}$ is a negligible 
contribution to the inclusive semileptonic signals. 
\figc~\ref{fig:dimpact} also shows the signed $d_0$ distributions of $\dplus$ and 
$\dzero$.

\begin{thebibliography}{99}
\bibitem{theory}
Predictions of ${\cal B}(\lbsemil)$ and ${\cal B}(\lbhad)$ are both obtained 
by combining the predicted decay widths with the world averages of  
$\left|V_{cb}\right| = (41.2 \pm 1.1)\times 10^{-3}$ and \Lb\ lifetime 
$\tau_{\Lb} = 1.383{+0.049 \atop -0.048}$ ps from the PDG~\cite{pdg}.
\bibitem{coor}
The CDF~II detector uses a cylindrical coordinate system in which $\phi$ is 
the azimuthal angle, $r$ is the radius from the nominal beam line, and $z$ 
points in the proton beam direction, with the origin at the center of 
the detector. The transverse $\rphi$, or $x$-$y$ plane, is the plane 
perpendicular to the $z$ axis.
\bibitem{eta}
The pseudorapidity, $\eta$, is defined as $-\ln\tan(\theta/2)$, where 
$\theta$ is the polar angle to the proton beam.
\bibitem{svtbeamline}
The $z$ position of the beam line at the second trigger level is 
obtained by averaging the detector $z$ positions of all the tracks found by 
the silicon vertex trigger. The $x$ and $y$ positions are obtained by fitting 
the track impact parameter, $d$, with respect to the detector origin, as a 
function of the track azimuthal angle, $\phi$: 
\(d = -x \sin(\phi) + y \cos(\phi)\). The fitting is done for each event.
\bibitem{axialstereo}
  The central outer tracker (COT) has 48 axial and 48 stereo sense wire layers.
\bibitem{SVTtrk}
 A displaced track found by the silicon vertex trigger (SVT) has a 
transverse 
impact parameter within 120~$\mu$m--~1000~$\mu$m and a transverse momentum 
greater than 2 \gevc. 
 The transverse impact parameter is computed at the trigger 
level using information from the silicon vertex detector (SVX~II), while the 
transverse momentum is computed by combining the information reconstructed in 
the SVX~II and in the central outer tracker (COT).
\bibitem{cmuchi2}
The matching $\chi^2$ has one degree of freedom. It is calculated from the 
difference between the track and the muon stub positions in the $\rphi$ plane 
and the expected amount of multiple scattering for a track of given \pt\ obtained from the \geant~\cite{geant} simulation of CDF detector material.
\bibitem{bgen}
  The \bgen\ is a CDF program which generates a single $b$ hadron according to 
  the input \pt\ and rapidity spectra.
\bibitem{ratio}
The bin-by-bin ratios are derived by taking the ratio of generator-level 
$M_{\Lamc\mu}$ histograms from the FF and PHSP simulation samples.
\bibitem{cerror}
The uncertainties on the measured branching fractions of $c$-baryon decays are 
obtained from the PDG~\cite{pdg}. A 5$\%$ uncertainty is assigned to the 
estimated branching fractions of those unmeasured decays. 
Because the excited $c$ baryons decay mainly via the 
isospin-conserving strong interaction, their branching fractions can be 
inferred from Clebsch-Gordan coefficients. An example of the deviation from 
the isospin invariance, in the $c$-meson system, due to additional 
electromagnetic decay, 
$\dstar\rightarrow \dplus \gamma$, is found to be $\approx5\%$ of the 
branching fraction of the $\dstar\rightarrow \dplus\pi^0$ decay.
\bibitem{DELPHI}
The branching fraction of the \lblcpim\ decay was measured to be 5.6$\pm$3.1\% 
by the DELPHI collaboration~\cite{Abdallah:2003gn}. However, given that the 
uncertainty is large, the branching fraction is not used in this analysis.
\bibitem{physbinnew}
  The contribution of $b\bar{b}$/$c\bar{c}$ background may be inferred from 
Appendix~\ref{sec:bbcc}; it is expected to be tiny and may be ignored. The 
false-muon background is considered. However, this background occurs at the 
level of $\approx 4$\%\ in both the numerator and denominator 
and cancels to first order.  If the contribution is ignored in the numerator 
or denominator, a 4\% fractional uncertainty is still small compared with 
the statistical and other systematic uncertainties.
\bibitem{alpha}
 Here, the denominator of kinematic acceptance is the number of events in 
which the $b$ hadrons have $\pt>10$~\gevc\ [$\pt(H_b) > 10$~\gevc]. 
The kinematic requirements include requirements on the $\eta$, \pt, and 
transverse energy of electron, \pt\ of each daughter track of the $c$ hadron, 
significance of the decay length and impact parameter of the $c$ hadron, and 
the four-track invariant mass.
\bibitem{xsecrec}
  The ratio of \Lb\ to \Bd\ efficiencies is found to be independent of 
 \pt\ when the same requirements are made for both 
 $b$ hadrons~\cite{yile:lblcpi}. Therefore, a comparison of the shapes of 
 reconstructed \pt\ distributions is equivalent to a comparison of the shapes 
of differential cross-sections $\left(d\sigma/d\pt\right)$.
\bibitem{whylcpkpi}
 In Ref.~\cite{yile:lblcpi}, the \Lc\ is reconstructed using 
the three-body decay $\Lc \rightarrow pK^-\pi^+$. Therefore, 
the branching fraction ${\cal B}\left(\Lc \rightarrow pK^-\pi^+\right)$
 is required to extract \yile.
\bibitem{whynotlblcpipdg}
The PDG calculates ${\cal B}\left(\lbhad\right)$ 
using the ratio of production cross sections, $\rxsec=0.25\pm 0.04$, with the 
assumption that the \pt\ dependence of the \Lb\ and \Bd\ production is the 
same.  This analysis measures the \Lb\ \pt\ spectrum to be softer 
(more $b$ hadrons at lower \pt) than that of the \Bd\ and estimates the 
\rxsec\ to be 2.5 times larger than that used by the PDG.
\bibitem{uncertainty}
The uncertainties on the fragmentation fraction and the measured $b$-meson 
branching ratios are obtained from the Particle Data Group (PDG)~\cite{pdg}, 
while the uncertainties on the unmeasured $b$-meson branching ratios are 
assumed to be three times the uncertainties on the closest equivalent measured 
decays~\cite{yile:lblcpi}. The unmeasured branching ratios of \Lb\ decays have 
asymmetric systematic uncertainties of $+100\%$, $-50\%$~\cite{yile:lblcpi}.
\bibitem{detail}
  The PDG uncertainty on the \Lb\ relative branching fraction is dominated 
  by the following two sources. The uncertainty from the 
  world average of ${\cal B}\left(\Lc \rightarrow pK^-\pi^+\right)$, 
  which is used in the CDF measurement of \yile~\cite{yile:lblcpi}, is 
  \erlbrlcbr. The 
  uncertainty from the CDF~I measurement of \rxsec\ is \erlbrrxsec.
\bibitem{Bs}
 The value $0.21\pm0.04$ is used and $\Gamma(\Bs)/\Gamma(\Bd)$ is derived 
 assuming $\Gamma(\Bd)=\Gamma(\Bu)$.
\bibitem{support}
Ref.~\cite{Gibson:2006rt} reported $\Gamma(\Bu)/\Gamma(\Bd)=1.054\pm0.018(\rm{stat}){+0.025 \atop -0.045}(\rm{syst})\pm0.058({\cal B})$, which is consistent
 with unity. Ref.~\cite{Gibson:2006rt} also reported $\Gamma(\Bs)/\left[\Gamma(\Bu)+\Gamma(\Bd)\right]=0.160\pm0.005(\rm{stat}){+0.011 \atop -0.010}(\rm{syst}){+0.057 \atop -0.034}({\cal B})$, which is consistent 
with $0.21\pm0.04$.
\bibitem{b0error}
A 5$\%$ uncertainty is assigned to the estimated branching fractions 
	of the excited $c$-hadron decays while a 100$\%$ uncertainty is 
	assigned to the unobserved $b$-hadron decays to cover the wide range 
	of theoretical predictions.
\bibitem{Dzero}
  The $\Xi_b$ contribution could be cross-checked by using the D\O\ measurement of 
$\sigma(\Xi_b){\cal B}(\Xi_b^- \rightarrow J/\psi \Xi^-)/\sigma(\Lb){\cal B}(\Lb \rightarrow J/\psi \Lambda)$.
 Assuming ${\cal B}(\Xi_b^- \rightarrow J/\psi \Xi^-)={\cal B}(\Lb \rightarrow J/\psi \Lambda)$, 
and that the branching fraction and efficiency of $\Xi_b \rightarrow \Lc K \mu^- \bar{\nu}_{\mu}$ are the same as 
those of \dmitridecaythreefour, the contribution of $\Xi_b$ is estimated to be $\approx 0.35\%$.
\bibitem{whygen}
Note that simulations for extracting the $\eff_{\lbhad}$ and the numerator of 
\eq~(\ref{eq:bbcc2}) are performed at the generator level. Generator-level 
simulations are sufficient because this measurement is only sensitive to the 
\eff\ of background relative to the hadronic signal, not the absolute value. 
The results from a generator-level simulation and from a full simulation 
differ by only $\approx3\%$. 
Finally, the \eff\ of the $b\bar{b}$/$c\bar{c}$ background is quite 
small compared to those of the feed-in and false-muon backgrounds. A 
large simulation sample is required to achieve reasonably small 
uncertainties. The full simulations are time- and CPU-intensive 
and is therefore not used for the estimate of the $b\bar{b}$/$c\bar{c}$ 
background.
\bibitem{restrict}
The CDF $\sigma_{B^+}$ measurement~\cite{Abulencia:2006ps} is restricted to the
$B^+$ with \pt\ greater than 6 \gevc\ and rapidity ($y$) less than 1.0. Therefore, 
the denominator of the efficiency $\eff_{\lbhad}$ is the number of events in which 
the \Lb\ baryons have $\pt > 6$ \gevc\ and $|y|< 1.0$.
\end{thebibliography}
\end{document}